\documentclass[12pt]{iopart}
\usepackage{graphicx}
\usepackage{iopams}
\usepackage[colorlinks,citecolor=blue]{hyperref}
\usepackage{amstext}
\begin{document}

\title{Electron capture in stars}

\author{K Langanke$^{1,2}$, G Mart{\'\i}nez-Pinedo$^{1,2,3}$ and
  R.G.T. Zegers$^{4,5,6}$} 
\address{$^1$GSI Helmholtzzentrum f{\"u}r Schwerionenforschung,
D-64291 Darmstadt, Germany}
\address{$^2$Institut f{\"u}r Kernphysik (Theoriezentrum), Department
  of Physics, Technische Universit{\"a}t Darmstadt, D-64298 Darmstadt, Germany}
\address{$^3$Helmholtz Forschungsakademie Hessen f\"ur FAIR, GSI Helmholtzzentrum f\"ur Schwerionenforschung, 
D-64291 Darmstadt, Germany}
\address{$^4$ National Superconducting Cyclotron Laboratory, Michigan State University, East Lansing, Michigan 48824, USA}
\address{$^5$ Joint Institute for Nuclear Astrophysics: Center for the Evolution of the Elements, Michigan State University, East Lansing, Michigan 48824, USA}
\address{$^6$ Department of Physics and Astronomy, Michigan State University, East Lansing, Michigan 48824, USA}

\eads{\mailto{k.langanke@gsi.de}, \mailto{g.martinez@gsi.de},
  \mailto{zegers@nscl.msu.edu}} 

\begin{abstract}
  Electron captures on nuclei play an essential role for the dynamics
  of several astrophysical objects, including core-collapse and
  thermonuclear supernovae, the crust of accreting neutron stars in
  binary systems and the final core evolution of intermediate mass
  stars. In these astrophysical objects, the capture occurs at finite
  temperatures and at densities at which the electrons form a
  degenerate relativistic electron gas.

  The capture rates can be derived in perturbation theory where
  allowed nuclear transitions (Gamow-Teller transitions) dominate,
  except at the higher temperatures achieved in core-collapse
  supernovae where also forbidden transitions contribute significantly
  to the rates. There has been decisive progress in recent years in
  measuring Gamow-Teller (GT) strength distributions using novel
  experimental techniques based on charge-exchange reactions. These
  measurements provide not only data for the GT distributions of
  ground states for many relevant nuclei, but also serve as valuable
  constraints for nuclear models which are needed to derive the
  capture rates for the many nuclei, for which no data exist yet. In
  particular models are needed to evaluate the stellar capture rates
  at finite temperatures, where the capture can also occur on excited
  nuclear states.

  There has also been significant progress in recent years in the
  modelling of stellar capture rates. This has been made possible by
  advances in nuclear many-body models as well as in computer soft-
  and hardware. Specifically to derive reliable capture rates for
  core-collapse supernovae a dedicated strategy has been developed
  based on a hierarchy of nuclear models specifically adapted to the
  abundant nuclei and astrophysically conditions present at the
  various collapse conditions. In particular at the challenging
  conditions where the electron chemical potential and the nuclear $Q$
  values are of the same order, large-scale diagonalization shell
  model calculations have been proven as an appropriate tool to derive
  stellar capture rates, often validated by experimental data. Such
  situations are relevant in the early stage of the core collapse of
  massive stars, for the nucleosynthesis of thermonuclear supernovae
  as well for the final evolution of the core of intermediate-mass
  stars, involving nuclei in the mass range $A \sim 20-65$.

  This manuscript reviews the experimental and theoretical progress
  achieved recently in deriving stellar electron capture rates. It
  also discusses the impact these improved rates have on the various
  astrophysical objects.
\end{abstract}

\submitto{\RPP}
\maketitle

\section{Introduction}
Electron capture is one of the fundamental nuclear processes mediated
by the weak interaction.  In this reaction, a free proton or one bound
inside a nucleus is transformed into a neutron by capture of an
electron producing an electron neutrino. Electron captures on
nuclei play an important role in various dense astrophysical
environments and all three properties which characterize this process
(change of the nuclear charge, reduction of the number of electrons
and energy release by neutrinos) have important consequences in these
astrophysical environments~\cite{Bethe:1990}.  Stellar electron
captures, however, differ significantly from those which can be
studied in the laboratory. In the latter, the decay occurs within an
atom (or ion) by capturing an electron from the atomic cloud where
electrons in tightly bound orbitals are strongly preferred due to
their larger probability density at the nucleus. However, in the
high-density, high-temperature environments of stars the atoms are
strongly (like in our Sun) or completely ionized (in advanced stellar
burning stages or supernovae). Hence stellar capture rates differ from
laboratory rates and are, unfortunately, yet not directly
experimentally accessible and have to be modelled.

Supernovae are arguably the most important astrophysical sites in
which electron captures on nuclei play a decisive role. This includes
the core collapse of massive stars
\cite{Bethe.Brown.ea:1979,Fuller.Fowler.Newman:1982a,Janka.Langanke.ea:2007},
the final evolution of the ONeMg cores in intermediate-mass
stars~\cite{Doherty.Gil-Pons.ea:2017,Nomoto.Leung:2017b}, the crust
evolution of neutron stars in binaries
\cite{Haensel.Zdunik:1990,Schatz.Gupta.ea:2014} as well as the
nucleosynthesis occurring in thermonuclear (Type 1a) supernovae
\cite{Iwamoto.Brachwitz.ea:1999,Brachwitz.Dean.ea:2000}. In all
scenarios the densities at which electron captures play a role are in
excess of about $10^9$~g~cm$^{-3}$ and at finite temperatures which
range from $10^8$ K in electron-capture supernovae to above
$10^{10}$~K which are encountered in the advanced core collapse of
massive stars. Under these conditions electrons are characterized by a
relativistic Fermi gas with Fermi energies of MeV to tens of MeV. As a
consequence, electron captures can occur under these conditions also
on nuclei which, under laboratory conditions, are
stable~\cite{Bethe.Brown.ea:1979}.

At the relatively low electron energies the capture is dominated by
allowed Gamow-Teller (GT) transitions, with forbidden transitions
contributing at the higher densities/temperatures or in exceptional
cases~\cite{Bethe.Brown.ea:1979,Fuller.Fowler.Newman:1982a,Cooperstein.Wambach:1984,Langanke.Martinez-Pinedo:2000,Juodagalvis.Langanke.ea:2010}.
This observation has been the basis to recognize the importance of
electron captures in core-collapse supernova, but also of the decisive
progress that has been achieved in recent years to derive reliable
stellar capture rates. The pioneering work of Bethe, Brown, Applegate,
and Lattimer~\cite{Bethe.Brown.ea:1979} derived capture rates on
the basis of a single GT transition transforming an $f_{7/2}$ proton
into an $f_{5/2}$ neutron. This assumption was motivated by the
Independent Particle Model (IPM) structure of $^{56}$Fe which is quite
abundant during the early collapse phase. The important insight into
the collapse dynamics drawn in their pioneering work was that electron
capture is a very efficient cooling mechanism and that the entropy
stays low during the entire collapse.  As a consequence the
composition of the core is predominantly made by heavy nuclei rather
than being dissociated into free nucleons. The challenge of deriving
an improved set of stellar capture rates was taken up by Fuller,
Fowler and Newman who, in a series of papers, outlined the formalism
to determine stellar capture rates and applied it to calculate rate
tables for nuclei in the mass range $A=21-60$ at appropriate
temperature and density conditions in the core
\cite{Fuller.Fowler.Newman:1980,Fuller.Fowler.Newman:1982a,Fuller.Fowler.Newman:1982b,Fuller.Fowler.Newman:1985}. These
derivations were based again on the IPM, but considered experimental
data wherever available. Fuller noticed that, within the IPM, GT
transitions from $pf$ proton orbitals are Pauli blocked for nuclei
with $N \geq 40$ for which the $pf$ shell for neutrons is completely
filled \cite{Fuller:1982}.  Based on this observation Bruenn derived a
parametric description for stellar electron capture rates which
assumed vanishing capture rates for all nuclei with neutron numbers
$N>40$ \cite{Bruenn:1985}. Although Cooperstein and Wambach pointed
out that the Pauli blocking could be overcome at high temperatures and
by forbidden transitions \cite{Cooperstein.Wambach:1984}, the Bruenn
prescription has been the default for electron captures in supernova
simulations until the early 2000s (e.g. \cite{Bethe:1990}). On this
basis, simulations predicted that during the advanced collapse phase
for densities in excess of $10^{10}$~g~cm$^{-3}$ electron capture
proceeds on free protons rather than on nuclei. As free protons are
significantly less abundant than nuclei during the collapse, electron
capture and the associated core cooling was drastically throttled once
capture on nuclei was blocked.

During the last two decades the role played by electron captures for
the supernova dynamics has been decisively revised. This was made
possible by new theoretical insights, improved models and not the
least by the development of novel experimental techniques to determine
nuclear GT strength distributions. This breakthrough was made possible
by the observation that strongly forward-peaked cross sections in
charge-exchange reactions, mediated by the strong interaction, are
dominated by the spin-isospin operator needed to derive
weak-interaction GT transitions
\cite{Gaarde.Rapaport.ea:1981,Osterfeld:1992}.  The pioneering GT
measurements were performed at TRIUMF using the $(n,p)$ charge-exchange
reaction
\cite{Vetterli.Haeusser.ea:1990,Vetterli.Jackson.ea:1992,Helmer.Punyasena.ea:1997}.
Despite of its moderate energy resolution of about an~MeV, these
measurements clearly showed that the nuclear GT strength is
significantly more fragmented and also reduced in total strength
compared to the predictions of the IPM. These findings were
subsequently confirmed by measurements performed at KVI, Groningen
using the $(d,{}^2\textrm{He})$~\cite{frekers:2006,martinez-pinedo.liebendoerfer.frekers:2006,Frekers.Alanssari:2018} 
and at NSCL, Michigan State University by exploiting $(t,{}^3\textrm{He})$
charge-exchange reactions, respectively \cite{zegers:2006}.  With both
techniques, experimenters were successful to measure GT strength
distributions for many $pf$ shell nuclei with an energy resolution
nearly an order of magnitude better than being possible in the
pioneering TRIUMF experiments.  These measurements became an
indispensable constraint for nuclear models, which were developed in
parallel to the experimental progress.

Due to the strong energy dependence of phase space electron capture
rates are quite sensitive to the detailed fragmentation of the GT
strength if the Fermi energies of the electron reservoir and the
nuclear $Q$ value are of the same magnitude
\cite{Langanke.Martinez-Pinedo:2000,Langanke.Martinez-Pinedo:2003}. This
is the case during hydrostatic silicon burning and at the onset of the
collapse at core densities up to about $10^{10}$~g~cm$^{-3}$. Under
these conditions the core consists mainly of nuclei in the Fe-Ni mass
range, while $sd$ shell nuclei are also present during silicon burning
\cite{Heger.Woosley.ea:2001}. The method of choice to describe the
properties of these nuclei is the interacting shell model
\cite{Caurier.Martinez-Pinedo.ea:2005}. Due to advances in
computational capabilities and progress in software and an improved
understanding of the decisive ingredients of the residual interaction,
diagonalization shell model calculations became possible for the
complete $sd$ shell and for $pf$ shell nuclei at a truncation level
that guaranteed sufficiently converged results for the nuclear
quantities needed to derive reliable electron capture rates. This in
particular includes detailed description of the GT strength
distributions which, except for a slightly shell-dependent constant
factor, reproduced the total GT strength and its fragmentation quite
well \cite{Brown.Wildenthal:1988,Caurier.Langanke.ea:1999}. This
success was first used by Oda and collaborators to derive shell model
electron capture rates for $sd$ shell nuclei
\cite{Oda.Hino.ea:1994}. This was followed by the calculation of
individual capture rates for nuclei in the mass range $A=45$--65 based
on GT strength distributions derived in large-scale shell model
calculations
\cite{Langanke.Martinez-Pinedo:2000,Langanke.Martinez-Pinedo:2001}. Due
to the finite temperature of the astrophysical environment the shell
model calculations also include GT transitions occurring from thermally
excited nuclear states.  The shell model rates became the new
standards in supernova simulations for intermediate-mass nuclei. It
turned out to be quite relevant that the shell model rates for $pf$
shell nuclei are systematically and significantly smaller compared to
the prior rates based on the IPM
\cite{Langanke.Martinez-Pinedo:2000}. As a consequence, simulations
with the shell-model rates showed a noticeably slower deleptonization
and resulted in different Fe-core masses at the end of the
presupernova phase when the collapse sets
in~\cite{Heger.Woosley.ea:2001,Heger.Langanke.ea:2001}.

The Pauli blocking of the GT strength at the $N=40$ shell closure
exists in the IPM \cite{Fuller:1982}, but can be overcome by
correlations which move protons or neutrons into the next major shell
(the $sdg$ shell) \cite{Langanke.Kolbe.Dean:2001}.  To describe such
cross-shell correlations within the diagonalization shell model
requires usually model spaces with dimensions which are not feasible
with today's computers. However, such studies exist for $^{76}$Se (the
intermediate nucleus in the double-beta decay of $^{76}$Ge) showing
that its GT strength is small, but non-vanishing, even for the ground
state \cite{Zhi.Caurier.ea:2013}. This finding is in good agreement
with the experimental determination of the GT strength by the
$(d,{}^2$He) technique \cite{Grewe.Baeumer.ea:2008}. As a consequence
the stellar electron capture rate on $^{76}$Se is sizable, showing
that the assumption of neglecting the capture on nuclei with $N>40$ is
not justified.  To derive at the stellar capture rate for such nuclei
a hybrid model had already been proposed and applied prior to the
shell model studies of $^{76}$Se. This model is based on two steps
\cite{Langanke.Kolbe.Dean:2001,Langanke.Martinez-Pinedo.ea:2003}: At
first, the crucial cross-shell correlations are studied using the
Shell Model Monte Carlo approach
\cite{Johnson.Koonin.ea:1992,Koonin.Dean.Langanke:1997}, which is a
stochastical approach to the shell model allowing to calculate nuclear
properties at finite temperature considering correlations in
unprecedentedly large model spaces. These calculations have been
applied to determine partial occupation numbers for protons and
neutrons in the combined $pf$-$sdg$ shells and at finite temperature.
In the second step, these partial occupation numbers served as input
in RPA calculations of the GT and forbidden strength distributions and
subsequently the stellar capture rates. The use of the RPA for these
nuclei is justified as they dominate the core abundance only at higher
densities and temperatures where the Fermi energy of the electron gas
is noticeable larger than the $Q$ value of the respective nuclei
requiring only a reasonable reproduction of the total strength and its
centroid for a reasonable estimate of the rate. The hybrid model has
been applied to about 200 nuclei in the mass range $A=65-110$
\cite{Langanke.Martinez-Pinedo:2003}.  The studies clearly implied
that Pauli blocking of the GT strength is overcome by cross-shell
correlations at the temperature/density conditions at which these
nuclei are abundant
\cite{Langanke.Martinez-Pinedo:2003,Hix.Messer.ea:2003}. The SMMC
calculations also yield rather smooth trends in the partial occupation
numbers at the relevant temperature of about 1~MeV. Based on
observation a simple parametrization of the occupation numbers has
been derived which was the basis of RPA calculations of stellar
capture rates for another 2700 nuclei~\cite{Juodagalvis.Langanke.ea:2010}.

On the basis of the shell model calculations for $sd$ and $pf$ shell
nuclei, of the hybrid model for cross-shell $N=40$ nuclei and the
parametric study for the heavier nuclei an electron capture rate table
has been derived for core-collapse
conditions~\cite{Juodagalvis.Langanke.ea:2010}. The nuclear
composition of the core has been assumed to be in nuclear statistical
equilibrium (NSE)~\cite{hix.thielemann:1996}. When incorporated into
supernova simulations these rates had significant consequences for the
collapse dynamics.  In particular, the simulations show that capture
on nuclei dominate over capture on free protons during the entire
collapse. Furthermore, the dominating capture on nuclei leads to a
stronger deleptonization of the core and to smaller temperatures and
entropies, in comparison to the previous belief that capture on nuclei
would vanish due to Pauli
blocking~\cite{Langanke.Martinez-Pinedo:2003,Hix.Messer.ea:2003}.

As an alternative to the hybrid model, the temperature-dependent
Quasiparticle RPA model has been developed and applied to stellar
electron capture for selected nuclei by Dzhioev and
coworkers~\cite{Dzhioev.Vdovin.ea:2010}. This approach formally
improves the hybrid model as it describes correlations and strength
function calculations consistently within the same framework. In
contrast to the hybrid model it restricts correlations to the 2p-2h
level which due to the diagonalization shell model studies is not
completely sufficient to recover all cross-shell correlations. This
shortcoming is relevant for ground state strength functions, but gets
diminished with increasing temperatures. Satisfyingly both quite
different approaches yield similar capture rates in the
density/temperature regimes where nuclei with neutron gaps at $N=40$
and $N=50$ matter during the collapse~\cite{Dzhioev.Langanke.ea:2020}.

Electron capture also plays a role for the final fate of the O-Ne-Mg
cores of intermediate-mass
stars~\cite{Nomoto:1987,Takahara.Hino.ea:1989} and for the
nucleosynthesis occurring behind the burning flame during a
thermonuclear supernova~\cite{Parikh.Jose.ea:2013,Bravo:2019}.  In
these scenarios only $sd$- and $pf$-shell nuclei are relevant and
hence the respective diagonalization shell model rates can be
applied. For the dynamics of the O-Ne-Mg cores, however, also beta
decays are quite decisive for selected nuclei. The relevant rates can
also be calculated quite reliable within the shell model
(e.g. \cite{Oda.Hino.ea:1994,Suzuki.Toki.Nomoto:2016}. It has been pointed out
that the electron capture on $^{20}$Ne constitutes a very unusual case
as its rate is dominated by a second-forbidden
ground-state-ground-state transition in the relevant
density/temperature regime~\cite{Martinez-Pinedo.Lam.ea:2014}. As an
experimental milestone this transition has very recently been
experimentally determined with quite considerable consequences for the
fate of intermediate-mass stars~\cite{Kirsebom.Jones.ea:2019}.

Electron captures on selected nuclei play also a role during
hydrostatic stellar burning or during s-process nucleosynthesis. In
these environments, ions are not completely stripped from electrons so
that the capture predominantly occurs from bound (K-shell) electrons
(however, modified by screening from the surrounding plasma), but also
from `free' electrons out of the plasma. The description of these
capture processes requires a different treatment as described here. We
will not review these capture processes in this manuscript, but list a
few relevant references for the interested readers.  An important
example for capture during hydrostatic burning is the one on $^7$Be
which is a source of high-energy solar neutrinos. The respective solar
rate is derived in
\cite{Bahcall:1962,Bahcall.Moeller:1969,Johnson.Kolbe.ea:1992,Adelberger.Austin.ea:1998,Adelberger.Garcia.ea:2011}. Electron
  capture on $^7$Be is also important in evolved stars as it affects
  the abundance of $^7$Li in red giant branch and asymptotic giant
  branch stars~\cite{Simonucci.Taioli.ea:2013}.  During s-process
nucleosynthesis certain pairs of nuclei (like $^{187}$Rh-$^{187}$Os,
$^{205}$Tl-$^{205}$Pb) serve as potential
cosmochronometers~\cite{Clayton:1968}. These pairs are characterized
by very small $Q$ values against electron captures so that, in the
inverse direction, $\beta$ decay with an electron bound in the ionic
K-shell (or higher shells) becomes possible and even dominates the
decay.  Such bound-state $\beta$ decay strongly depends on the degree
of ionization and of corrections due to plasma screening, while the
competing electron capture process is often modified by contributions
due to thermally excited nuclear levels. The formalism to describe the
relevant electron capture, $\beta$ and bound-state $\beta$-decay rates
for the appropriate s-process temperature and density conditions is
derived in~\cite{Takahashi.Yokoi:1983}; detailed rate tables can be
found in \cite{Takahashi.Yokoi:1987}.  Application of these rates in
s-process simulations are discussed
in~\cite{Yokoi.Takahashi.Arnould:1985}.  For reviews of s-process
nucleosynthesis the reader is referred
to~\cite{Kappeler.Beer.Wisshak:1989,Kaeppeler.Gallino.ea:2011}.

In this review we will summarize the theoretical and experimental
progress achieved during the last two decades in describing stellar
electron captures on nuclei.  Section 2 is devoted to the experimental
advances describing the various techniques to measure Gamow-Teller
strength distributions.  Section 3 starts with some general remarks
defining the strategy how to derive the rates at the relevant
conditions, followed by some brief discussions of the adopted models
and the rates derived within these approaches.In Section 4 we
summarize the consequences of modern electron capture rates in
core-collapse supernovae, for the fate of O-Ne-Mg cores in
intermediate-mass stars and for the nucleosynthesis in thermonuclear
supernovae.

\section{Experimental techniques and progress}
\label{experiment}

To accurately estimate electron-capture rates on nuclei present in
stellar environments, it is key to have accurate Gamow-Teller strength
distributions from which the electron-capture rates can be
derived. Direct information about the Gamow-Teller strength
distribution can in principle be obtained from $\beta$-decay and
electron-capture measurements, but this provides only information
about transition strengths between ground states and a limited number
of final states. Moreover, since in most astrophysical phenomena
electron captures near the valley of stability and/or on neutron-rich
isotopes are most important, the $Q$ value for $\beta^+$/EC decay is
often negative and direct information is available only on the
Gamow-Teller transition strength from the ground state of the mother
to the ground state of the daughter that is derived from $\beta^-$
decay in the inverse direction, and only if the ground-state to
ground-state decay is associated with a Gamow-Teller
transition. Therefore, an indirect method is needed to gain
information about Gamow-Teller strength distributions and to benchmark
and guide the development of theoretical models for Gamow-Teller
strengths. Charge-exchange reactions~\cite{Osterfeld:1992,HAR01,Ichimura:2006,Fujita2011549,Frekers.Alanssari:2018} at
intermediate beam energies ($E_{b}\gtrsim 100$~MeV) have served as
that indirect method, as it is possible to extract the Gamow-Teller
strength distribution up to sufficiently high excitation energies to
perform detailed assessments of the validity of the theoretical models
employed. The remarkable feature of this method is that detailed
information about transitions mediated by the weak nuclear force can
be extracted from reactions with hadronic probes mediated by the
strong nuclear force. The methods and associated experimental
techniques are described in this section.    

It is important to note that only a limited number of charge-exchange
experiments can be carried out and that these experiments only provide
data on transitions from the ground state of the mother nucleus. Since
in many astrophysical scenarios a relatively large number of nuclei
play a significant role and, if the stellar environment is at high
temperature, transitions from excited states also play a role, it is
rarely possible to rely on experimental data of Gamow-Teller strength
distributions only. To make accurate estimates for electron-capture
rates in stars, theoretical nuclear models are necessary, which can be
tested against charge-exchange data where available. Another important
consideration is that electron captures in stars take place on stable
and unstable nuclei. Hence, the obtain information about the latter,
charge-exchange experiments with unstable nuclei are needed. As
described below, such experiments are challenging and relevant
techniques for performing charge-exchange experiments in inverse
kinematics are still in development, although good progress have been
made over the past decade.

For the purpose of extracting Gamow-Teller strength distribution of
relevance for electron captures in stars, charge-exchange experiments
in the $\beta^{+}/EC$ direction or $(n,p)$ direction are necessary and
the primary focus in this section. These experiments probe
proton-hole, neutron-particle excitations. However, charge-exchange
data in the $\beta^{-}$ or $(p,n)$ direction (neutron-hole,
proton-particle excitations) are important as well. Firstly, the
development of the techniques to extract Gamow-Teller strengths has
been primarily developed by using charge-exchange reaction in the
$\beta^{-}$ direction, starting with the pioneering work by
\cite{Taddeucci1987125}. Many detailed studies have been performed by
using the ($^{3}$He,$t$) reaction. benefiting in part from the fact
that for mirror nuclei the $\beta^+$ decay of the neutron-deficient
nucleus and the $(p,n)$-type reaction on the mirror neutron-rich
nucleus populate states with the same isospin. This allows for
detailed comparisons of Gamow-Teller strengths through $\beta$ decay
and charge-exchange reactions \cite{Fujita.Rubio.Gelletly:2011}.

Secondly, for certain astrophysical phenomena,
detailed information about the Gamow-Teller strengths in the
$\beta^{-}$ direction are needed. Thirdly, by assuming
isospin-symmetry, information about Gamow-Teller strengths in the
$\beta^{+}/EC$ direction can sometimes be derived from data in the
$\beta^{-}$ direction. Finally, the theoretical models used to
calculate Gamow-Teller strength distributions in the $\beta^{+}$
direction usually rely on the same parameters of the nuclear
interaction as those calculated in the $\beta^{-}$ direction. Hence,
by comparing the results of models against data from charge-exchange
experiments in the $\beta^{-}$ direction, additional information about
the strengths and weaknesses of those models is obtained. The summed
Gamow-Teller strengths in the $\beta^{+}$ and $\beta^{-}$ directions
are connected through a sumrule, first developed by Ikeda, Fujii and
Fujita \cite{Ikeda1963271}:
\begin{equation}
\label{eq:sumrule}
S_{\beta^{-}}(GT)-S_{\beta^{+}}(GT)=3(N-Z)
\end{equation}
Although experimentally, only about 50--60\% of the sum-rule strength
is observed in the Gamow-Teller resonance at excitation energies
below $\sim 20$~MeV \cite{GAA81,GAA85}, referred to as the
``quenching'' phenomenon~\cite{Osterfeld:1992}, it allows one
to obtain information about the strength in the $\beta^{+}/EC$
direction from the measurement in the $\beta^{-}$ direction. However,
as the electron-capture rates that are derived from the Gamow-Teller
strength strongly depends on the strength distribution and not just
the magnitude of the strength, measurement of the strength in the
$\beta^{-}$ direction is of limited use for detailed evaluations of
the strength distribution. This is especially true for nuclei with
increasing neutron number for fixed atomic number as $S_{\beta^{-}}$
becomes increasingly larger than $S_{\beta^{+}}$. On the other hand,
the Ikeda sum rule is a very useful constraint for the total GT
strength for the cross section calculation of charged-current
$(\nu_e,e^-)$ reactions for neutron-rich nuclei
\cite{Langanke.Kolbe:2001,Balasi.Langanke.Martinez-Pinedo:2015}.

\subsection{The extraction of Gamow-Teller strengths from
  charge-exchange data} 
\label{}  
The extraction of the Gamow-Teller strength distribution from
charge-exchange reaction data at intermediate beam energies is based
on the proportionality between the Gamow-Teller transition strength
$B$(GT) for small linear momentum transfer, $q\approx 0$, expressed
through the following relationship \cite{Taddeucci1987125}:
\begin{equation}
\label{eq:dsigma}
\left[\frac{d \sigma}{d \Omega}(q,\omega)\right]_{GT} =
F(q,\omega)\hat{\sigma}B(\textrm{GT}), 
\end{equation}
in which $\frac{d \sigma}{d \Omega}(q,\omega)$ is the measured
differential cross section for a transition associated with energy transfer $\omega=Q_{gs}-E_{x}$ and linear momentum transfer $q$. 
$Q_{gs}$ is the ground-state reaction $Q$ value that is negative for a transition that requires energy. $E_{x}$ is
the excitation energy of the final nucleus. $B$(GT) is the
Gamow-Teller transition strength and represents the same matrix
elements as probed in $\beta$ and EC decay transitions between the
same initial and final states. The condition that $q=0$ requires that
the cross section is extracted at or close to a center-of-mass
scattering angle of zero degrees and that an extrapolation is required
based on a calculation to correct for the finite reaction $Q$
value. This extrapolation is represented by the factor
$F(q,\omega)$. The factor $\hat{\sigma}$ is the so-called unit cross
section, which depends on the reaction kinematics, the nuclei involved
in the interaction and the properties of the nucleon-nucleon ($NN$)
interaction. In the Eikonal approximation \cite{Taddeucci1987125},
these components are factorized:
\begin{equation}
    \label{eq:unit}
    \hat{\sigma}=K N |J_{\sigma\tau}^{2}|.
\end{equation}
In this factorization, $K$ is a calculable kinematic factor, $N$ is a
distortion factor, and $J_{\sigma\tau}$ is the volume integral of the
spin-transfer, isospin-transfer $\sigma\tau$ component of the $NN$
interaction \cite{LOV81}. The distortion factor accounts for the
distortion of the incoming (outgoing) particle by the mean-field of
the target (residual) nucleus and can be estimated by taking the ratio
of a distorted-wave impulse or Born approximation calculation to a
plane-wave calculation \cite{Taddeucci1987125}. The strength of the
method to extract Gamow-Teller distributions from charge-exchange data
is that the details of the components that make up the unit cross
section do not need to be known since the unit cross section is
conveniently calibrated by using transitions for which the $B$(GT) is
known from $\beta$-decay experiments. Once established for one or a
few transitions for given nucleus and charge-exchange reaction, the
proportionality can be applied to all transitions identified as being
associated with $\Delta L=0$ and $\Delta S=1$, except for the
extrapolation to $q=0$ through the factor $F(q,\omega)$ of
Eq. \ref{eq:dsigma}. This extrapolation carries a relatively small
uncertainty. Calibrations against known transitions from $\beta$ decay
are not always possible. Therefore, mass-dependent parametrizations
of the unit cross sections have been successfully developed for the
$(p,n)$/$(n,p)$~\cite{Sasano:2009} and
$(^{3}\textrm{He},t)$/$(t,{}^{3}\textrm{He})$~\cite{zegers07,Perdikakis:2011}
reactions, which provide a convenient way to extract Gamow-Teller
strength distributions for such cases.

\begin{figure}[htbp]
  \centering
  \includegraphics[width=0.7\linewidth]{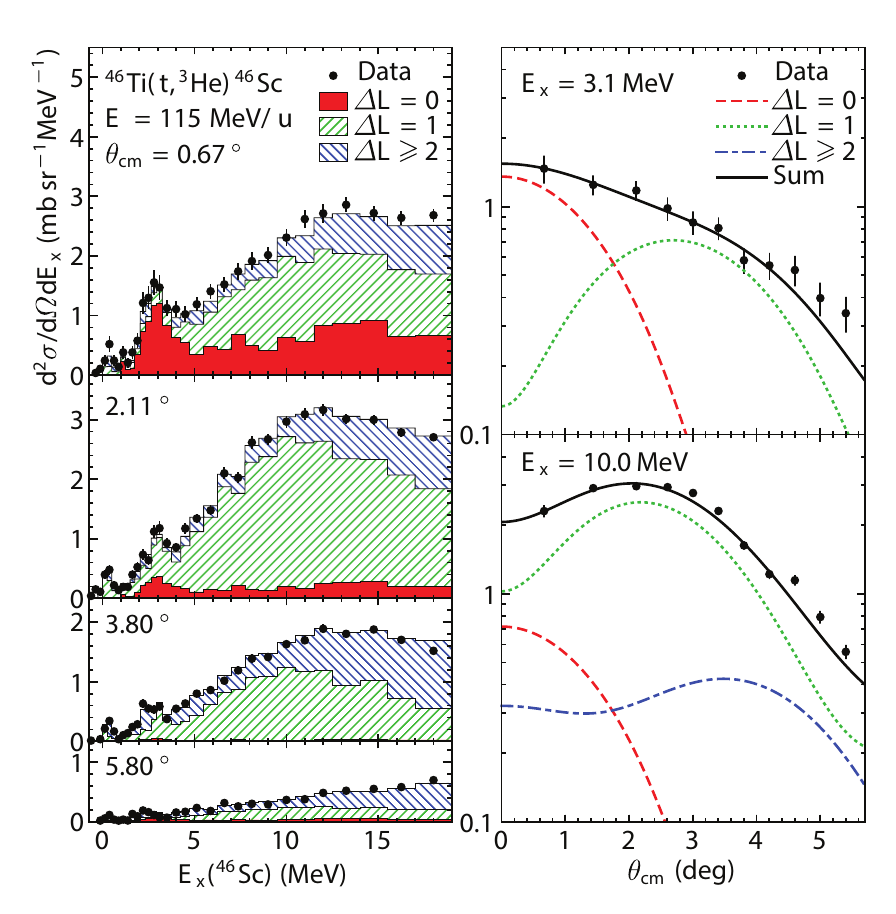}
  \caption{An example of the MDA for the $^{46}$Ti$(t,{}^{3}\textrm{He})$
    reaction at 115 MeV/$u$. On the left-hand side, differential cross
    sections at 4 scattering angles are shown. On the right-hand side,
    the MDA analyses for two excitation energy bins (at 3.1 MeV and 10
    MeV) in $^{46}$Sc are shown. At 3.1 MeV (10.0 MeV), the
    $\Delta L=0$ ($\Delta L=1$) contribution is strongest. The stacked
    colored histogram on the left-hand side indicate the contributions
    from the different angular momentum transfers based on the MDA.
    (from Ref. \cite{PhysRevLett.112.252501}).}
\label{fig:mda}
\end{figure}

In order to use Eq. \ref{eq:dsigma} and extract the Gamow-Teller
strength distribution from measured differential cross sections, one
must first identify the contributions to the experimental spectra that
are associated with monopole ($\Delta L=0$) and spin-transfer
($\Delta S=1$). This is performed by investigating the differential
cross sections as a function of scattering angle, since excitations
that are associated with increasing units of angular momentum
transfers have angular distributions that peak at larger scattering
angle. Therefore, through a process called multipole decomposition
analysis (MDA)~\cite{Moinester:1989}, in which the measured
differential cross sections of a particular peak or data in an
excitation-energy bin is fitted by a linear combination of calculated
angular distributions for different units of $\Delta L$, the
$\Delta L=0$ contribution to the cross section is extracted. An
example for the $^{46}$Ti$(t,{}^{3}\textrm{He})$ reaction is shown in Fig.
\ref{fig:mda}. Since the $\Delta L=0$, $\Delta S=0$ contribution is
almost completely associated with the excitation of the isobaric
analog state, it does not contribute to the $\Delta L=0$ yield for
$(n,p)$-type reactions for nuclei with $N\geq Z$, as the isospin of
states in the final nucleus always exceed that of the target. The
Fermi sum rule of $S_{-}-S_{+}=(N-Z)$ is nearly fully exhausted by the
excitation of the isobaric analog state in the $\beta^{-}$ $(p,n)$
direction.

For $(n,p)$-type reactions, at excitation energies $\gtrsim 10$ MeV,
contributions to the $\Delta L=0$ yield arise from the excitation of
the isovector giant monopole resonances (IVGMR) and isovector spin
giant monopole resonance (IVSGMR) \cite{HAR01}. In charge-exchange
reactions with beam energies of $\gtrsim 100$ MeV, the IVSGMR
dominates. Since the angular distribution of the IVSGMR is very
similar to that of Gamow-Teller excitations, the two are not easily
separable experimentally. Only through a comparison between $(n,p)$
and $(t,{}^{3}\textrm{He})$ data it is possible to disentangle the two
contributions \cite{PhysRevLett.108.262503}. Since the transition
density for the IV(S)GMR has a node near the nuclear surface, a
cancellation occurs for the $(n,p)$ probe that penetrates relatively
deeply into the nuclear interior, whereas such a cancellation does not
occur for the peripheral $(t,{}^{3}\textrm{He})$
reaction~\cite{AUE98,AUERBACH1989184}. Hence, the excitation of the
IV(S)GMR is enhanced for the latter probe. As this comparison between
probes is generally not available, the extraction of Gamow-Teller
strengths for the purpose of estimating electron-capture rates and
bench marking the theory is usually limited to excitation energies up
to about 10~MeV.

Since the extracted Gamow-Teller strengths from the charge-exchange
data are calibrated against known weak transitions strengths, the
uncertainties introduced by the need to extract absolute cross
sections through careful beam intensity normalizations and target
thickness measurements are absent. If phenomenological relationships
between the unit cross section and mass number are utilized~\cite{zegers07} to determine the unit cross section, usually a
measurement with a target for which the unit cross section has been
well established is included in an experiment, so that a relative
normalization can be performed, rather than relying on an absolute
cross section measurement that is usually more uncertain. This helps
to reduce experimental systematic uncertainties to about 10\%~\cite{zegers07}.

\begin{figure}[htbp]
  \centering
  \includegraphics[width=0.7\linewidth]{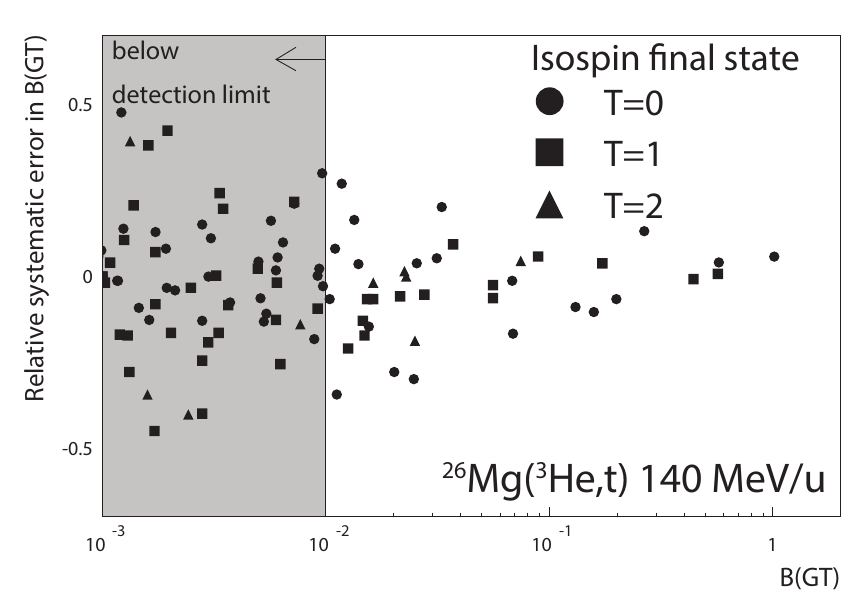}
  \caption{Results from a theoretical study to estimate the magnitude
    of the uncertainty in the proportionality between Gamow-Teller
    strength and differential cross sections for the
    $^{26}$Mg$(^{3}\textrm{He},t)$ reaction at 140 MeV/$u$ due to the effects
    of the tensor-$\tau$ component of the $NN$
    interaction. Transitions to final states in $^{26}$Al with isospin
    $T=0,1,$ and 2 are included. The uncertainty increases with
    decreasing $B$(GT). The detection limit of 0.01 is
    indicated. (from Ref. \cite{zegers:2006}).} 
\label{fig:breaking}
\end{figure}

The main remaining uncertainties in the extraction of Gamow-Teller
strengths arise from effects that perturb the proportionality of
Eq. \ref{eq:unit}. It has been shown \cite{zegers:2006,HITT2006264}
that the leading cause for the perturbation of the proportionality is
due to the interference between $\Delta L=0$, $\Delta S=1$ and
$\Delta L=2$, $\Delta S=1$ amplitudes that both contribute to
$\Delta J=1$ transitions in which the parity does not change. This
interference is mediated by the tensor-$\tau$ component of the $NN$
force~\cite{LOV81,Osterfeld:1992}. The uncertainty introduced by this effect
depends on the magnitude of the Gamow-Teller transition strength and was estimated~\cite{zegers:2006} to be $\approx
0.03-0.035\ln({B(\textrm{GT})})$, 
which amounts to an uncertainty of about 20\% for $B$(GT)=0.01. The
results of this study are shown in Fig. \ref{fig:breaking}. A $B$(GT)
of 0.01 is close to the detection limit in charge-exchange
experiments. It has been shown that this uncertainty estimate is not
strongly dependent on the nucleus studied \cite{HITT2006264}. It is
also clear that the systematic deviation fluctuates around 0, and
after integrating over many states, the uncertainty in the summed or
average transitions strength is small. 

\subsection{Probes}
The extraction of Gamow-Teller strengths from charge-exchange
reactions in the $\beta^{+}$ direction for the purpose of constraining
electron-capture rates has primarily been performed with three probes:
the $(n,p)$, $(d,{}^{2}$He), and $(t,{}^{3}\textrm{He})$ reactions. In this
subsection, a brief overview of these three probes and experimental
methods will be provided.

\subsubsection{$(n,p)$ reaction}

Although $(n,p)$ charge-exchange reactions have been performed at a
variety of facilities, the pioneering work at TRIUMF has been
particularly impactful for the purpose of testing theoretical models
used to estimate electron-capture rates for astrophysical
simulations. The nucleon charge-exchange facility at
TRIUMF~\cite{HEL87,YEN87} utilized the $(p,n)$ reaction on a $^{7}$Li
target to produce neutrons of about 200~MeV associated with
transitions to the ground and first excited state of $^{7}$Be that
were subsequently impinged on the reaction target of interest. The
setup used a segmented target chamber, which allowed for the insertion
of several targets simultaneously. Events induced by reactions on
different targets were disentangled through the analysis of
hitpatterns in multi-wire proportional chambers placed in-between the
targets. Usually, one of the targets was a CH$_{2}$ target, so that
the well-known $^{1}$H$(n,p)$ reaction could be utilized to perform
absolute normalizations of the neutron beam intensity. Protons
produced in the $(n,p)$ reaction were momentum analyzed in the
medium-resolution spectrometer (MRS). Measurements at different
scattering angles were utilized to determine the differential cross
sections as a function of center-of-mass angles, facilitating the
multipole decomposition analysis and extraction of Gamow-Teller
strength from the proportionality between strength and differential
cross section discussed above. A wide variety of experiments were
performed for the purpose of extracting Gamow-Teller strengths for
astrophysical purposes, primarily on stable nuclei in the $pf$ shell
(see e.g.~\cite{Alford.Helmer.ea:1990, ALF91, A_alf93, A_vet87,
  A_vet89, ELK94, WIL95}). The excitation energy resolutions achieved
varied between 750~keV and 2~MeV, depending on the experiment. In
Fig.~\ref{fig:probes}, three examples of the extracted $\Delta L=0$
contributions for the $^{60,62,64}$Ni$(n,p)$ reactions are shown,
displaying a concentration of Gamow-Teller strength at low excitation
energies, with a long tail up to higher excitation energies.

\subsubsection{$(d,{}^2\mathrm{He})$ reaction}

The $(d,{}^2\textrm{He})$ reaction has become one of the most powerful
probes to study the Gamow-Teller strengths in the $\beta^{+}$
direction. This probe was first developed for the purpose of
extracting Gamow-Teller strengths at RIKEN by using a 260~MeV deuteron
beam \cite{PhysRevC.47.648}, followed by the development of this probe
at Texas A\&M~\cite{PhysRevC.52.R1161} by using a 125~MeV deuteron
beam. In these experiments, a resolution of 500--700 keV could be
achieved, and the beam intensities were limited due to the background
from deuteron break-up reactions. The method was perfected in
experiments with the Big-Bite Spectrometer at KVI in combination with
the EuroSuperNovae (ESN) detector \cite{RAKERS2002253} and using
deuteron beams of $\sim 170$ MeV. Owing to the use of data signal
processing, two-proton coincidence events could be selected online,
strongly reducing the background from deuteron break-up reactions and
making it feasible to run at higher incident beam rates. In addition,
the excitation energy resolution was improved to values of typically
150 keV. A recent overview of the $(d,{}^2\textrm{He})$ program at KVI can be
found in Ref.~\cite{Frekers.Alanssari:2018}.

\begin{figure}[htbp]
  \centering
  \includegraphics[width=0.8\linewidth]{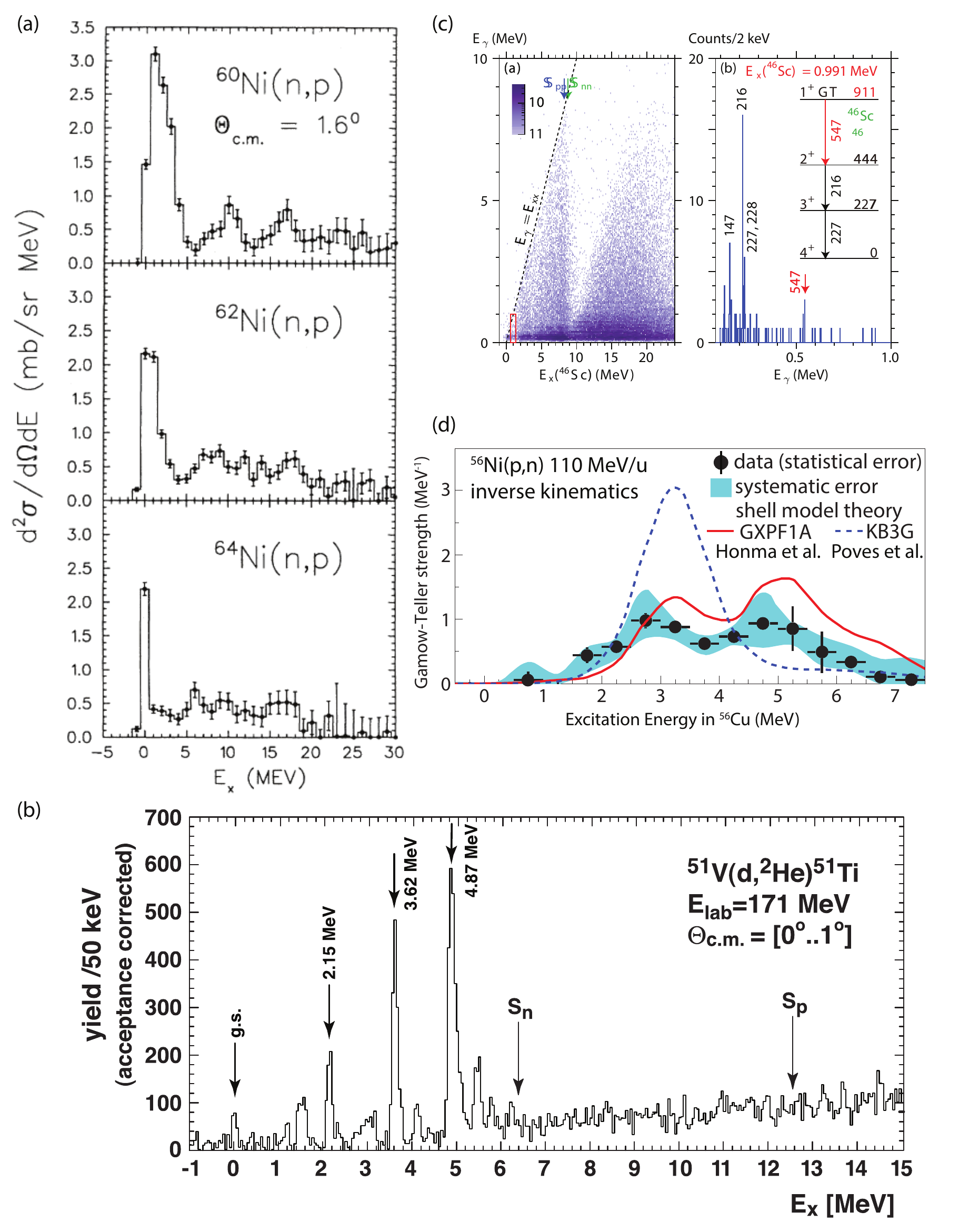}
  \caption{(a) Differential cross sections associated with
    $\Delta L=0$ for the $^{60,62,64}$Ni($n,p$) reaction at 198 MeV
    (from Ref. \cite{WIL95}). (b) Differential cross
    sections at forward scattering angles for the
    $^{51}$V($d$,$^{2}$He) reaction at 170 MeV. Owing to the
    high-resolution, individual transitions are well resolved (from
    Ref. \cite{A_bau03}). (c) left: $\gamma$ energy versus excitation
    energy for the $^{46}$Ti($t$,$^{3}$He+$\gamma$) reaction (see also
    Fig. \ref{fig:mda}). right: by gating on the $^{46}$Sc
    excitation-energy range around 0.991 MeV, the decay by a very weak
    $1^{+}$ state can be identified, sufficient for estimating the
    Gamow-Teller transition strength to this state (from
    Ref. \cite{PhysRevLett.112.252501}). (d) Extracted Gamow-Teller
    strength distribution from the $^{56}$Ni($p,n$) reaction at 110
    MeV/$u$, performed in inverse kinematics. Two sets of shell
    model-calculations with different interactions are super imposed
    (from Ref. \cite{Sasano.Perdikakis.ea:2011}).  }
\label{fig:probes}
\end{figure}

The use of the $(d,{}^2\textrm{He})$ probe requires that the momentum
vectors of the two protons from the unbound $^{2}$He must be measured
with high accuracy in order to reconstruct the momentum of the
$^{2}$He particle created in the $(d,{}^2\textrm{He})$ reaction, as
well as the relative energy $\epsilon$ between the two protons. On the
other hand, the determination of the relative energy makes it possible
to enhance the spin-transfer nature of the probe. As the incident
deuterons are primarily in the $^{3}S_{1}$ state, a spin-transfer
$\Delta S$ of 1 is ensured if the outgoing protons couple to the
$^{1}S_{0}$ state, which can be accomplished by reconstructing the
relative energy between the protons and selecting events that have
small relative energies (typically smaller than 1 MeV). This removes
transitions associated with $\Delta S=0$ from the spectra and makes it
easier to isolate the $\Delta S=1$ Gamow-Teller transitions.

A variety of $(d,{}^2\textrm{He})$ experiments were performed at KVI
with the goal to extract Gamow-Teller strengths for testing
theoretical models used to estimate electron-capture rates of interest
for astrophysical simulations, see e.g. Refs. \cite{RAK04,A_bau03,
  A_bau05, A_hag04, A_hag05, A_pop07, GRE08}. Because of the high
resolution achieved, detailed studies of the Gamow-Teller strength
distribution could be performed, including for nuclei for which it was
difficult to obtain the targets, such as $^{50,51}$V \cite{A_bau03,
  A_bau05}, as shown in Fig. \ref{fig:probes}(b). Clearly, the
excellent resolution achieved makes it possible to extract very
detailed information about the Gamow-Teller strength distribution.

\subsubsection{$(t,{}^3\mathrm{He})$ reaction}
The use of the $(t,{}^3\textrm{He})$ reaction has the disadvantage
that it is complicated to generate tritium beams. Although tritium has
been used to produce primary beams (see
e.g. Ref. \cite{PhysRevC.73.014616}), experiments performed for the
purpose of extracting Gamow-Teller strength distributions for
astrophysical purposes utilized secondary tritium beams. These
experiments are performed at NSCL with the S800 Spectrometer
\cite{bazin03}. A primary $^{16}$O beam is impinged on a thick
Beryllium production target to produce a secondary tritium beam of 345
MeV \cite{HITT2006264}. Because the momentum spread of the secondary
beam is large (typically 0.5\%), the dispersion-matching technique
\cite{FUJITA200217} is utilized to achieve excitation-energy
resolutions ranging from $200-350$ keV. At present, the beam
intensities are limited to about $10^{7}$ particles per second, but
with the completion of the Facility for Rare Isotope Beam (FRIB), the
beam intensities will increase significantly.

In addition to the good excitation-energy resolution that can be
achieved with the $(t,{}^3\textrm{He})$ reaction, it has the advantage
that the inverse $(^{3}\textrm{He},t)$ reaction is studied in great detail
and with excellent resolution at comparable beam energies
\cite{Fujita2011549, Perdikakis:2011,Frekers.Alanssari:2018}. This makes it possible to
utilize the dependence of unit cross section on mass number determined
from $(^{3}\textrm{He},t)$ data for extracting Gamow-Teller strengths from
$(t,{}^3\textrm{He})$ experiments \cite{zegers:2006, zegers07,
  Perdikakis:2011}.

As for the $(n,p)$ and $(d,{}^2\textrm{He})$ reactions, the
$(t,{}^3\textrm{He})$ reaction has been used to study a variety of
nuclei to test theoretical models used in the estimation of
electron-capture rates in astrophysical scenario, e.g.
Refs. \cite{PhysRevC.101.014308, Titus.Ney.ea:2019,
  Zamora.Zegers.ea:2019, PhysRevC.92.024312, PhysRevC.90.025801,
  PhysRevLett.112.252501,
  PhysRevC.80.014313,PhysRevC.74.034333}. Since the electron-capture
rate is very sensitive to the transitions to the lowest-lying final
states in the daughter nucleus, especially at low stellar densities,
the $(t,{}^3\textrm{He})$ probe was combined with the high-resolution
detection of $\gamma$-rays in the Gamma-Ray Energy Tracking In-beam
Nuclear Array (GRETINA) \cite{PASCHALIS201344}. This has made it
possible to extract Gamow-Teller transition strengths of as low as
0.01 \cite{PhysRevLett.112.252501}, as shown in
Fig. \ref{fig:probes}{c} for the
$^{46}$Ti$(t,{}^{3}\textrm{He}+\gamma)$ reaction, for which the
  $B$(GT) for the transition to the first
  $1^{+}$ state at 0.991 MeV could only be determined due to the
  measurement of the decay $\gamma$ rays.

In recent years, the focus of the experiments has shifted from
nuclei in the $pf$-shell to nuclei near
$N=50$~\cite{PhysRevC.101.014308, Titus.Ney.ea:2019,
  Zamora.Zegers.ea:2019} given their relevance for electron capture
  rates during the collapse of massive stars (see section~\ref{sec:core-coll-supern}).

\subsubsection{$(p,n)$ reaction and isospin symmetry}
For nuclei with $N>Z$, the Gamow-Teller transition strength in the
$\beta^{+}$ direction can also be extracted from $(p,n)$ reactions
under the reasonable assumption that isospin-symmetry breaking effects
are small. Hence, states with isospin $T_{0}+1$ populated from a
$(n,p)$-type reaction for a nucleus with ground-state isospin of
$T_{0}$, have analogs in the $(p,n)$-type reaction on that same
nucleus.  By measuring the $(p,n)$-type reaction and identifying the
$T_{0}+1$ states in the spectrum, the Gamow-Teller transition
strengths of relevance for estimating electron-capture rates can be
extracted.  Unfortunately the excitation of states with higher
isospin is suppressed compared to states with lower isospin
\cite{Bohr.Mottelson:1969}, and the $T_{0}+1$ states sit on a strong
background of states with isospin $T_{0}-1$ and $T_{0}$, which are
also excited in a $(p,n)$-type reaction on a nucleus with isospin
$T_{0}$. Still for nuclei near $N=Z$ Gamow-Teller strengths have been
extracted from $(p,n)$ data for the purpose of testing theoretical
models used to estimate electron-capture rates in nuclei
\cite{ANA08,Cole.Anderson.ea:2012}.

For nuclei with $N=Z$ and assuming isospin symmetry, the Gamow-Teller
strength distribution in the $\beta^{+}$ and $\beta^{-}$ directions
are identical and a $(p,n)$-type measurements can be used to directly
obtain the Gamow-Teller strength distribution of relevance for the
electron-capture rates. This feature was used to extract the
Gamow-Teller strength distribution from $^{56}$Ni. By using a novel
method to perform a $(p,n)$ experiment in inverse kinematics
\cite{Sasano.Perdikakis.ea:2011,sasano12}, the Gamow-Teller strength
distribution in $^{56}$Cu was extracted (see Fig. \ref{fig:probes}(d),
which is the same as the Gamow-Teller strength distribution from
$^{56}$Ni to $^{56}$Co. In this experiment, the excitation-energy
spectrum in $^{56}$Cu was reconstructed by measuring the recoil
neutron from the $(p,n)$ reaction when the $^{56}$Ni beam was impinged
on a liquid hydrogen target. Since it is important to measure the
reactions at small linear momentum transfer to main the
proportionality of Eq.~\ref{eq:dsigma}, the relevant recoil neutrons
have very low energies and were detected in a neutron-detector array
developed especially for that purpose \cite{PERDIKAKIS2012117}. With
this method, it became possible to measure $(p,n)$ reaction in inverse
kinematics on any unstable nucleus and it was recently used to study
$^{132}$Sn~\cite{PhysRevLett.121.132501}.

\subsection{$(n,p)$-type charge-exchange reactions on unstable isotopes} 
Since many of the nuclei that undergo electron captures in stellar
environments are unstable, it is important to develop experimental
techniques to perform $(n,p)$-type charge-exchange experiments in
inverse kinematics. This poses a significant challenge. A neutron
target is not available and all candidate reactions have a light
low-energy charged particle as the recoil, which is not (easily)
detectable as it interacts with the target material. Therefore, unlike
in the $(p,n)$ reaction, the recoil particle is not readily available
for the precise reconstruction of the excitation energy and scattering
angle of the reaction. If the excitation-energy of the residual
nucleus after the charge-exchange reaction is below the nucleon
separation energy, a precise measurement of the momentum and
scattering angle of the residual can be sufficient to reconstruct the
event kinematics. This method was used to extract the low-lying
$\beta^{+}$ Gamow-Teller strength distributions from unstable nuclei
$^{12}$B and $^{34}$Si through the
$(^{7}\mathrm{Li},{}^{7}\mathrm{Be})$ reaction in 
inverse
kinematics~\cite{PhysRevLett.104.212504,PhysRevLett.108.122501}.
Unfortunately, 
this probe is very difficult to use for studying Gamow-Teller strength
distributions in unstable nuclei heavier than $^{34}$Si. If the
excitation energy of the residual exceeds the nucleon separation
energy, it is necessary to measure the decay nucleon in addition to
measuring the residual and achieving the necessary energy and angular
resolutions to reconstruct the event kinematics becomes challenging
because of the strong forward kinematic boost of the laboratory
reference frame \cite{PhysRevLett.104.212504}.

Most recently, efforts have been initiated to utilize the
$(d,{}^2\textrm{He})$ reaction in inverse kinematics to study
$(n,p)$-type charge-exchange reactions on unstable isotopes. In such
experiments, the rare-isotope beam is impinged on an active-target
time projection chamber in which deuteron gas serves both as the
target and the detector medium~\cite{AYYAD2020161341}. The tracks from
the two protons originating from the unbound $^{2}$He particle can be
used to reconstruct the momentum of the $^{2}$He particle, from which
the excitation energy and scattering angle of the charge-exchange
reaction can be determined. The unique two-proton event signature is
also very helpful to separate the $(d,{}^2\textrm{He})$ reaction from
other types of reactions that occur in the time projection chamber and
that have much higher cross sections. If successful, the method will
be equally powerful for the extraction of Gamow-Teller strengths in
the $\beta^{+}$ direction as the $(p,n)$ reaction in inverse
kinematics is for the extraction of Gamow-Teller strengths in the
$\beta^{-}$ direction.

\section{Strategy and model to calculate stellar electron capture rates} 

During their long lasting lives stars balance gravitational
contraction thanks to the energy gained from nuclear fusion reactions
in their interior. Massive stars develop a sequence of core burning
stages (started by hydrogen burning via the CNO cycle, then followed
by helium, carbon, neon, oxygen and the finally silicon
burning). During this evolution the density $\rho$ and temperature $T$
in the core increases gradually and has reached values in excess of
$10^9$~g~cm$^{-3}$ and $10^9$~K, respectively, at the end of silicon
core burning. At these high temperatures nuclear reactions mediated by
the strong and electromagnetic force are in equilibrium with their
inverse reactions. This situation is called Nuclear Statistical
Equilibrium (NSE) and determines the nuclear composition for given
values of temperature, density and the proton-to-neutron ratio
(usually defined by the proton-to-nucleon ratio $Y_e$). Once NSE is
reached the star cannot generate energy from nuclear fusion reactions
anymore.  Hence the core looses an important source of pressure
against gravitational contraction.  This situation is reached in the
core produced by silicon burning. This core is usually called Fe core
because it is made of nuclei in the Fe-Ni mass range which are favored
under the core density and temperature conditions and for a $Y_e$
value only slightly smaller than 0.5.  However, the electrons, present
in the core to balance the charges of protons, form a highly
degenerate relativistic gas and can balance the gravitational
contraction of a stellar mass up to the famous Chandrasekhar limit
$M_{Ch} = 1.44 (Y_e)^2 M_\odot$ with the solar mass denoted by
$M_\odot$. Once this limiting mass is exceeded to continued silicon
burning or, as we will see below, due to electron captures, the
electron gas cannot stabilize the core anymore. The core collapses
under its own gravity.

\begin{figure}[htb]
  \centering
  \includegraphics[width=0.70\linewidth]{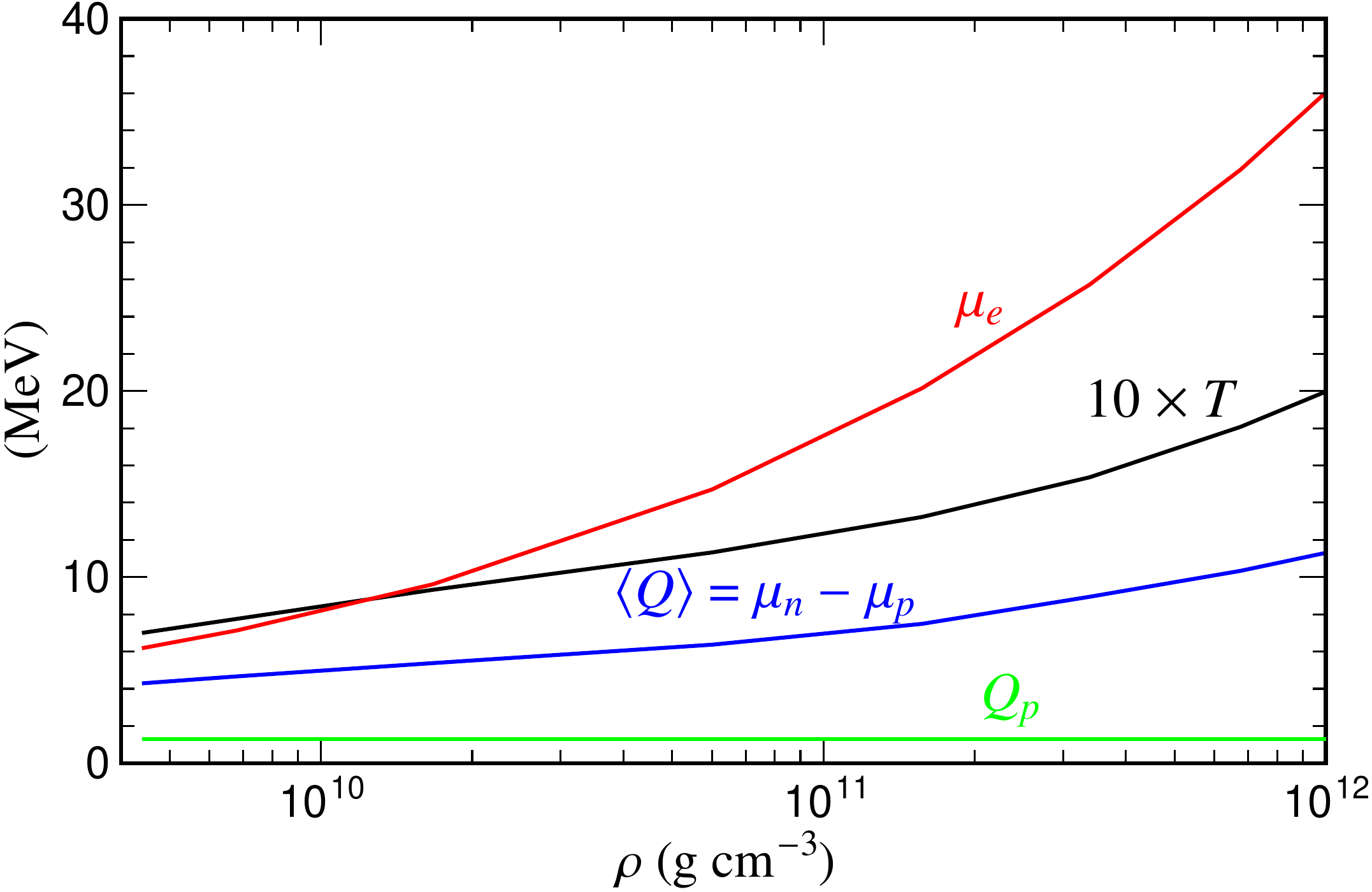}
  \caption{Various energy scales related to electron captures on
    nuclei and protons as function of density during the
    collapse. Shown are the temperature, $T$, the chemical potential
    of electrons, $\mu_e$, the $Q$ value for electron capture on
    protons (constant) and the average $Q$ value for electron capture
    on nuclei approximated as the difference in chemical potential of
    neutrons and protons (adapted
    from~\cite{martinez-pinedo.liebendoerfer.frekers:2006}).}
  \label{fig:energyscales}
\end{figure}

\begin{figure}[htb]
  \centering
  \includegraphics[angle=270,width=0.7\linewidth]{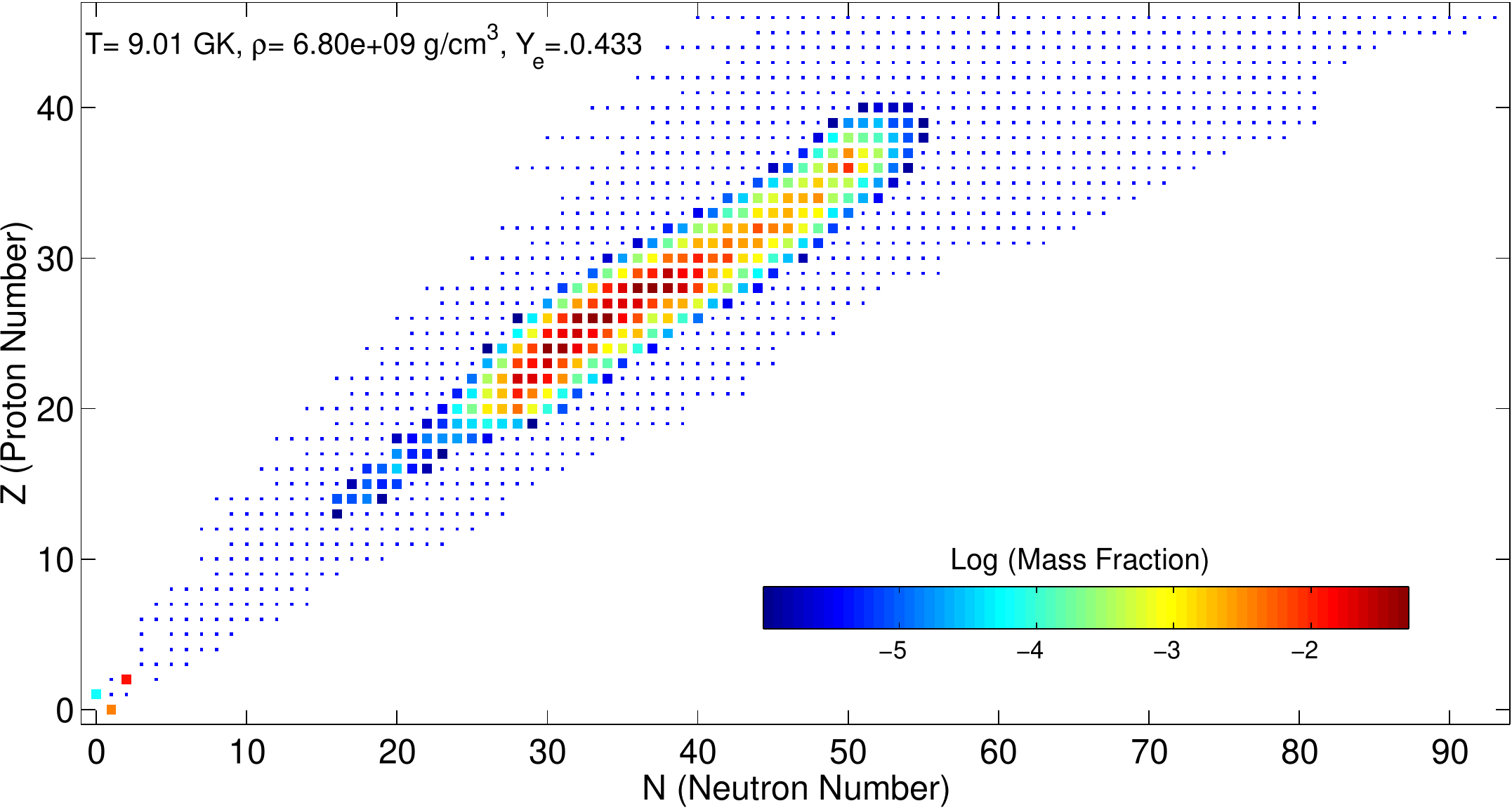}\\
  \includegraphics[angle=270,width=0.7\linewidth]{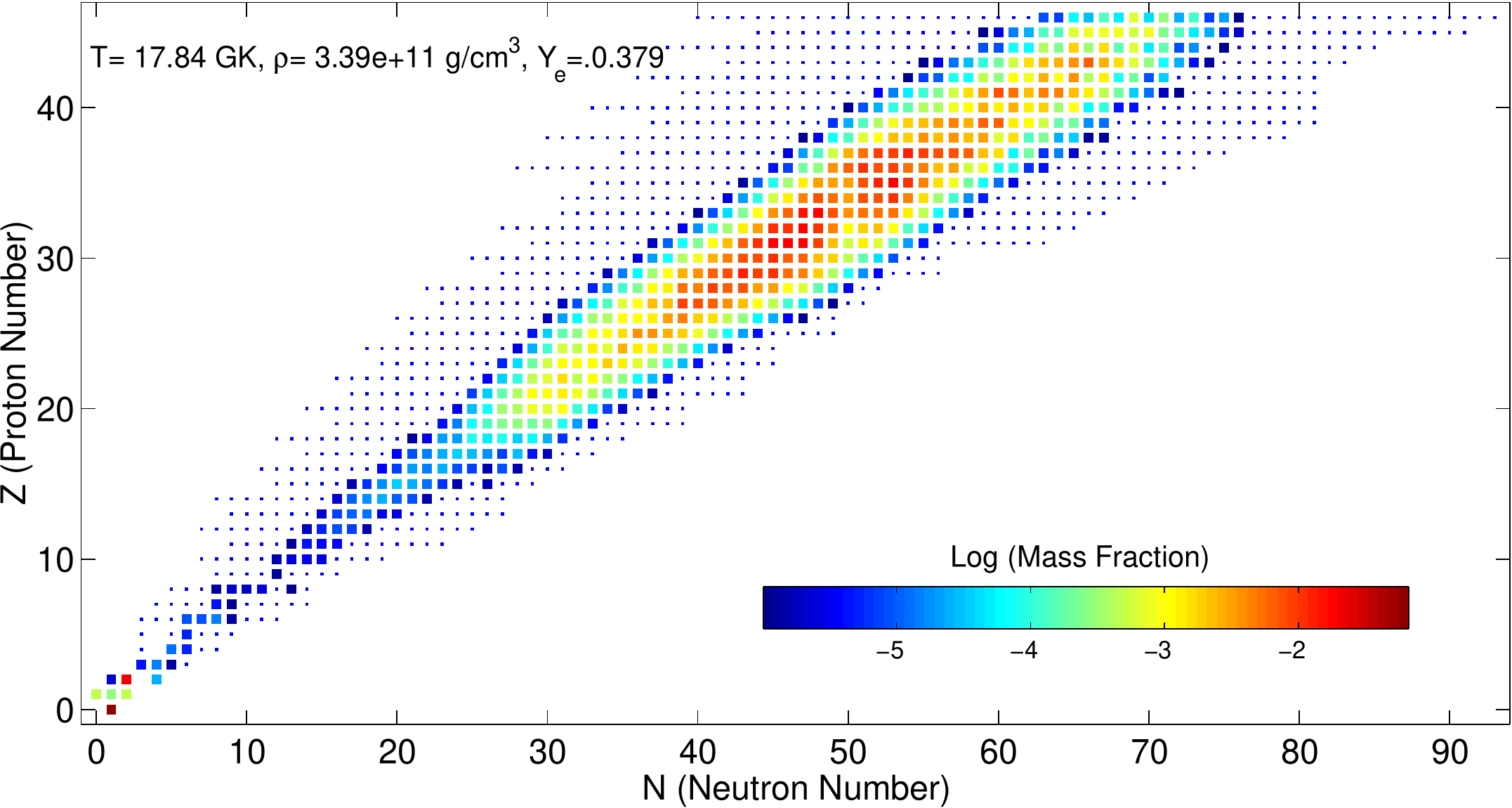}
  \caption{Mass fraction of nuclei in Nuclear Statistical Equilibrium
    at conditions which resemble the presupernova stage (top) and the
    neutrino trapping phase (bottom) of core-collapse simulations
    (courtesy of W.R. Hix).\label{fig:NSEabund}}
\end{figure}

It is important to note that $Y_e$ can be modified by charge-changing
reactions which, however, can only be mediated by the weak
interaction. Such reactions (electron capture, beta decay) are not in
equilibrium under the early collapse conditions (as for example the
neutrinos produced by the processes can leave the star and hence are
not available to initiate inverse reactions) and can change the
nuclear composition. It is also very important to note that under
core-collapse supernova conditions, i.e. at sufficiently high
densities, electron capture and beta decay do not balance each other.
The reason for this unbalance lies in the fact that the electron Fermi
energy, which scales like $\rho^{1/3}$, grows noticeably faster than
the $Q$ values of the nuclei present in the core (see
Fig.~\ref{fig:energyscales}).  As a consequence, the electron capture
rates are accelerated, while beta decays on the opposite are throttled
due to Pauli blocking of the final electron states.  Hence electron
captures win over beta decays with three very important
consequences. First, electron captures reduce the number of electrons
and hence the degeneracy pressure which the electron gas can stem
against the gravitational collapse. Second, the neutrinos produced in
the capture process can leave the star nearly unhindered during the
early phase of the collapse. They carry energy away which serves as an
effective cooling mechanism and keeps the core temperature and entropy
low. As consequence of the low entropy heavy nuclei exist during the
entire collapse phase. The situation changes when the collapse reaches
densities of order $10^{12}$~g~cm$^{-3}$ where the diffusion time
scale due to coherent scattering on nuclei becomes longer than the
collapse time of the core. Neutrinos are then effectively trapped in
the core which until bounce collapses as a homologous unit.  Third,
electron capture reduces $Y_e$ and makes the core composition more
neutron-rich. The NSE composition is driven to heavier nuclei with
larger neutron excess (see Fig. \ref{fig:NSEabund}).  This effect is
the reason why nuclei with valence protons and neutrons in different
major shells become relevant for electron captures, introducing the
Pauli unblocking as mentioned above.

Electron capture plays an important role for the dynamics of the core
collapse of massive stars for core densities between
$10^9$~g~cm$^{-3}$ and
$10^{12}$~g~cm$^{-3}$. Fig. \ref{fig:energyscales} shows the evolution
of crucial energy scales for this density regime. The strongest
growing quantity is the electron chemical potential $\mu_e$ which
increases from 6~MeV to about 40~MeV. As nuclei get increasingly more
neutron rich due to continuous electron captures, the average electron
capture $\langle Q\rangle$ value of the nuclear composition present at
the various stages of collapse grows too, but this increase is
noticeably smaller from about 4~MeV to 12~MeV. At all stages the
average nuclear $\langle Q\rangle$ value is larger than for free
protons (1.29~MeV). Finally the temperature in the core also grows
during the collapse, from about $T=0.8$~MeV to $T=2.0$~MeV. The
comparison of these different energy scales allows us to derive a
strategy how to determine electron capture rates at the various stages
of collapse.

\begin{figure}[htb]
  \centering
  \includegraphics[width=\linewidth]{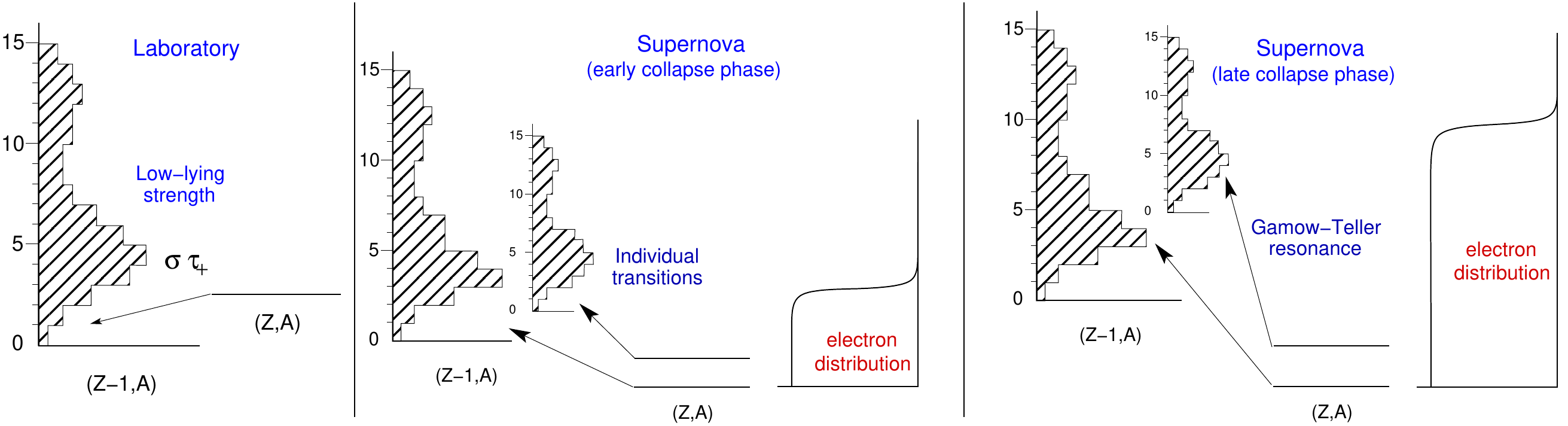}
  \caption{Sketch of electron capture conditions at different
    conditions: a) in the laboratory (left) where the electron is
    captured from an atomic orbital, b) in the early collapse phase
    (middle) where the electron is captured from a Fermi-Dirac (FD)
    distribution with an electron chemical potential of order the nuclear $Q$ value,
    and c) later in the collapse at higher densities where the
    electron is captured from a FD distribution with a chemical potential
    which are noticeably larger than the nuclear $Q$ values. It is important that, with
    increasing core density, the electron chemical potential grows faster
    than the average nuclear $Q$ value. Electron
    captures in the star (middle and left) can also proceed from
    thermally excited states where the temperature, respectively
    average nuclear excitation energy, is increasing during the
    collapse.}
\label{fig:GT-scheme}
\end{figure}

Fig. \ref{fig:GT-scheme} depicts the consequences which the different
behavior of the energy scales has for the electron capture process. We
schematically compare the situation in the laboratory with the one in
the early stage of the collapse where $\mu_e \approx \langle Q\rangle$
and at an advanced stage with $\mu_e \gg \langle Q\rangle$. In the
laboratory the daughter nucleus must be more bound than the decaying
nucleus ($Q <0$). In our schematic sketch of the GT strength
distribution we indicate that the strongest GT transitions at a few
MeV excitation energies are not accessible in laboratory electron
captures. The situation changes completely in the stellar interior, as
the capture occurs from a degenerate electron gas. In the early
collapse phase (middle diagram) electron Fermi energy and nuclear $Q$
value are similar (for example the $Q$ value of the abundant
$^{56}$Fe is 4.20~MeV) which makes the calculation of the rate
quite sensitive to the reproduction of the low-lying GT strength
distribution. An additional complication arises from the fact that the
stellar environment has a finite temperature. Hence the capture can
also occur from thermally excited nuclear states which can have
different GT strength distributions than the ground state.  The
nuclear composition at this stage of the collapse is dominated by
nuclei of the Fe-Ni range.  This is a fortunate situation as
diagonalization shell model calculations for $pf$ shell nuclei are now
feasible and have been proven to reproduce GT strength distributions
and energy levels quite well.  Thus, diagonalization shell model is
the method of choice to determine the capture rates for $pf$ shell
(and $sd$ shell) nuclei.

Due to continuous electron captures the nuclei abundant in the core
composition become more neutron rich and heavier. The right panel of
Fig.~\ref{fig:NSEabund} shows the NSE distribution for the conditions
reached around the onset of neutrino trapping. The most abundant
nuclei correspond to nuclei with valence protons in the $pf$ shell,
while the valence neutrons occupy orbitals in the $sdg$ shell.  Hence
the description of cross-shell correlations is the challenge to
determine capture rates for these nuclei. We also note that at the
higher densities more nuclei contribute to the NSE abundances.  This
is an effect of the slight increase of core entropy as
  neutrino-trapping sets in and of the decrease of the relative
  differences of nuclear binding energies as the composition moves to
  heavier neutron-rich nuclei. The right part of
Fig. \ref{fig:GT-scheme} describes the energy situation encountered at
higher densities in the collapse (a few $10^{10}$~g~cm$^{-3}$ and
above). At first, the electron chemical potential is now significantly
larger than the average nuclear $Q$-value. For example, the
neutron-rich nuclei $^{66}$Fe (with $N=40$) and $^{82}$Ge ($N=50$)
have Q-values of 13.8~MeV and 13.0~MeV, respectively.  Furthermore the
temperature has grown to about $T=1$~MeV. At such temperatures the
average nuclear excitation energy, estimated in the Fermi gas model as
$E^*= A\,T^2/8$ is 8.3~MeV for $^{66}$Fe and 10.2~MeV for $^{82}$Ge
and the capture, on average, occurs from excited states, making it
even easier for electron capture to overcome the $Q$ value.  Under
these conditions calculations of stellar capture rates for the
abundant nuclei on the basis of the diagonalization shell model are
not appropriate nor possible.  At first, diagonalization shell model
studies of nuclear GT strength distributions for the relevant
cross-shell nuclei is not feasible due to model space restrictions
yet. Moreover, there are simply too many thermally excited nuclear
states in the mother nucleus which can contribute to the capture
process.  However, the detailed reproduction of the GT strength
distribution --- as required at lower densities where $pf$ shell
nuclei dominate --- is not needed at the advanced conditions of the
collapse. At first, the fact that the electron Fermi energy and the
average nuclear excitation energy are together noticeably larger than
the average nuclear $Q$ value makes the capture rate less sensitive to
the detailed reproduction of the GT strength distribution. Thus it
suffices if the total GT strength and its centroid are well
described. This is possible within the Random Phase
approximation (RPA).  Second, due to the exponential increase of the
level density with excitation energy, there will be many states which
contribute to the capture so that some averaging is expected over the
GT strength functions. However, there are two further demands which
have to be considered.  The Pauli unblocking of the GT strength
requires the consideration of multi-particle-multi-hole
correlations. These correlations are not expected to be the same at
the higher excitation energies than for the ground state. A many-body
method which accounts for both of these effects is the Shell Model
Monte Carlo approach which allows the calculation of average nuclear
properties at finite temperature considering all many-body
correlations in unprecedentedly large model
spaces~\cite{Koonin.Dean.Langanke:1997}.  Hence a hybrid model has
been proposed to calculate stellar electron capture rates for heavy
nuclei: In the first step partial proton and neutron occupation
numbers are determined within the SMMC, which, in the second step,
become the input of RPA calculations of the GT and forbidden strength
distributions from which finally the capture rates are
evaluated~\cite{Langanke.Martinez-Pinedo.ea:2003}.  An alternative
method to the hybrid model is the temperature-dependent Quasiparticle
RPA approach which treats the ground state and thermally excited
states consistently on the level of 2p-2h correlations
\cite{Dzhioev.Vdovin.ea:2010}. This approach has also been used to
describe astrophysically important neutrino-nucleus reactions at
finite temperatures
\cite{Dzhioev.Vdovin.ea:2014,Dzhioev.Vdovin.Wambach:2015} (for a
review on this subject see~\cite{Balasi.Langanke.Martinez-Pinedo:2015}).

\subsection{Capture rates for nuclei with $A<65$}
The method of choice to determine electron capture and beta decay
rates for medium mass nuclei is the diagonalization shell model. As
the shell model allows for the description of individual states and
their properties, within the chosen model space, the stellar electron
capture rate can be determined on the basis of the state-by-state
formalism from states in the parent nucleus at energy $E_i$ to final
states in the daughter nucleus at $E_f$. This formalism explicitly
considers that the stellar interior has finite temperature $T$ Thus
beta decays and electron captures can occur from excited nuclear
levels, where the thermal nuclear ensemble is described by a Boltzmann
distribution. Beta-decay $\lambda_\beta$ and electron capture rates
$\lambda_{ec}$ can be derived in perturbation theory and the
respective formulas and derivations are presented
in~\cite{Fuller.Fowler.Newman:1980,Langanke.Martinez-Pinedo:2001}. Analytical
approximations are provided in~\cite{Martinez-Pinedo.Lam.ea:2014}. In
the derivation of the weak-interaction rates only Gamow-Teller
transitions are included (with an important exception for $^{20}$Ne,
as discussed below).

\subsubsection{$pf$ shell nuclei}
The first derivation of stellar weak interaction rates for the
$pf$-shell nuclei relevant for core-collapse supernovae has been
presented in Ref.~\cite{Langanke.Martinez-Pinedo:2000}. The
calculations are based on diagonalization shell model calculations
considering either all correlations in the complete $pf$ shell or at a
truncation level which basically guaranteed convergence of the
low-energy spectra and the GT strength distributions which are the
essential quantities to calculate electron capture and beta decay
rates.  The GT strength functions were determined using the Lanczos
method. Hence it represents the strength for individual states at low
energies, while at moderate excitation energies the GT strength is not
completely converged and gives the average value for a rather small
energy interval.  We note that the shell model gives in general a good
account of nuclear properties in the $pf$ shell if appropriate
residual interactions including monopole corrections are used (see
Ref.~\cite{Caurier.Martinez-Pinedo.ea:2005} and references therein).
Ref.~\cite{Caurier.Langanke.ea:1999} presented detailed studies of the
GT strength distributions and validated the method by comparison to
the charge-exchange data available at that time.  In fact, good
agreement with data was found, if the shell model GT distributions
were reduced by a constant factor $(0.74)^2$
(\cite{Langanke.Dean.ea:1995,Martinez-Pinedo.Poves.ea:1996b}).  The
origin of this renormalization (often called quenching of GT strength)
is caused by the fact that shell model calculations performed within a
single shell miss short-range correlations which shift GT strength to
significantly higher energies
\cite{Caurier.Poves.Zuker:1995,Wakasa.Sakai.ea:1997}. Modern many-body
techniques which are able to account for these short-range
correlations recover indeed most of the GT
renormalization~\cite{Gysbers.Hagen.ea:2019}.

\begin{figure}[htbp]
  \centering
  \includegraphics[width=0.70\linewidth]{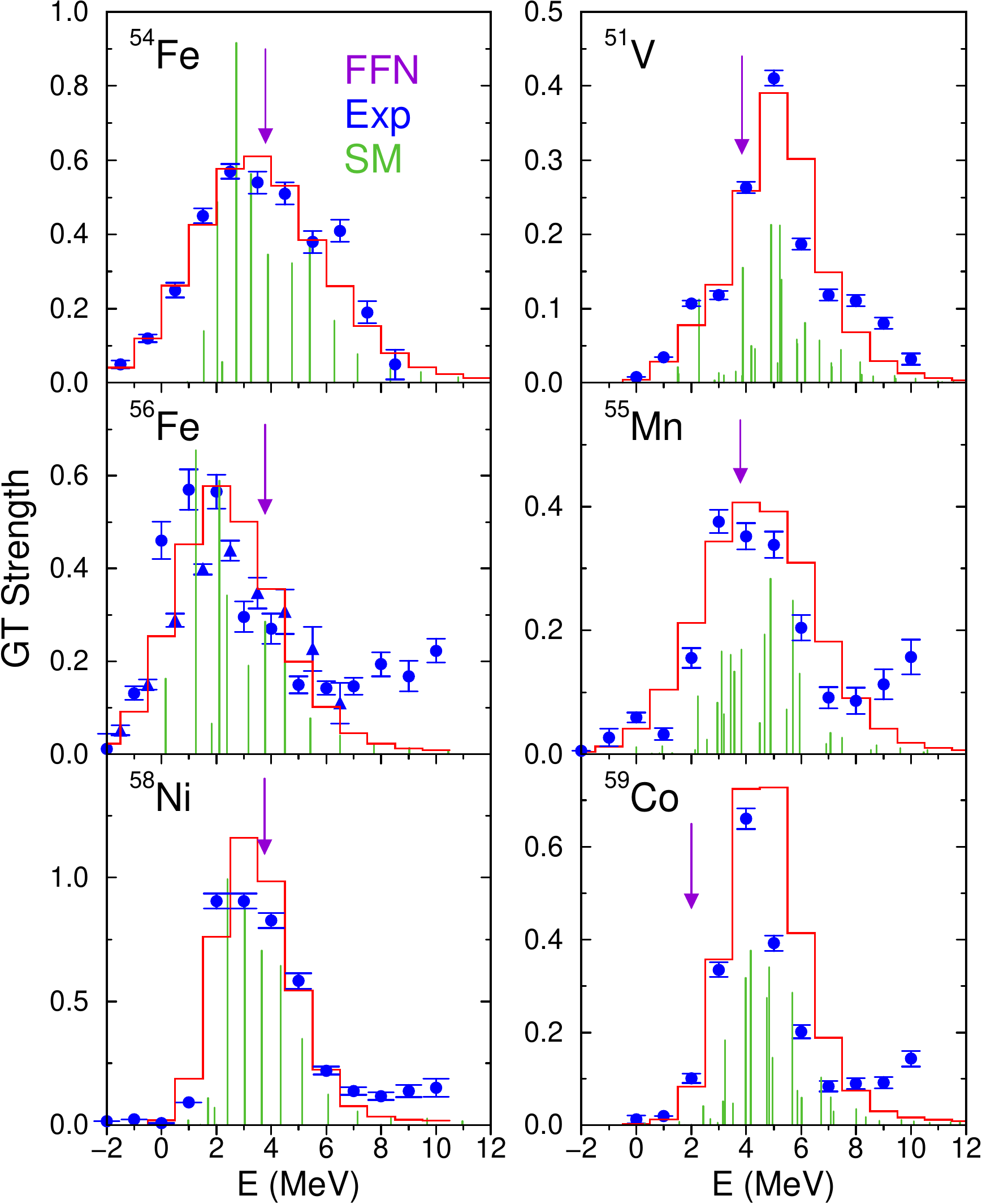}
  \caption{Comparison of experimental and shell model GT strength
    distributions for several $pf$ shell nuclei. The data are derived
    from $(n,p)$ charge-exchange experiments~\cite{A_vet89,ELK94,A_alf93}. The shell model results
    are given as histograms and folded with the experimental energy
    resolution. The energies at which the FFN evaluation placed the GT
    strengths are shown as arrows.}
\label{fig:GT-npdata}
\end{figure}

Fig. \ref{fig:GT-npdata} compares the shell model GT$_+$ strength
distributions with the experimental data derived from $(n,p)$
charge-exchange reactions and the energy position at which the FFN
rates assumed the total GT$_+$ strength to reside. The fragmentation
of the GT strength is quite obvious. It is even more visible in
high-resolution data determined by the $(d,{}^2\textrm{He})$ and
$(t,{}^3\textrm{He})$ techniques, e.g. see the data for
$^{51}$V$(d,{}^2\textrm{He})$ in 
panel b of fig.~\ref{fig:probes}. The data for nickel
isotopes showed that the KB3 residual interaction, used in
Refs.~\cite{Caurier.Langanke.ea:1999, Langanke.Martinez-Pinedo:2000},
had some shortcomings in describing low-energy details of the GT
strength function~\cite{Suzuki.Honma.ea:2011}. These are better
reproduced using an alternative residual interaction
(GXPF1J~\cite{Honma.Otsuka.ea:2005}) (see panel d of Fig.~\ref{fig:probes}).

\begin{figure}[htbp]
  \centering
  \includegraphics[width=0.70\linewidth]{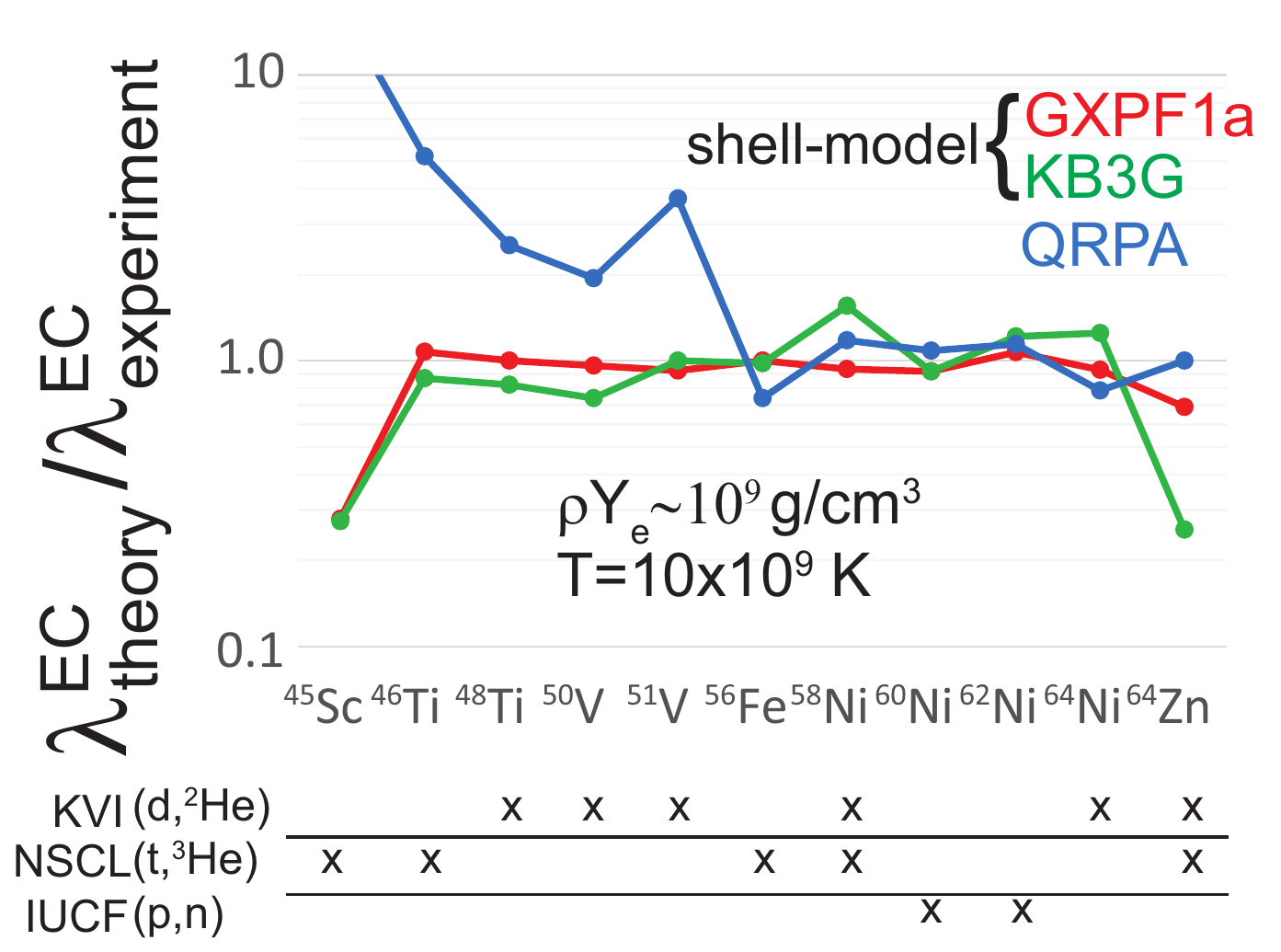}
  \caption{Comparison of electron capture rates for $pf$ shell nuclei
    calculated from GT$_+$ distributions derived experimentally and
    from shell model calculations with two different residual
    interactions and within the QRPA approach.  The astrophysical
    conditions represent a situation at which $pf$ shell nuclei
    dominate the core composition.  The rates presented are originally
    from~\cite{Cole.Anderson.ea:2012} and used existing data from
    $(d,{}^2\textrm{He})$, $(t,{}^3\textrm{He})$, and $(p,n)$
    experiments, as discussed in Section \ref{experiment}. For the
    purpose of this review, they are supplemented with later results
    from the $(t,{}^3\textrm{He})$ reactions on $^{45}$Sc
    \cite{PhysRevC.92.024312}, $^{46}$Ti
    \cite{PhysRevLett.112.252501}, and $^{56}$Fe
    \cite{PhysRevC.90.025801}.}
\label{fig:FP-comparison}
\end{figure}

Fig. \ref{fig:FP-comparison} compares electron capture rates
calculated for all $pf$ shell nuclei, for which experimental GT$_+$
distributions have been measured, with the predictions from the shell
model on the basis of two residual interactions (KB3G~\cite{Poves.Sanchez-Solano.ea:2001} 
and GXFP1a~\cite{Honma.Otsuka.ea:2002}). The chosen astrophysical conditions correspond
to the presupernova stage of the collapse at which the $pf$ shell
nuclei dominate the abundance distribution.  The GXPF1a rates giving a nearly
perfect reproduction, except for $^{45}$Sc. The KB3G rates are
slightly worse than those based on the GXPF1a interaction, but still
very good, except for $^{45}$Sc and $^{64}$Zn. On the other hand, the
rates based on the QRPA calculations, with their restricted account of
correlations, can deviate from the data and shell model rates by up to
a factor of 10 for light $pf$ nuclei, although for the heavier $pf$
shell nuclei the rates based on the QRPA calculations do well at this
stellar density.

The rates presented in Fig. \ref{fig:FP-comparison} have been
calculated solely from the ground state GT distribution. This assumes
that the GT distributions of excited mother states is the same as for
the ground state, shifted only by the respective excitation
energy. This assumption often is called Brink-Axel hypothesis
\cite{Brink:1955,Axel:1968} It cannot be strictly valid as it does not
allow for deexcitations. As we will see below it is also not
appropriate for nuclei at shell closures.
Ref.~\cite{Langanke.Martinez-Pinedo:1999} discusses the validation of
this hypothesis.  A modification of the Brink-Axel hypothesis for high
temperatures is proposed in \cite{Misch.Fuller.Brown:2014}. A novel
method to calculate electron capture rates for excited nuclear states
based on the Projected Shell Model has been proposed
in~\cite{Tan.Liu.ea:2020}. 

Ref.~\cite{Langanke.Martinez-Pinedo:2000} calculated stellar beta
decay and electron capture rates for more than 100 $pf$ shell nuclei
in the mass range $A=45$--64.  These calculations approximated the
state-by-state formalism discussed above by considering the low-energy
states and their GT distributions explicitly. These contributions were
supplemented by the considerations of `back-resonances'. These are GT
transitions calculated for the inverse reaction and then inverted by
detailed balance~\cite{Fuller.Fowler.Newman:1982a,Fuller.Fowler.Newman:1982b}.  The calculated energies and GT
transition strengths had been replaced by experimental data whenever
available.  A detailed table of the weak interaction rates for the
individual nuclei and for a fine grid of astrophysical conditions at
which $pf$ shell nuclei are relevant have been published in
\cite{Langanke.Martinez-Pinedo:2001}.  The rate table is publicly
available and is incorporated in several leading supernova codes.  A
procedure how to interpolate between the grid points in temperature,
density and $Y_e$ value is discussed
in~\cite{Langanke.Martinez-Pinedo:2001}, based on the work
of~\cite{Fuller.Fowler.Newman:1985}. 

\begin{figure}[htbp]
  \centering
  \includegraphics[width=0.70\linewidth]{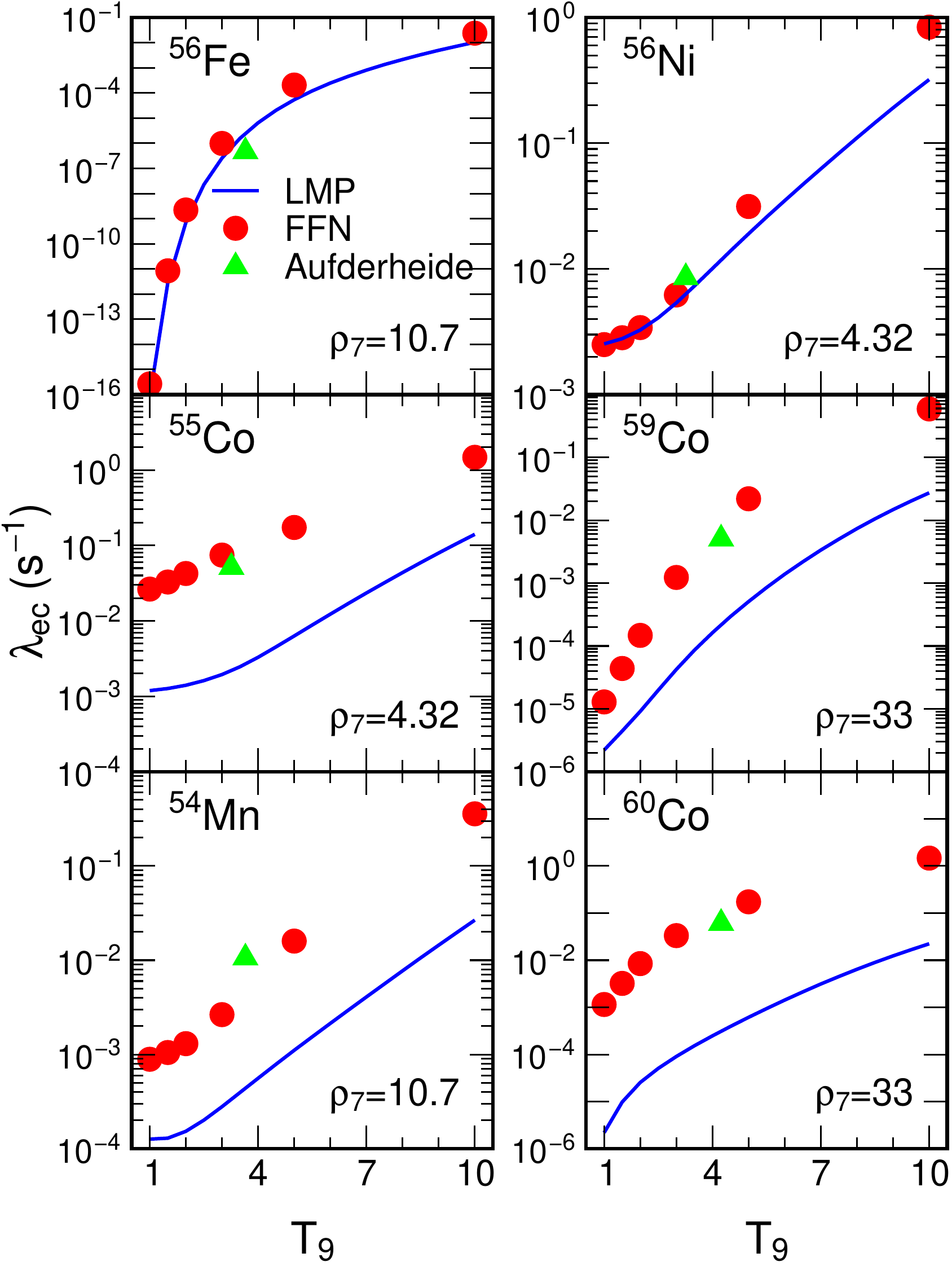}
  \caption{Comparison of the FFN and shell model rates for selected
    nuclei as function of temperature (in $10^9$ K) and at densities
    (in $10^7$~g~cm$^{-3}$) at which the nuclei are relevant to the
    capture process in simulations which used the FFN rates. The
    triangles refer to shell model estimates derived on the basis of
    rather strong truncations
    \cite{Aufderheide.Fushiki.ea:1994a}.\label{fig:FFNvsSM}} 
\end{figure}

Fig. \ref{fig:FFNvsSM} compares the shell model and FFN electron
capture rates for several nuclei.  The chosen nuclei represent the
most abundant even-even, odd-$A$ and odd-odd nuclei for electron
captures as been identified by simulations on the basis of the FFN
rates at the respective astrophysical conditions during early collapse
(presupernova phase). The shell model rates are systematically smaller
than the FFN rates with quite significant consequences for the
presupernova evolution, as discussed below.  The reasons for these
differences is mainly due to the treatment of nuclear pairing which
had been empirically considered in the FFN calculations. This leads in
particular to the drastic changes observed for odd-odd nuclei. The
shell model rates also considered experimental data which were not
available at the time when the FFN rates were derived.  The
differences between the FFN and shell model beta decay rates are
smaller than for electron capture and do not show a systematic trend~\cite{Martinez-Pinedo.Langanke.Dean:2000,Langanke.Martinez-Pinedo:2000}.

\subsubsection{$sd$ shell nuclei}
\label{sec:sd-shell-nuclei}

Beta decays and electron capture on $sd$ shell nuclei (mass numbers
$A=17$--39) can occur during silicon burning in massive
stars~\cite{Heger.Woosley.ea:2001}. The processes are, however, of
essential importance for the fate of the O-Ne-Mg core which develops
at the end of hydrostatic burning in intermediate mass stars. This was
the motivation for Oda et al.~\cite{Oda.Hino.ea:1994} to derive at
stellar beta decay and electron capture rates for $sd$ shell nuclei
covering the relevant astrophysical conditions (temperatures
$10^8$--$10^{9}$ K and densities $10^8$--$10^{10}$~g~cm$^{-3}$).  The rate
evaluations used the state-by-state formalism defined above.  The
spectra of the nuclei and the respective Gamow-Teller strength
distributions for ground states and excited states were determined by
diagonalization in the $sd$ shell using the Brown-Wildenthal USD
residual interaction which had been proven before to give a quite
reliable account of nuclear properties for $sd$ shell nuclei. Like for
the nuclei in the $pf$ shell the Gamow-Teller strength distributions
were renormalized by a constant factor. The rates have been made
available in table form for a grid of temperature-density-$Y_e$
points.

\begin{figure}[htbp]
  \centering
  \includegraphics[width=0.48\linewidth]{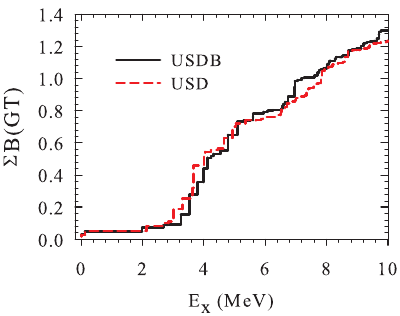}%
  \hspace{0.01\linewidth}%
  \includegraphics[width=0.48\linewidth]{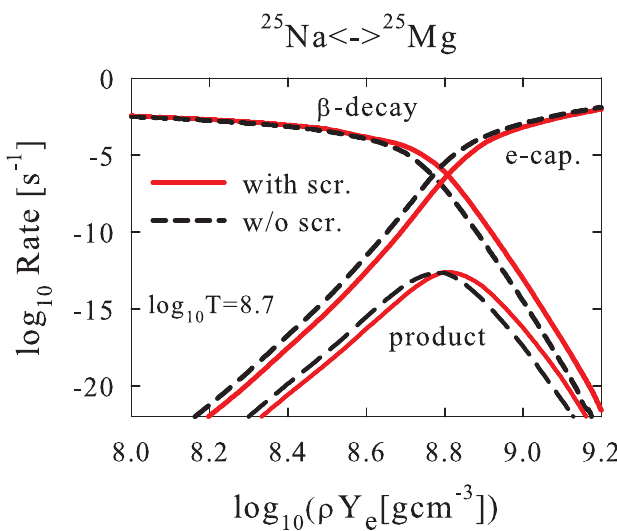}%
  \caption{(left) Cumulative GT strength for $^{25}$Mg calculated in
    $sd$ shell model studies with two different interactions. The
    ground state strength is known experimentally. (right) Beta decay
    and electron capture rates for the URCA pair $^{25}$Na and
    $^{25}$Mg as function of density and a specific temperature.  The
    curve labelled 'product' is given by the sum of the two rates and
    identifies the density at which the URCA pair operates most
    efficiently. The rates are given with and without screening
    corrections. (from \cite{Suzuki.Toki.Nomoto:2016})}
\label{fig:A25}
\end{figure}

More recently, an updated rate table has been published by Suzuki
which is based on the USDB residual interaction (a modified version of
the USD interaction) and additional experimental
information~\cite{Suzuki.Toki.Nomoto:2016}. These modern shell model
rates differ not too much from those of
Ref.~\cite{Oda.Hino.ea:1994}. However, they are given on a finer mesh
of temperature and density.  This finer grid is particularly required
for the study of the core evolution of stars in the mass regime
8--10~$M_\odot$.  Of particular importance are the URCA pairs
($^{23}$Ne-$^{23}$Na, $^{25}$Na-$^{25}$Mg and potentially
$^{27}$Mg-$^{27}$Al) which have $Q$ values against electron captures
which are reached during core contraction at densities around
$10^9$~g~cm$^{-3}$. As the environment also has a finite temperature
of order $10^8$--$10^9$~K, which smears the electron chemical
potential and implies the presence of thermally excited states, it is
possible that both electron captures and beta decays occur between the
pairs of nuclei. The neutrinos produced in both processes carry energy
away making the URCA pairs an efficient cooling mechanism. The
operation of URCA pairs is restricted to a relatively narrow density
range requiring the knowledge of weak interaction rates on a rather
fine density-temperature grids. Such rates have been provided in
\cite{Toki.Suzuki.ea:2013,Suzuki.Toki.Nomoto:2016}. Fig. \ref{fig:A25}
compares the GT$_+$ strength for the $^{25}$Mg ground state as
calculated with the USD~\cite{Oda.Hino.ea:1994} and
USDB~\cite{Suzuki.Toki.Nomoto:2016} interactions. The transition to
the $^{25}$Na ground state is known experimentally. We note the rather
close agreement between the two calculations.  \ref{fig:A25} shows the
beta decay and electron capture rates calculated on the basis of the
$sd$ shell model. With increasing density, the electron chemical
potential grows which reduces the beta decay rates due to Pauli final
state blocking and increases the electron capture rates. At the URCA
density $\log (\rho Y_e)$ = 8.81 both rates match. The product of beta
decay and electron capture rates indicates the density range at which
the URCA pair operates.  Screening effects induced by the
astrophysical environment shift the URCA density to slightly larger
values (see below).

While the URCA pairs cool the core, electron capture on the two
abundant nuclei $^{24}$Mg ($Q=6.03$~MeV) and $^{20}$Ne ($Q=7.54$~MeV)
heat it.  (The third abundant nucleus $^{16}$O has such a high
$Q$-value that electron capture does not occur at the densities
achieved during the evolution of the ONeMg core).  Electron captures
on these nuclei set in once the core density is large enough for the
electron chemical potential to overcome the respective $Q$ value.
(Due to its lower $Q$ value this occurs first on $^{24}$Mg.)  At these
densities the electron captures on the daughter nuclei $^{24}$Na and
$^{20}$F, respectively, occurs then instantaneously with noticeably
larger capture rates, as the odd-odd daughter nuclei have
significantly smaller $Q$-values against electron capture due to
pairing effects.  As $\mu_e > Q$, the capture often leads to excited
states in the final nuclei $^{24}$Ne and $^{20}$O which de-excite via
gamma emission heating the environment.

\begin{figure}[htbp]
  \centering
  \includegraphics[width=0.70\linewidth]{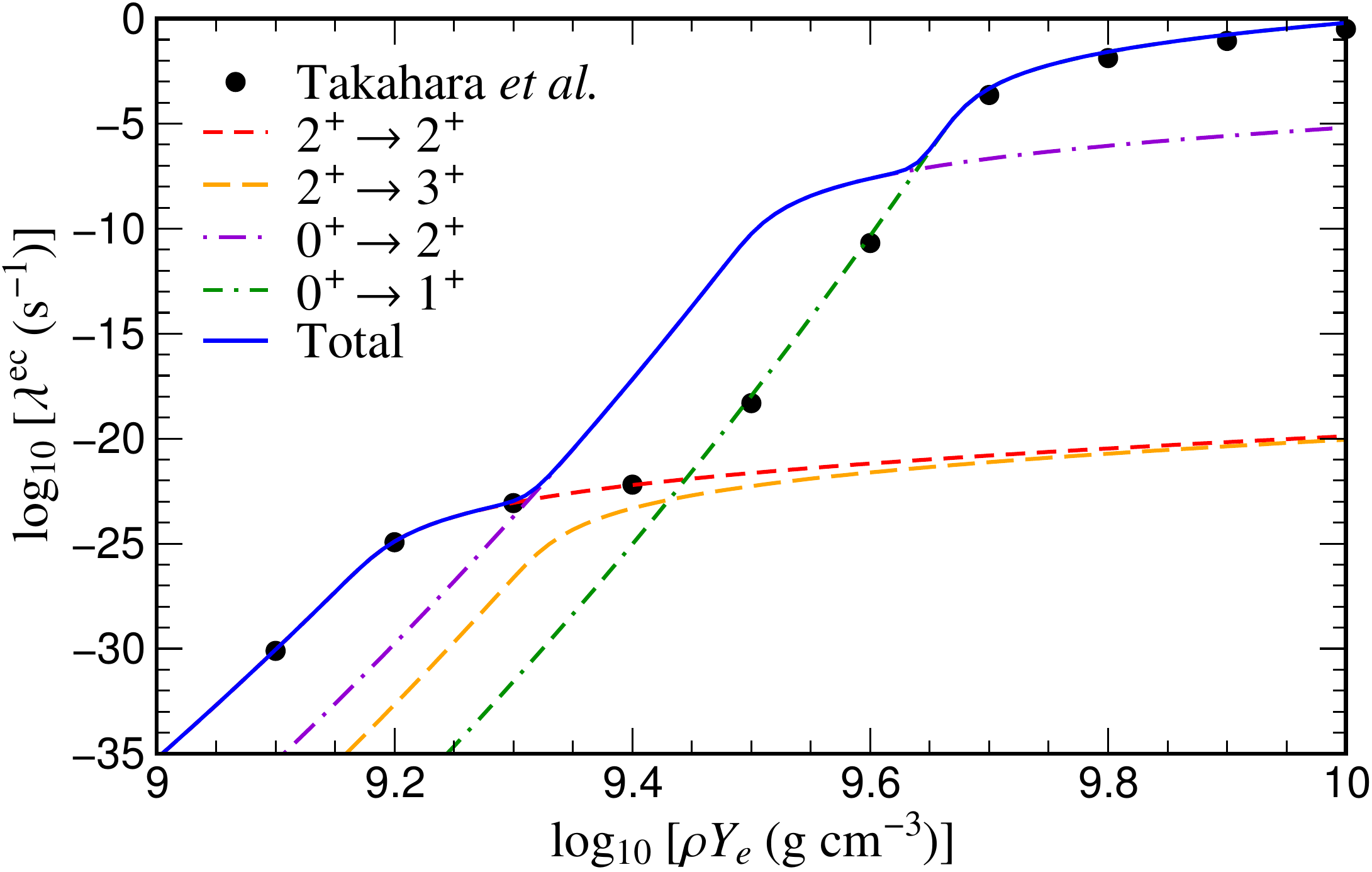}
  \caption{Electron capture rate for $^{20}$Ne as function of density
    and for a specific temperature. The rate labelled 'Takahara et al'
    was evaluated from GT distributions calculated within the shell
    model \cite{Takahara.Hino.ea:1989}. The rate is broken down into
    the individual state-by-state contributions where the energies and
    transition strengths are all taken from experiment. The label
    `$0^+ \rightarrow 2^+$' identifies the contribution from the
    second forbidden ground-state-to-ground-state transition whose
    strength has been measured by Kirsebom {\it et
      al.}. \cite{Kirsebom.Jones.ea:2019}. This transition dominates
    the capture rate at the densities most relevant for the core
    evolution of intermediate-mass stars. (from
    \cite{Kirsebom.Jones.ea:2019}).}
\label{fig:ne20-rate}
\end{figure}

The electron capture rates for $^{20}$Ne and $^{24}$Mg and their
daughters have been determined on the basis of shell model
calculations by Takahara et al.~\cite{Takahara.Hino.ea:1989} and Oda
et al. \cite{Oda.Hino.ea:1994}. These rates have been the default
values until recently in studies of the core evolution of intermediate
mass stars. The $^{24}$Mg capture rate has been updated in
Ref.~\cite{Martinez-Pinedo.Lam.ea:2014} using experimental data which
became available in the meantime leading to rather small
modifications. This is different for the electron capture rate on
$^{20}$Ne which can be considered a milestone and an exception.  At
first, Ref. \cite{Martinez-Pinedo.Lam.ea:2014} showed that all
relevant Gamow-Teller contributions to the rate could be derived from
experiment using data from $(p,n)$ charge-exchange
measurements~\cite{Anderson.Tamimi.ea:1991} (applying isospin
symmetry) and from beta decays of $^{20}$F (see
Fig.~\ref{fig:ne20-rate}).  Furthermore, the authors noticed that, due
to the relatively low temperatures of a few $10^8$~K, the, at the time
unknown, $^{20}$Ne-$^{20}$F ground-state-to-ground-state might
contribute to the capture rate just at the relevant densities, despite
that it is highly suppressed due to angular momentum mismatch. The
strength of this second forbidden transition has recently been
measured in a dedicated experiment at the IGISOL facility in
Jyv\"askyl\"a~\cite{Kirsebom.Jones.ea:2019,Kirsebom.Hukkanen.ea:2019}
and it was found large enough to increase the $^{20}$Ne capture rate
by several orders of magnitude as is shown in
Fig.~\ref{fig:ne20-rate}. We emphasize that the electron capture rate
on $^{20}$Ne in the temperature-density range important for
intermediate mass stars is now completely determined by
experiment. This is quite an achievements and shows the great
opportunities offered by modern Radioactive Ion Beam (RIB)
facilities. That a second forbidden transition essentially contributes
to an astrophysical electron capture rate is exceptional and due to
the low temperature of the environment and the peculiar structure
  of $^{20}$Ne.  In core-collapse supernovae the temperatures are an
order of magnitude higher at the same densities making allowed
Gamow-Teller transitions the dominating contributor to electron
capture rates.

In the astrophysical environment the weak interaction processes are
modified due to screening effects. The screening corrections for
electron capture have been developed
in~\cite{Juodagalvis.Langanke.ea:2010}, the extension to beta decays
is given in~\cite{Martinez-Pinedo.Lam.ea:2014}. There are two
important effects induced by the astrophysical environment. At first,
screening enlarges (reduces) the energy threshold for electron
captures (beta decays).  Second, it reduces the electron chemical
potential. Both effects together reduce the electron capture rates,
while they enhance beta decay rates.  Rate modifications due to
screening are are relatively mild of order a factor of 2.  The effects
for the URCA pair $^{25}$Na-$^{25}$Mg are shown in
Fig.~\ref{fig:A25}. Modifications of the electron capture rates
during the collapse of a massive star are discussed in
\cite{Juodagalvis.Langanke.ea:2010} and exemplified in their Fig. 10.
Many tabulations of electron capture rates
(e.g. Refs. \cite{Fuller.Fowler.Newman:1980,Fuller.Fowler.Newman:1982a,Fuller.Fowler.Newman:1982b,
  Fuller.Fowler.Newman:1985,Langanke.Martinez-Pinedo:2001,Oda.Hino.ea:1994,
  Nabi.Klapdor-Kleingrothaus:1999b,Pruet.Fuller:2003} do not include
screening corrections. Ref. \cite{Juodagalvis.Langanke.ea:2010}
presents a formalism how these rates can be approximately corrected
for screening effects.

Weak-interaction rates based on diagonalization shell model exist for
nuclei in the mass range $A=17$--65, with the exception of
$A=39$--44. Studies of these nuclei require the inclusion of
correlations across the $Z,N=20$ shell closures and hence large model
spaces enabling allowed Gamow-Teller and also forbidden
transitions. Steps in performing such demanding calculations have been
taken so that a shell model evaluation also for this mass range
appears to be in reach. Weak-interaction rates for $A=39$--44 were
provided by Fuller et al in their seminal work based on the IPM, but
also by Nabi and Klapdor-Kleingrothaus within the framework of the
QRPA~\cite{Nabi.Klapdor-Kleingrothaus:1999b}. The latter reference
gives electron capture rates for a wider range of nuclei.

\subsection{Electron capture on nuclei with $A > 65$}

Shell model studies of nuclei with mass numbers $A\ge 65$ require an
accurate description of cross-shell correlations. The associated model
spaces make diagonalization shell model calculations in general
unfeasible.  It is fortunate that by the time nuclei with $A \ge 65$
dominate the core composition the density, and accordingly the
electron chemical potential, has grown sufficiently that the capture
rates are mainly sensitive to the total GT strength and its
centroid. For these nuclei a hybrid model
\cite{Langanke.Kolbe.Dean:2001,Langanke.Martinez-Pinedo.ea:2003} has
been proposed to evaluate the stellar capture rates. In this model the
rates are calculated within an RPA approach in appropriately large
model spaces using partial occupation numbers. These occupation
numbers are calculated within the Shell Model Monte Carlo (SMMC)
method \cite{Johnson.Koonin.ea:1992,Koonin.Dean.Langanke:1997} and
hence consider the relevant multi-particle-multi-hole configurations
required to properly describe the cross-shell correlations which are
relevant for nuclei in this mass range. Moreover, the SMMC determines
the nuclear properties at finite temperature as is appropriate for the
astrophysical environment. The RPA approach is known to reproduce the
strength and centroids of collective excitations. It does, however,
usually not give a full account of the fragmentation of the strength
which, as explained above, might not be needed at the astrophysical
conditions at which the heavy nuclei studied by the hybrid model
appear during the collapse.

The SMMC calculations of the partial occupation numbers have been
performed in large model spaces ($pf$-$sdg$ for nuclei with neutron
numbers $N \le 61$ and $pf_{5/2}$-$sdg$-$h_{11/2}$ for even heavier
nuclei) using adjusted pairing+quadrupole interaction to avoid the
infamous sign problem~\cite{Koonin.Dean.Langanke:1997}. The hybrid
model has been validated in~\cite{Juodagalvis.Langanke.ea:2010} and
applied to about 250 nuclei in the mass range
$A=66$--120~\cite{Langanke.Martinez-Pinedo.ea:2003,Juodagalvis.Langanke.ea:2010}.

\begin{figure}[htbp]
  \centering
  \includegraphics[width=0.7\linewidth]{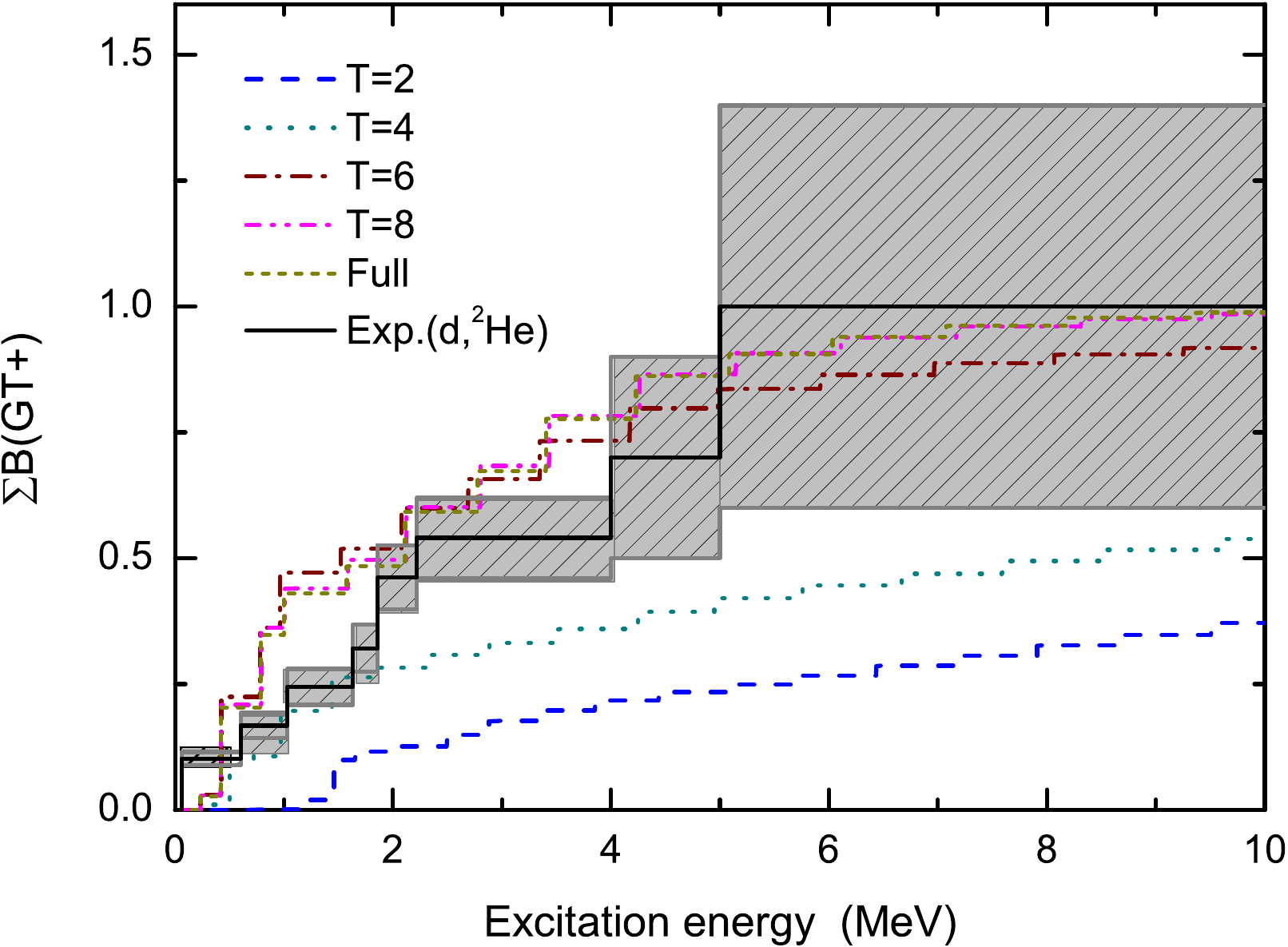}
  \caption{Comparison of experimental GT strength distribution for
    $^{76}$Se (shown as running sum) with results obtained by shell
    model diagonalization using the RG residual interaction and
    different levels of truncations
    (from~\cite{Zhi.Langanke.ea:2011}).\label{fig:76Se-GT}}
\end{figure}

In this context, a special nucleus is $^{76}$Se with $Z=34$ and
$N=42$. Thus, its GT$_+$ strength vanishes in the Independent Particle
Model (and in the Bruenn parametrization used in supernova simulations
prior to 2003~\cite{Bruenn:1985}). The GT strength has been
experimentally determined using the $(d,{}^2\textrm{He})$
charge-exchange technique at Groningen~\cite{Grewe.Baeumer.ea:2008}
proving that cross-shell correlations indeed unblock the GT strength
(see Fig.~\ref{fig:76Se-GT}). Diagonalization shell model
calculations, performed in different model spaces and with different
residual interactions, are able to describe the low-energy spectra of
$^{76}$Ge and $^{76}$Se and also the GT strength
(Fig. \ref{fig:76Se-GT}). These shell model calculations showed that
cross-shell correlations are a relatively slowly converging process
requiring the inclusion of multi-particle-multi-hole
configurations. For example, the consideration of only 2p-2h
configurations does not suffix to pull enough GT strength to low
energies (Fig. \ref{fig:76Se-GT}).

\begin{figure}[htbp]
  \centering
  \includegraphics[width=0.48\linewidth]{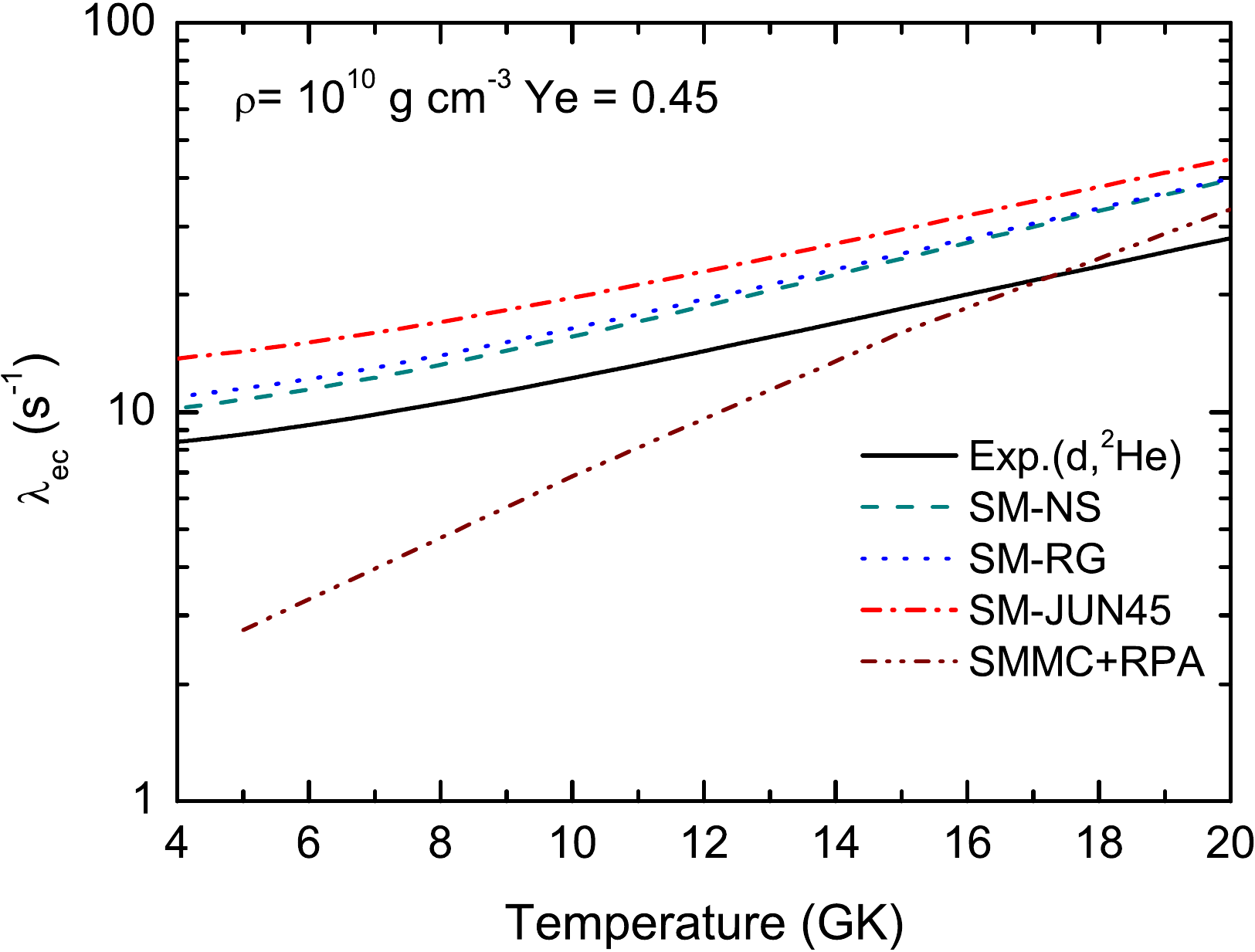}%
  \hspace{0.01\linewidth}%
  \includegraphics[width=0.48\linewidth]{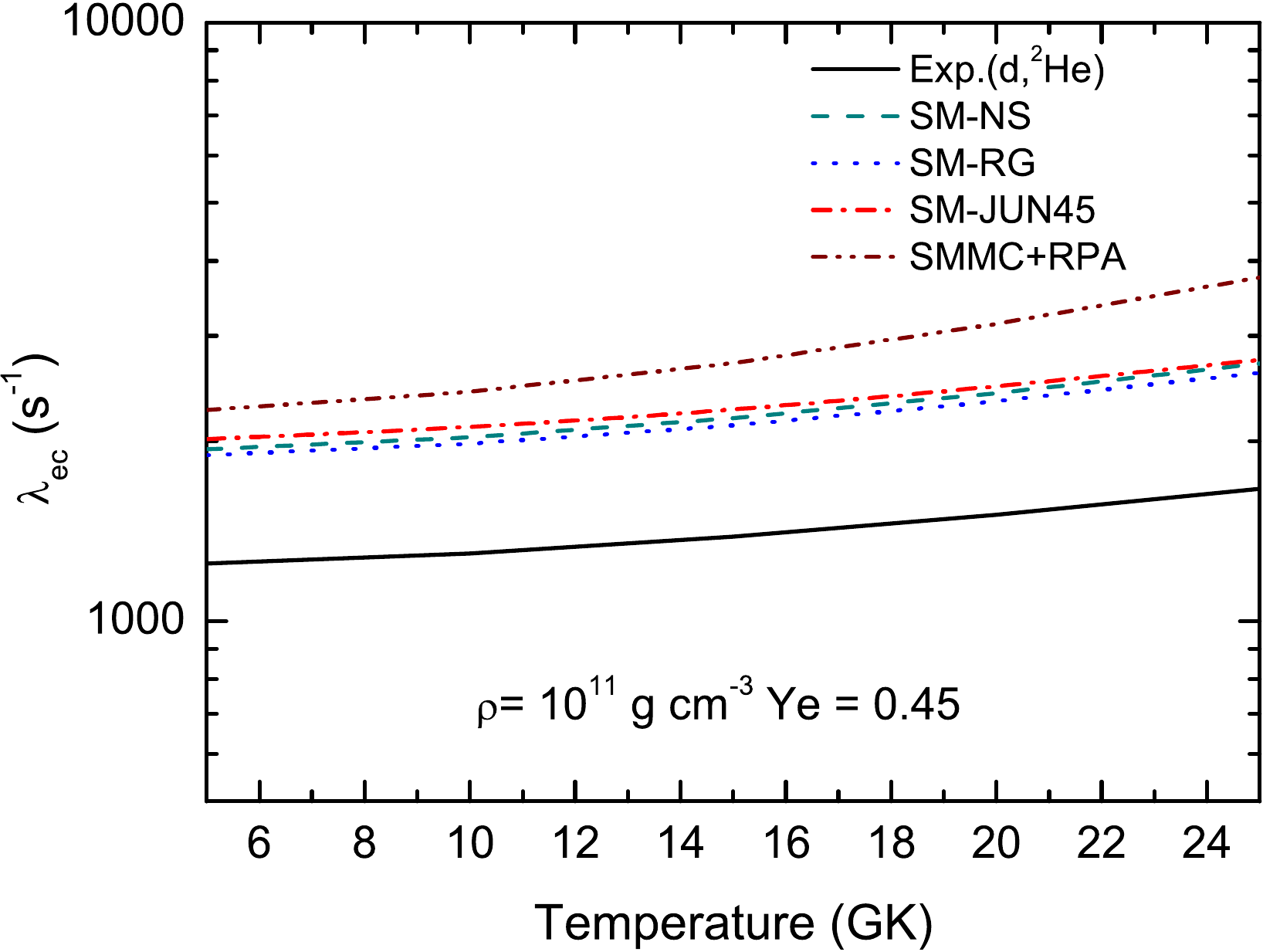}%
  \caption{Electron capture rates on $^{76}$Se at
    $\rho =10^{10}$~g~cm$^{-3}$ (left) and \mbox{$\rho=10^{11}$}~g~cm$^{-3}$
    (right) as function of temperatures. The rates have been
    calculated from the experimental ground state data~\cite{Grewe.Baeumer.ea:2008} and within diagonalization shell
    model approaches using different residual interactions. The
    results labelled `SMMC+RPA' have been obtained within the hybrid
    model.  (from \cite{Zhi.Langanke.ea:2011}).}
\label{fig:76Se-rate}
\end{figure}

We note that $^{76}$Se, being an odd-odd nucleus, is never very
abundant during core collapse. Nevertheless, Fig. \ref{fig:76Se-rate}
compares the electron capture es calculated from the experimental and
diagonalization shell model (for different interactions and model
spaces) GT distributions with those obtained in the hybrid model for
two different core densities and for various
temperatures~\cite{Zhi.Langanke.ea:2011}.  The lower density
corresponds to presupernova conditions, where electron capture is
dominated by $pf$ shell nuclei. The rates calculated from the data and
the shell model GT strength distributions agree quite well. The hybrid
model rates agree with the other rates within a factor of 3 for the
range of temperatures given, but they show a distinct different
$T$-dependence. This is related to the fact that the hybrid model does
not resolve the fragmentation of the GT strength, which is
particularly important at low temperatures and densities. In fact, at
the higher density, the agreement between all rates is quite
satisfactory. Under these conditions the electron chemical potential
is noticeable larger than the capture $Q$-value, making the rate less
sensitive to details of the GT distribution. The hybrid model
calculation considers also forbidden multipoles whose contributions
increase with temperature, but are relatively
small~\cite{Zhi.Langanke.ea:2011}. We note that the shell model and
experimental rates are solely determined from the ground state GT
distribution, while the hybrid model considers finite temperature
effects in the calculation of the occupation numbers. These turn out
to be not so relevant as the $N=40$ gap is already strongly overcome
by correlations in the ground state. This will be different for the
$N=50$ gap, discussed below.

The Thermal Quasiparticle RPA (TQRPA)
model~\cite{Takahasi.Umezawa:1975} is an alternative approach proposed
to calculate electron capture (and neutrino-nucleus reaction) rates at
finite
temperatures~\cite{Dzhioev.Vdovin.ea:2010,Dzhioev.Vdovin.Stoyanov:2019}. Like
the SMMC, also the TQRPA is based on an equilibrium statistical
formalism and treats the many-nucleon system in a heat bath and a
particle reservoir in the grand canonical ensemble. The method can be
understood as a proton-neutron QRPA approach extended to finite
temperatures and allows to determine temperature-dependent spectral
functions which are the basis to evaluate weak-interaction rates
within this model~\cite{Dzhioev.Vdovin.ea:2010}. Further extensions
allow to use
Skyrme~\cite{Paar.Colo.ea:2009,Dzhioev.Vdovin.Stoyanov:2019} and
relativistic functionals~\cite{Niu.Paar.ea:2011} to describe the
thermal state and its excitation considering 2p-2h correlations.

Compared to the hybrid model, the TQRPA has the advantage to be
formally consistent in treating the many-body problem. In contrast,
the two parts of the hybrid model have a different complexity in
dealing with the many-body states. It turns, however, to be important
that the SMMC considers multi-particle-multi-hole correlations as will
be discussed below.

\begin{figure}[htbp]
  \centering
  \includegraphics[width=0.8\linewidth]{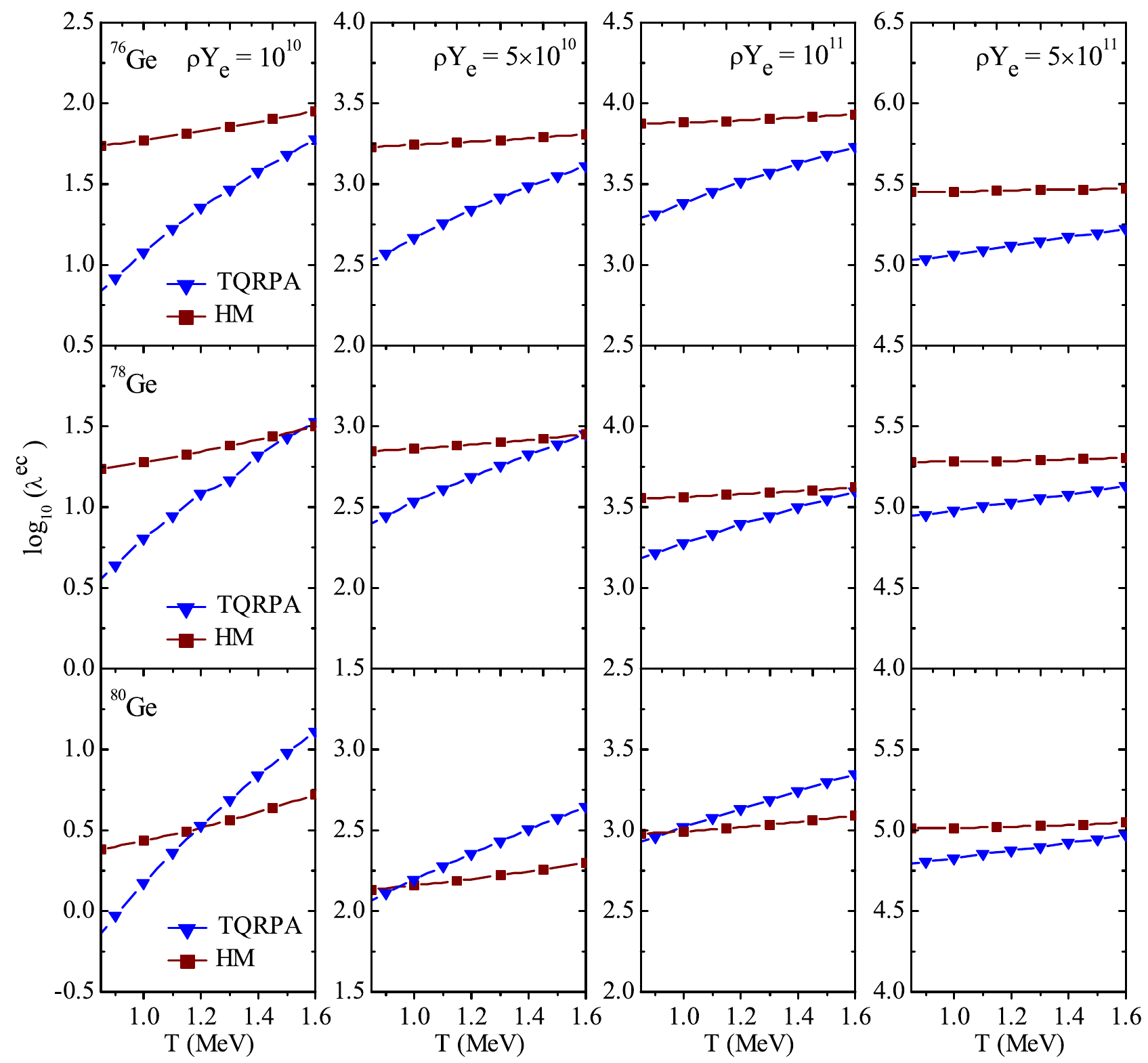}
  \caption{Comparison of electron capture rates for $^{76,78,80}$Ge
    for different densities $\rho Y_e$ and as function of temperature,
    calculated within the TQRPA and hybrid model (HM).  (from
    \cite{Dzhioev.Vdovin.ea:2010}).}
\label{fig:Ge-rates}
\end{figure}

The TQRPA approach has been used to calculate electron capture at
finite temperatures for selected Fe and Ge isotopes~\cite{Dzhioev.Vdovin.ea:2010} and for nuclei at the $N=50$ shell
closure~\cite{Dzhioev.Vdovin.Stoyanov:2019,Dzhioev.Langanke.ea:2020}. The
differences between the two models become illustrative in
Fig.~\ref{fig:Ge-rates} which compares the electron capture rates
calculated in both approaches for various neutron-rich Ge isotopes at
different densities and temperatures. In general, as the electron
chemical potential grows with density and temperature, the rates
increase as well where the sensitivity is larger to density than to
temperature. The rates decrease with increasing neutron numbers. This
has two reasons. Foremost, the $Q$ value increases, but also the
occupation of the $g_{9/2}$ neutron orbital grows decreasing the
unblocking of $pf$ shell neutron orbitals. Neutron-rich Ge isotopes
appear in the core composition at temperatures $ \ge 1$~MeV and
densities $\rho Y_e \ge 10^{11}$~g~cm$^{-3}$ and both models predict
quite sizable capture rates for these conditions. There are, however,
differences between the two models.  In general, the hybrid model
capture rates are larger than those obtained in the TQRPA, most
evidently at lower densities.  Furthermore, the TQRPA model shows a
steeper rise of the capture rates with temperature than the hybrid
model. These facts are foremost related to the increased unblocking
probabilities in the hybrid model due to many-body correlations which
result in larger GT strength at lower excitation energies. The
differences in the rates become smaller with increasing density and
temperature. This is mainly due to the growing electron chemical
potential which makes the rate less sensitive to the details of the GT
strength distribution.  Secondly, forbidden transitions contribute
increasingly with growing density and temperature. These contributions
are not subject to blocking effects.

\begin{figure}[htbp]
  \centering
  \includegraphics[width=0.70\linewidth]{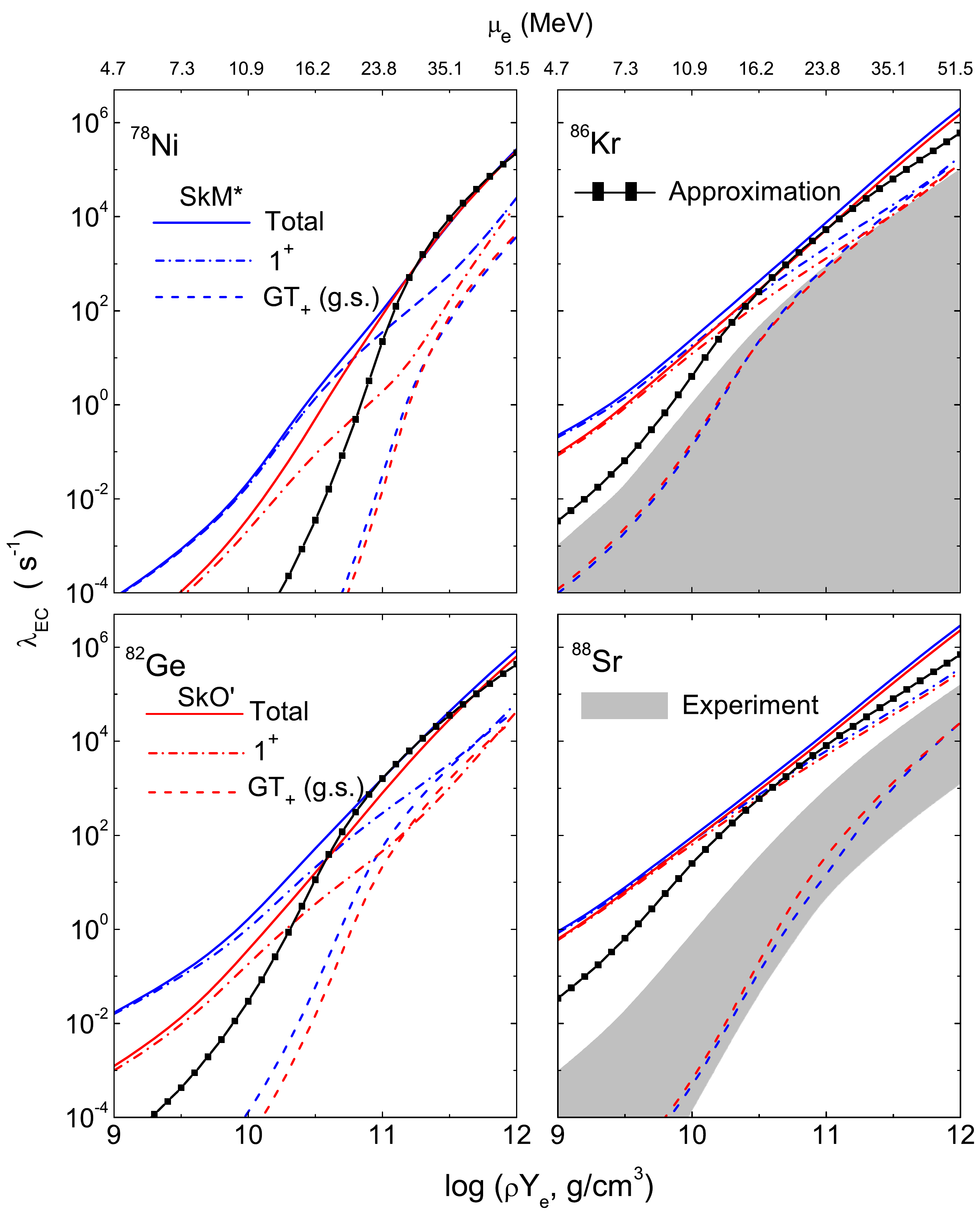}
  \caption{TQRPA electron capture rates for selected $N=50$ nuclei
    calculated at $T=0$ and at $T=1$ MeV and as function of
    density. The upper axis shows the corresponding electron chemical
    potential. The calculations have been performed for two Skyrme
    interactions: SkM$^*$ (blue lines) and SkO' (red lines).  The
    calculated total capture rates include also contributions from
    forbidden transitions; the GT contribution is presented
    individually. The shaded area is the rate obtained from the
    experimental ground state GT distribution (taken from~\cite{Zamora.Zegers.ea:2019}. The thick line labelled
    `Approximation' represents the 'single state approximation'
    adopted from \cite{Langanke.Martinez-Pinedo.ea:2003}.  (from
    \cite{Dzhioev.Langanke.ea:2020}).}
\label{fig:N50-rates}
\end{figure}

We have seen that many-body correlations overcome the $N=40$ shell
closure already in the ground state and unblock the GT contribution to
the capture rate. But what happens at the magic number $N=50$? In
fact, measurements of the GT$_+$ distribution for $^{86}$Kr ($Z=36$
and $N=50$)~\cite{Titus.Ney.ea:2019} and for $^{88}$Sr ($Z=38, N=50)$
shows only very little strength, mainly located at excitation energies
between 8--10~MeV~\cite{Zamora.Zegers.ea:2019}. This points to a
rather strong blocking of GT transitions at $N=50$.  Electron capture
rates, calculated from the experimental ground state data, are indeed
significantly lower than expected from
systematics~\cite{Zamora.Zegers.ea:2019}. The results for $^{86}$Kr
and $^{88}$Sr are surprising, given that a significant amount of GT
strength ($\sim 0.7$ units) was observed for
$^{90}$Zr~\cite{YAKO2005193} even though, based on transfer reaction
experiments~\cite{PFEIFFER1986381}, the proton $0g_{9/2}$ occupation
number for $^{88}$Sr and $^{90}$Zr are comparable: 0.7 and 1.0,
respectively. A high-resolution experiment for $^{90}$Zr will be
necessary to better understand these results.

In the collapsing core, $N=50$ nuclei (e.g.  $^{82}$Ge and $^{78}$Ni
with $Y_e$ values of 0.39 and 0.34, respectively) are very abundant at
densities in excess of about
$10^{11}$~g~cm$^{-3}$~\cite{Janka.Langanke.ea:2007,0004-637X-816-1-44}
and at temperatures $T > 1$ MeV. At these high 
temperatures the average nuclear excitation energy is about
$\langle E \rangle = 10$~MeV, which is larger than the $Z=28$ proton
gap and the $N=50$ neutron gap.  This implies that the capture at the
stellar temperatures occurs on average on states with important
many-body correlations across the two gaps, in this way unblocking the
GT contribution to the capture rate. This is indeed born out in TQRPA
calculations performed for $N=50$ nuclei between $^{78}$Ni and
$^{88}$Sr. The obtained capture rates are shown in
Fig.~\ref{fig:N50-rates}.  Satisfyingly the TQRPA calculations finds
no GT strength in the $^{86}$Kr and $^{88}$Sr ground states at low
energies, in agreement with observation. In fact the TQRPA capture
rates, calculated solely from the $T=0$ GT distributions, agree with
those obtained from the experimental GT distributions for both nuclei
(see Fig.~\ref{fig:N50-rates}). The TQRPA calculation shows, however,
a strong thermal unblocking of the GT strength as protons are moved
into the $g_{9/2}$ orbital and neutrons out of the $pf$ shell. This
leads to a strong increase in the capture rate for all nuclei (see
Fig.~\ref{fig:N50-rates}). Thermal unblocking of the GT strength has
the largest effect at small electron chemical potentials $\mu_e$ (low
densities), while its relative importance decreases with growing
$\mu_e$. With increasing density contributions from forbidden
transitions become more important and dominate the rate for densities
of order $\rho Y_e > 10^{11}$~g~cm$^{-3}$, hence at the conditions
where $N=50$ nuclei are abundant in the collapse. The capture rates
also increase with increasing proton numbers, i.e. from $^{78}$Ni to
$^{88}$Sr. This has two reasons: the growing $Q$ value with neutron
excess and the increased promotion of protons into the $g_{9/2}$
orbital.

In summary, GT measurements for nuclei which become relevant in the
high density/temperature environment during supernova collapse are
indispensable to constrain nuclear models and to create trust in
them. However, they cannot directly been used to determine the stellar
capture rate as thermal unblocking effects modify the rates under such
conditions noticeably. This is in particular true at shell closures,
i.e. for $N=50$ nuclei. For these nuclei forbidden transitions might
be as relevant as GT transitions and should be experimentally
constrained as well.

\begin{figure}[htbp]
  \centering
  \includegraphics[width=0.70\linewidth]{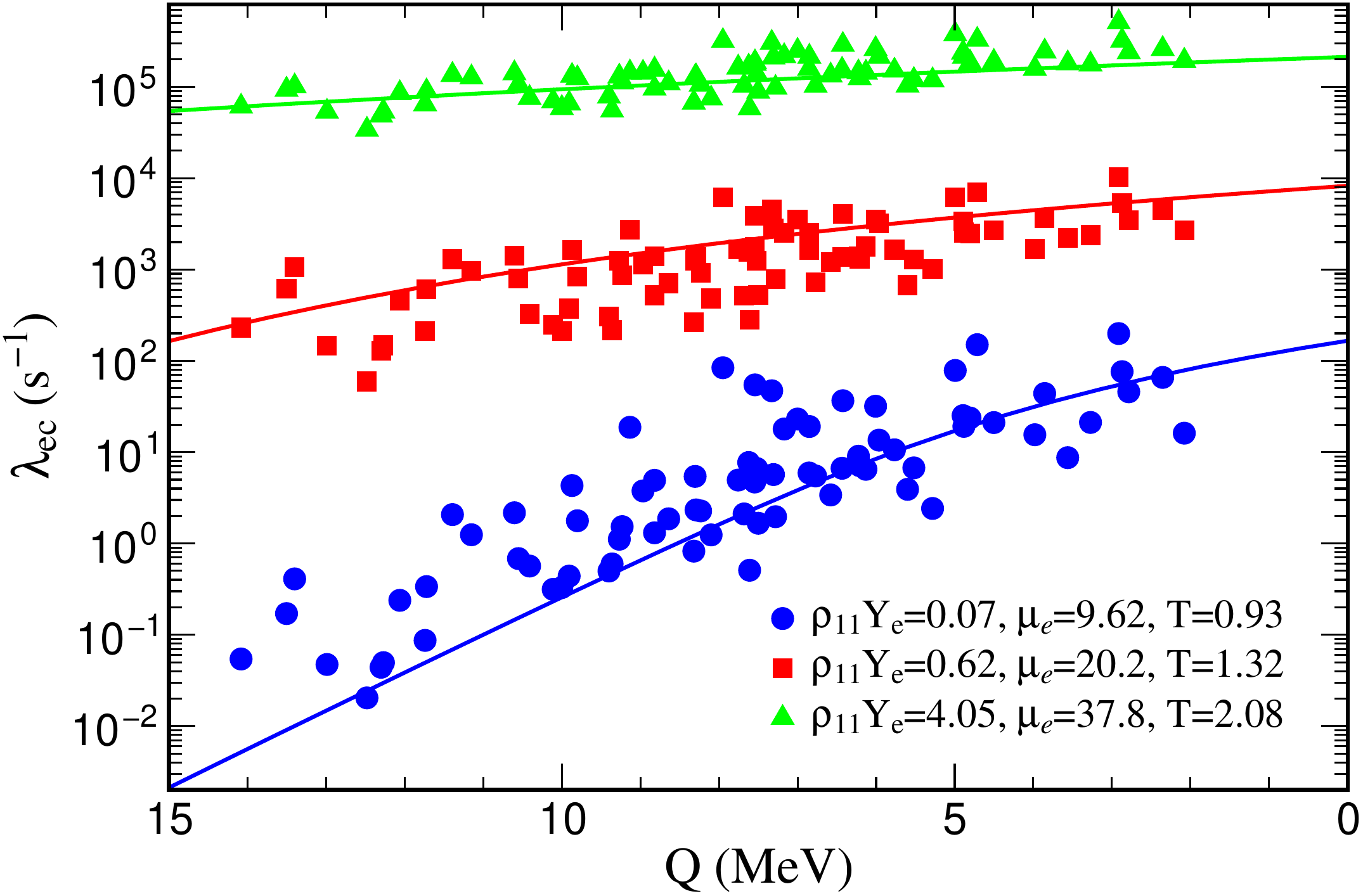}
  \caption{Electron capture rates on nuclei, for which individual
    shell model rates exist, as function of $Q$ value for 3 different
    stellar conditions. Temperatures are measured in MeV, density in
    $10^{11}$~g~cm$^{-3}$. The solid lines represent the rates
    obtained from the single-state approximation
    \ref{eq:single-state}. (from
    \cite{Langanke.Martinez-Pinedo.ea:2003}).}
\label{fig:single-fit}
\end{figure}

Fig.~\ref{fig:N50-rates} also shows the rate estimated by a
parametrization put forward in
Ref.~\cite{Langanke.Martinez-Pinedo.ea:2003}. This simple
parametrization assumes that the capture proceeds through a single
transition from an excited state in the parent nucleus at $E_i$ to a
state in the daughter nucleus at $E_f$ with $\Delta E = E_f - E_i$
(single-state approximation). Then the capture rate can be written as~\cite{Fuller.Fowler.Newman:1985}
\begin{equation}
  \label{eq:single-state}
  \lambda = \frac{\ln(2) B}{K} \left(\frac{T}{m_e c^2}\right)^5
  \left[F_4(\eta) + 2 \chi F_3(\eta) +\chi^2 F_2 (\eta)\right]
\end{equation}
where $\chi = (Q+\Delta E)/T$, $\eta=(\mu_e - Q - \Delta E)/T$,
$K=6146$~s and $B$ represents a typical (Gamow-Teller plus forbidden)
matrix element. The quantities $F_k$ are the relativistic Fermi
integrals of order $k$.  $Q$ is the ground state ground state
$Q$-value that is positive for capture in protons and neutron-rich
nuclei.  This approximation was used in
Refs.~\cite{Langanke.Martinez-Pinedo.ea:2003,Hix.Messer.ea:2003} to
estimate the rates of the many heavy nuclei which are abundant at
larger densities and for which no rates existed at that time.  The two
parameters (energy position and GT strength) were fitted to the rates
of about 200 nuclei for which individual $pf$ shell model and hybrid
model rates were available.  Fig.~\ref{fig:single-fit} compares the
shell model rates with the single-state
approximation~(\ref{eq:single-state}) using $B=4.6$ and
$\Delta E=2.5$~MeV. We note that the approximation does not consider
nuclear structure effects (or a dependence on the average excitation
energy) which result in quite a significant scatter of the shell model
rates with respect to the single-state rate. For the reasons discussed
above, the fluctuations get noticeably reduced with increasing
density.  It is worth noting that there is no systematic difference
between the approximation and the shell model rates so that
differences might at least partially cancel out. At the intermediate
density the single-state approximation shows some tendency to
overestimate the rate. In conclusion, the approximation in its simple
form~(\ref{eq:single-state}) should not be used at low densities say
below a few $10^{10}$~g~cm$^{-3}$. In this density regime the nuclear
composition is largely dominated by nuclei for which shell model rates
exist. The general trend seen in Fig.~\ref{fig:single-fit} is also
borne out in Fig.~\ref{fig:N50-rates} where the approximation badly
fails at low densities, but gives reasonable agreement at
$\rho Y_e > 10^{11}$ g~cm$^{-3}$. Ref.~\cite{0004-637X-816-1-44}
compares the shell model and single-state rates at slightly different
astrophysical conditions.

The single-state parametrization has been adopted for heavy nuclei in
supernova simulations which systematically studied the influence of
nuclear ingredients (electron capture rates, Equation of State, mass
models) on the collapse dynamics
\cite{0004-637X-816-1-44,Pascal.Giraud.ea:2020} (see below),
where Ref. ~\cite{Pascal.Giraud.ea:2020} used the improved
single-state parametrization of \cite{Raduta.Gulminelli.Oertel:2017}
(see below).

\subsection{Rate Tables}

Most supernova codes now use the rate table as provided by Juodagalvis
\emph{et al.}~\cite{Juodagalvis.Langanke.ea:2010}.  This table defines
electron capture rates on a grid of the three important parameters
characterizing the astrophysical conditions during collapse:
temperature, density, $Y_e$ value.  The rate evaluation assumes the
core composition to be given by nuclear statistical equilibrium, hence
it does not provide rates for individual nuclei.

The rate table is based on the hierarchical strategy defined
above. For the nuclei with $A < 65$ the shell model rates of Oda
\emph{et al.}~\cite{Oda.Hino.ea:1994} ($sd$ shell) and of Langanke and
Martinez-Pinedo~\cite{Langanke.Martinez-Pinedo:2001} ($pf$ shell) have
been adopted. This guarantees a reliable and detailed reproduction of
the GT strengths for the important nuclei at collapse conditions where
$\mu_e \sim Q$. The rates for nuclei in the range $A=39-44$ have been
taken from Fuller, Fowler and Newman
\cite{Fuller.Fowler.Newman:1982a}. For the heavier nuclei the table
adopts the rates from hybrid model calculations. For about 200 nuclei
in the mass range $A = 65-110$ these were calculated by using SMMC
partial occupation numbers in RPA calculations.  For a few nuclei in
this mass regime and for even heavier nuclei, in total about 2700
nuclei, the rates were evaluated on the basis of a parametrization of
the occupation numbers, derived in accordance with the SMMC studies,
and RPA response calculations. In this way the most relevant nuclear
structure input, like shell gaps, are accounted for.  Screening
corrections due to the astrophysical environment have been
incorporated into the rates.

Weak-interaction rates for $sd$ shell nuclei are important for the
core evolution of intermediate mass stars. Rates for individual nuclei
for the relevant density and temperature regime are given
in~\cite{Oda.Hino.ea:1994} and updated in~\cite{Suzuki.Toki.Nomoto:2016}.

Nuclei in the mass range $A = 45$--65 are essential for the early
phase of core collapse supernovae and for the nucleosynthesis in
thermonuclear (Type Ia) supernovae. The weak-interaction rates for
these $pf$ shell nuclei are individually given
in~\cite{Langanke.Martinez-Pinedo:2001}.  The rates are not corrected
for screening, which, however can be accounted for using the formalism
developed in \cite{Juodagalvis.Langanke.ea:2010}.

We note that at specific astrophysical conditions (e.g. during silicon
burning), at which the $sd$ and $pf$ shell nuclei are relevant, the
temperature is in general not high enough to establish an NSE
composition. Hence the knowledge of individual rates is essential.

To make it easier to incorporate complete sets of electron-capture
rates in astrophysical simulations, a library of rates was created
\cite{0004-637X-816-1-44,0954-3899-45-1-014004,weakrlib} based on the
rate tables for specific mass regions described above and on the
single-state approximation for nuclei where rates based on microscopic
calculations are not available. This library is incorporated in the
weak-rate library NuLib \cite{OCO15}.

\section{Electron captures in astrophysical applications}

\subsection{Core-collapse supernovae}
\label{sec:core-coll-supern}

Simulations of the evolution of massive stars distinguish two distinct
phases motivated by their specific needs and requirements. 1) During
hydrostatic burning energy released by nuclear reactions in the star's
interior are essential to balance gravity. The densities are low
enough that neutrinos, produced in weak interactions, can leave the
star unhindered transporting energy away. This loss has to be
considered in the energy balance, but a detailed treatment of neutrino
transport is not required. However, the simulations have to
incorporate a detailed network of nuclear reactions to follow the
nuclear energy production and the change in composition. This stellar
evolution period lasts to the so-called presupernova phase when the
core density has reached values of about $10^{10}$~g~cm$^{-3}$ and the
inner part of the iron core collapses with velocities in excess of
1000~km~s$^{-1}$~\cite{Woosley.Weaver:1995,Heger.Woosley.ea:2001}.

The final models obtained by the stellar evolution codes become the
input for the supernova codes in which the gravitational collapse of
the iron core and the explosion are simulated. The astrophysical
conditions relevant during these simulation lead to two important
changes compared to stellar evolution. The temperatures are
sufficiently high ($T >$ a few GK) so that the nuclear composition can
be well approximated by an NSE distribution, without the need to
follow a complicated network of nuclear reactions. On the other hand,
the involved densities require a detailed bookkeeping of
neutrinos. This is achieved by Boltzmann transport. An additional
complication arises from the fact that the assumption of spherical
symmetry, which holds approximately during hydrostatic stellar
evolution, is not valid during the core collapse and explosion. This
requires multidimensional treatments which is extremely challenging
and computationally demanding. Reviews about the recent impressive
progress in supernova modelling can be found
in~\cite{Kotake.Sumiyoshi.ea:2012,Burrows:2013,Janka.Melson.Summa:2016,Mueller:2020}. These
codes consider electron capture via the rates provided by~\cite{Juodagalvis.Langanke.ea:2010}.

\begin{figure}[htbp]
  \centering
  \includegraphics[width=0.32\linewidth]{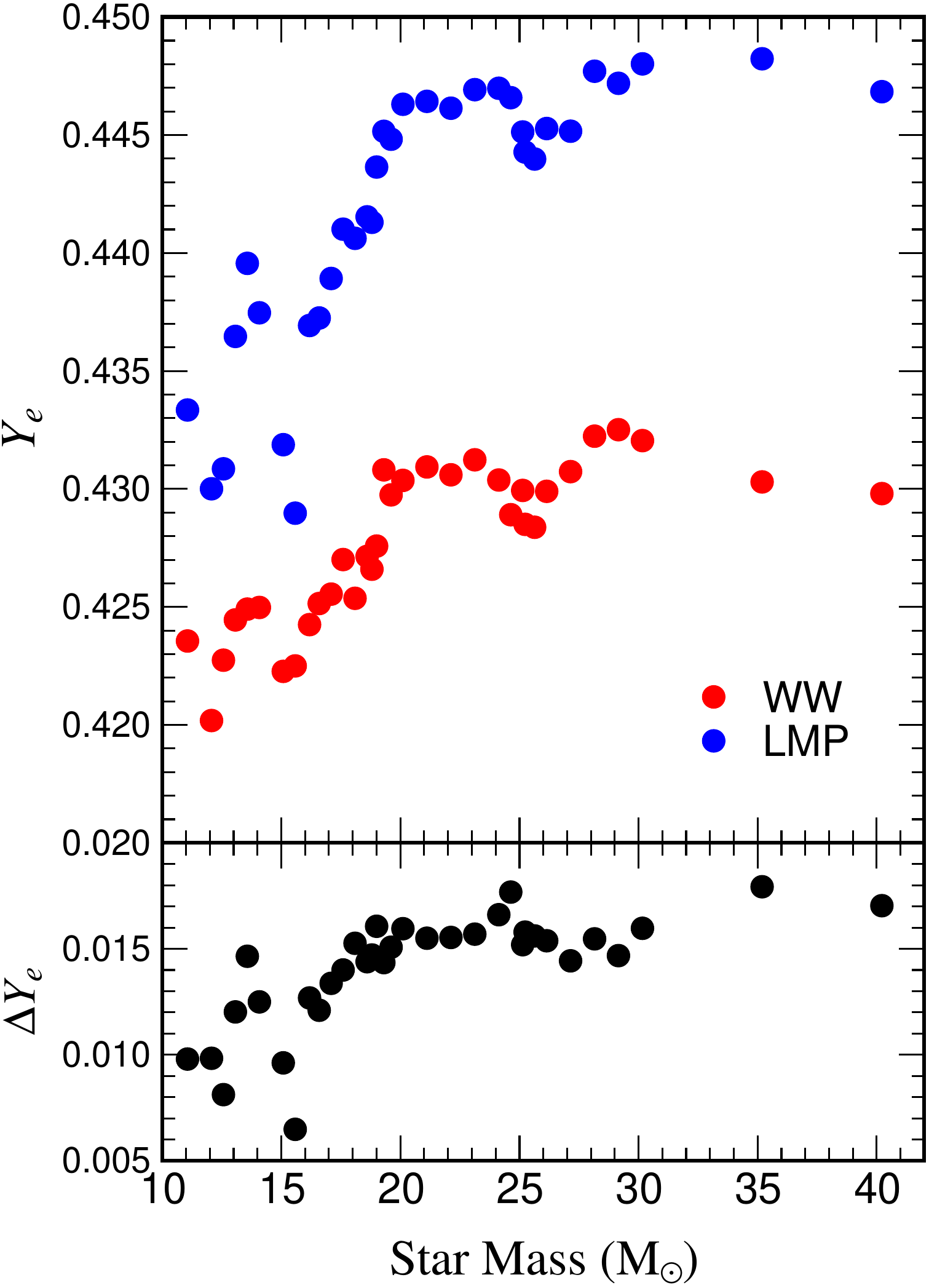}%
  \hspace{0.01\linewidth}%
  \includegraphics[width=0.32\linewidth]{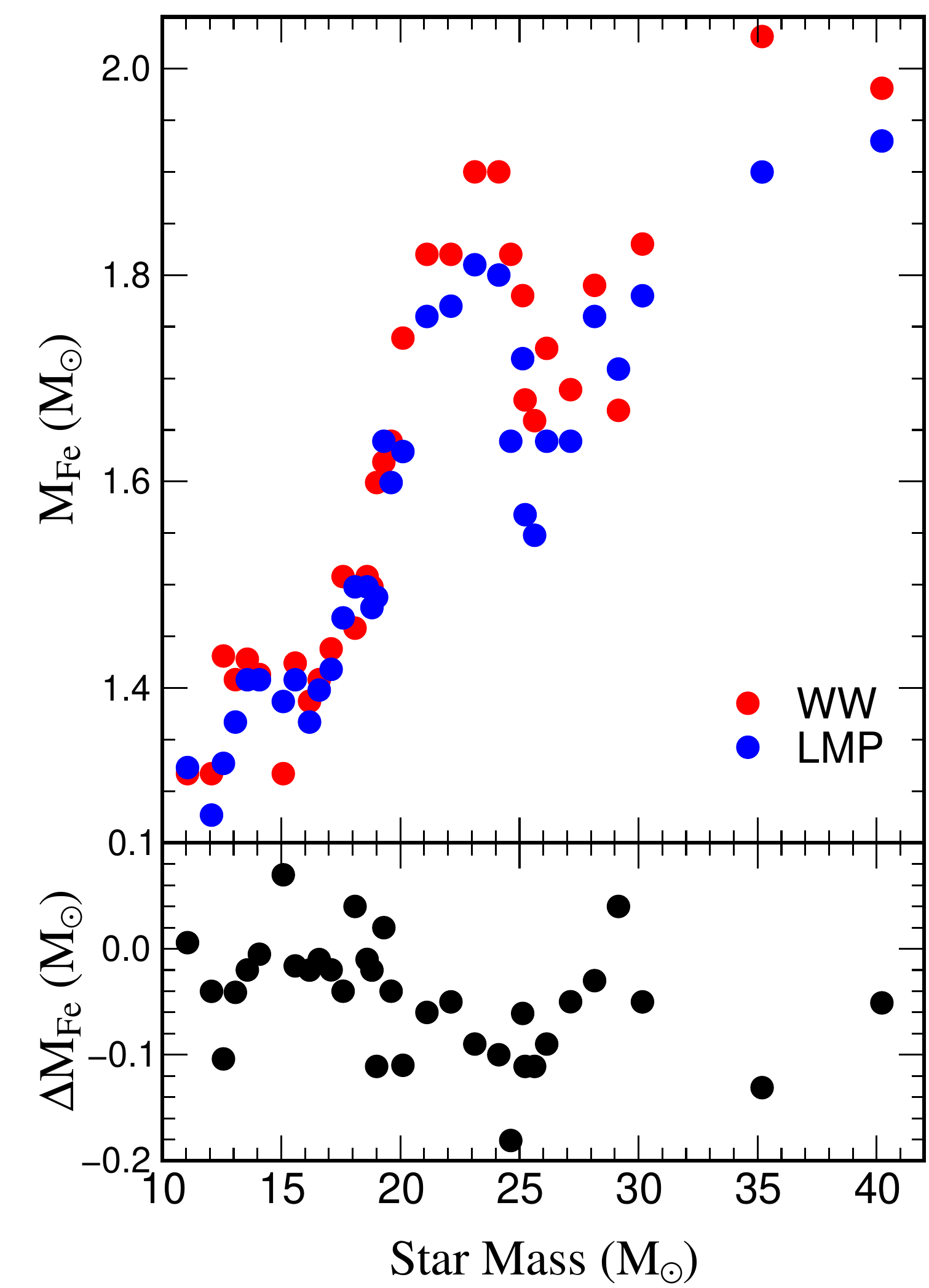}%
  \hspace{0.01\linewidth}%
  \includegraphics[width=0.32\linewidth]{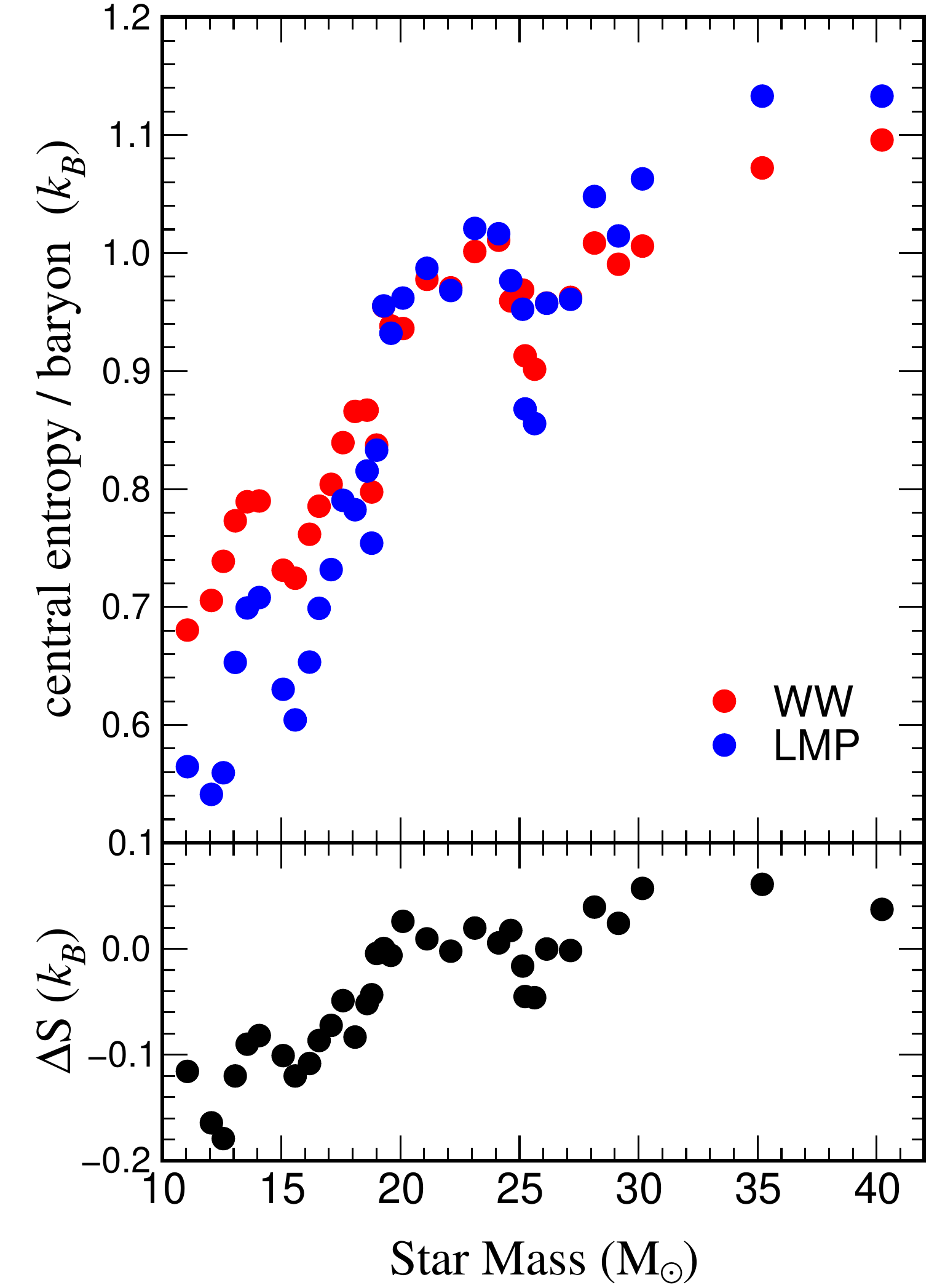}
  \caption{Comparison of the center values of $Y_e$ (left), the iron
    core sizes (middle) and the central entropy (right) for
    $11-40 M_\odot$ stars between the models using the FFN rates (WW
    models \cite{Woosley.Weaver:1995}) and models which used the shell
    model weak interaction rates (LMP
    \cite{Heger.Langanke.ea:2001}). The lower panels show the changes
    in the 3 quantities between the WW and LMP models.}
  \label{fig:presupernova}
\end{figure}
  
Heger \emph{et al.} have investigated which effect the diagonalization
shell model rates have on the presupernova evolution for stars in the
mass range
$M =
13$--40~M$_\odot$~\cite{Heger.Woosley.ea:2001,Heger.Langanke.ea:2001}. To
this end they repeated calculations of Weaver and Woosley
\cite{Woosley.Weaver:1995}, keeping the stellar physics as much as
possible, but replacing the weak interaction rates for $pf$ shell
nuclei by those of Ref. \cite{Langanke.Martinez-Pinedo:2001} (LMP
rates). Fig. \ref{fig:presupernova} summarizes which consequences the
shell model rates have on three quantities which are relevant for the
following collapse. The central $Y_e$ value is larger by
$\Delta Y_e = 0.01$--0.015 at the onset of collapse. This has two
reasons. First, the shell model rates are noticeably smaller than the
FFN rates (Fig.~\ref{fig:FFNvsSM}), hence reducing
leptonization. Second, during silicon burning $\beta$ decays can
compete with electron captures. Although this does not occur by
specific URCA pairs, but rather by an ensemble of nuclei, the effect
is the same: the star is additional cooled, while the $Y_e$ is kept
constant. The study confirmed that $\beta$ decays become increasingly
Pauli blocked with growing density and can be safely neglected during
collapse. Fig. \ref{fig:presupernova} also indicates that the iron
core masses are generally smaller with the LMP rates. However, this is
not a continuous effect and shows variations among the models with
different stellar masses. Finally, the LMP rates lead to presupernova
models with lower core entropy for stars with $M<20$~M$_\odot$. For the
more massive stars, the effect is not unique; stars with
$M=30$--40~M$_\odot$ show an increased core entropy. We mention that
lower (larger) core entropy implies less (more) free protons in the
nuclear composition, which, however, is overwhelmingly dominated by
nuclei.

The continuous electron capture drives the NSE composition of the core
more neutron-rich and towards heavier nuclei. At densities in access
of a few $10^{10}$~g~cm$^{-3}$ the composition is dominated by nuclei
with $Z<40$ and $N>40$ for which, for a long time, it was assumed that
electron captures vanish (e.g.~\cite{Bruenn:1985}) due to Pauli
blocking of the GT strength. As a consequence the capture process at
the later of the collapse continued solely on free protons, which are,
however, less abundant than heavy nuclei by orders of magnitude. As we
have discussed above, the GT strength at the $N=40$ shell gap is
unblocked by multi-nucleon correlations.  Furthermore, the blocking at
the $N=50$ shell closure, which results in a strong reduction in the
experimental ground state GT strength, is overcome at the
finite-temperature core conditions by thermal excitations.
  
Arguable the most important result reported in
Refs. \cite{Langanke.Martinez-Pinedo.ea:2003,Hix.Messer.ea:2003} is
the fact that electron capture proceeds on nuclei rather than on free
protons during the entire collapse, in contrast to previous belief
(e.g.~\cite{Bethe:1990}.  These findings are based on supernova
simulations performed independently by the Garching and Oak Ridge
groups which both adopted the hybrid model capture rates for more than
100 nuclei in the mass range $A=65$--110, supplemented by the shell
model rates for $pf$ shell nuclei. For the heavy nuclei, the capture
rates were estimated by the single-state approximation. The capture
rate on free protons was taken from~\cite{Bruenn:1985}.

\begin{figure}[htbp]
  \centering \hspace{-0.01\linewidth}%
  \centering \includegraphics[width=0.32\linewidth]{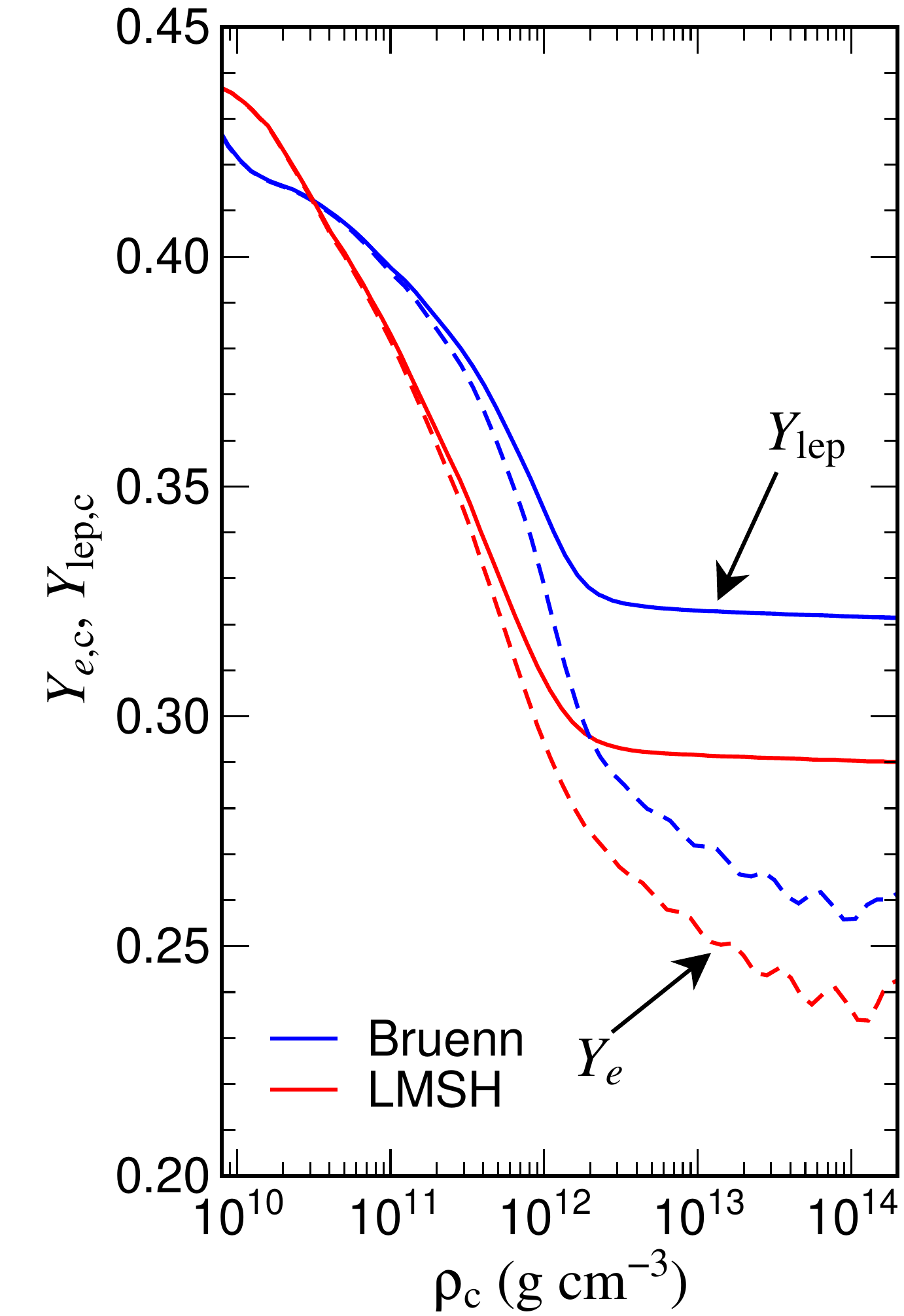}%
  \hspace{-0.01\linewidth}%
  \includegraphics[width=0.32\linewidth]{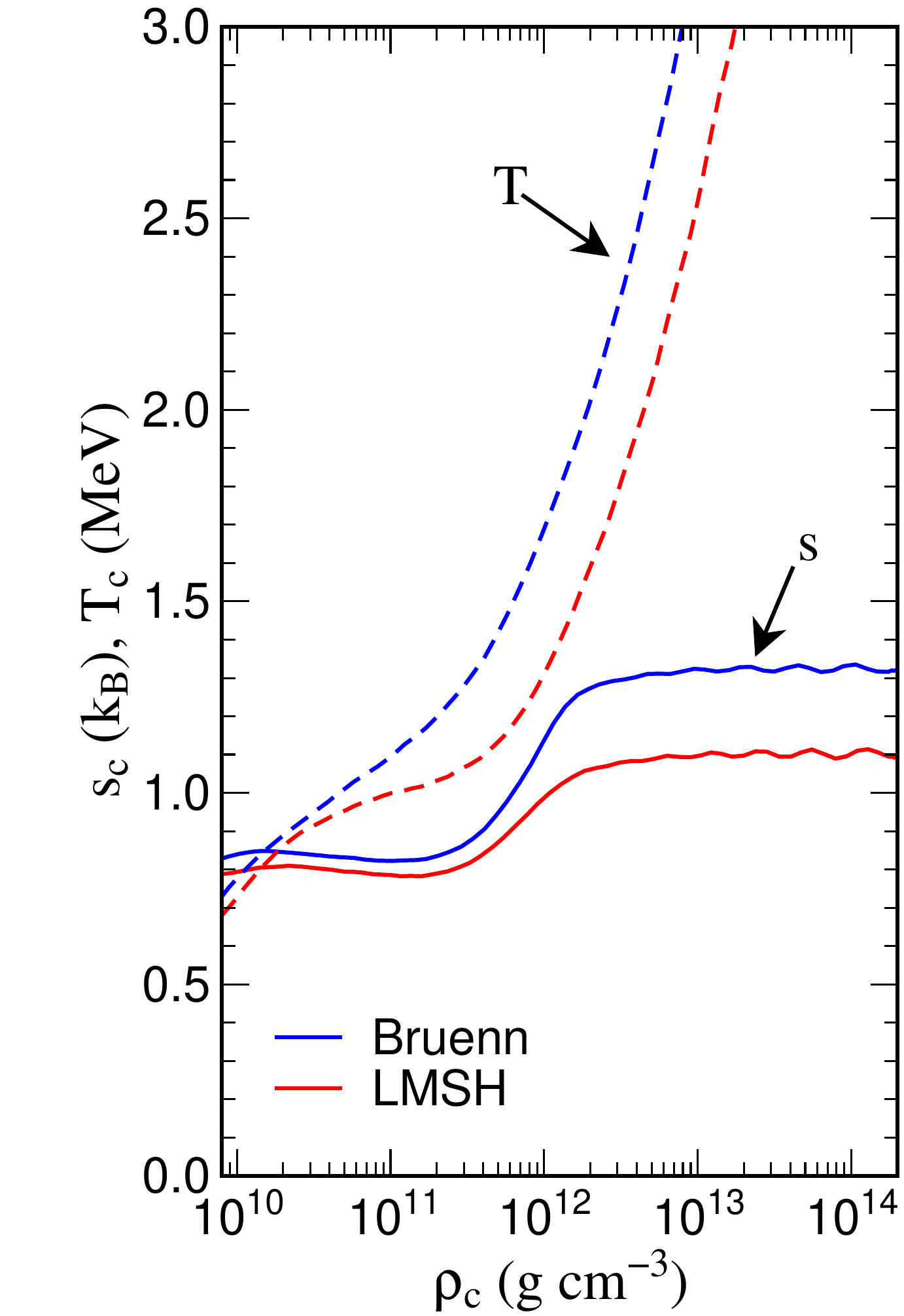}%
  \hspace{0.01\linewidth}%
  \includegraphics[width=0.32\linewidth]{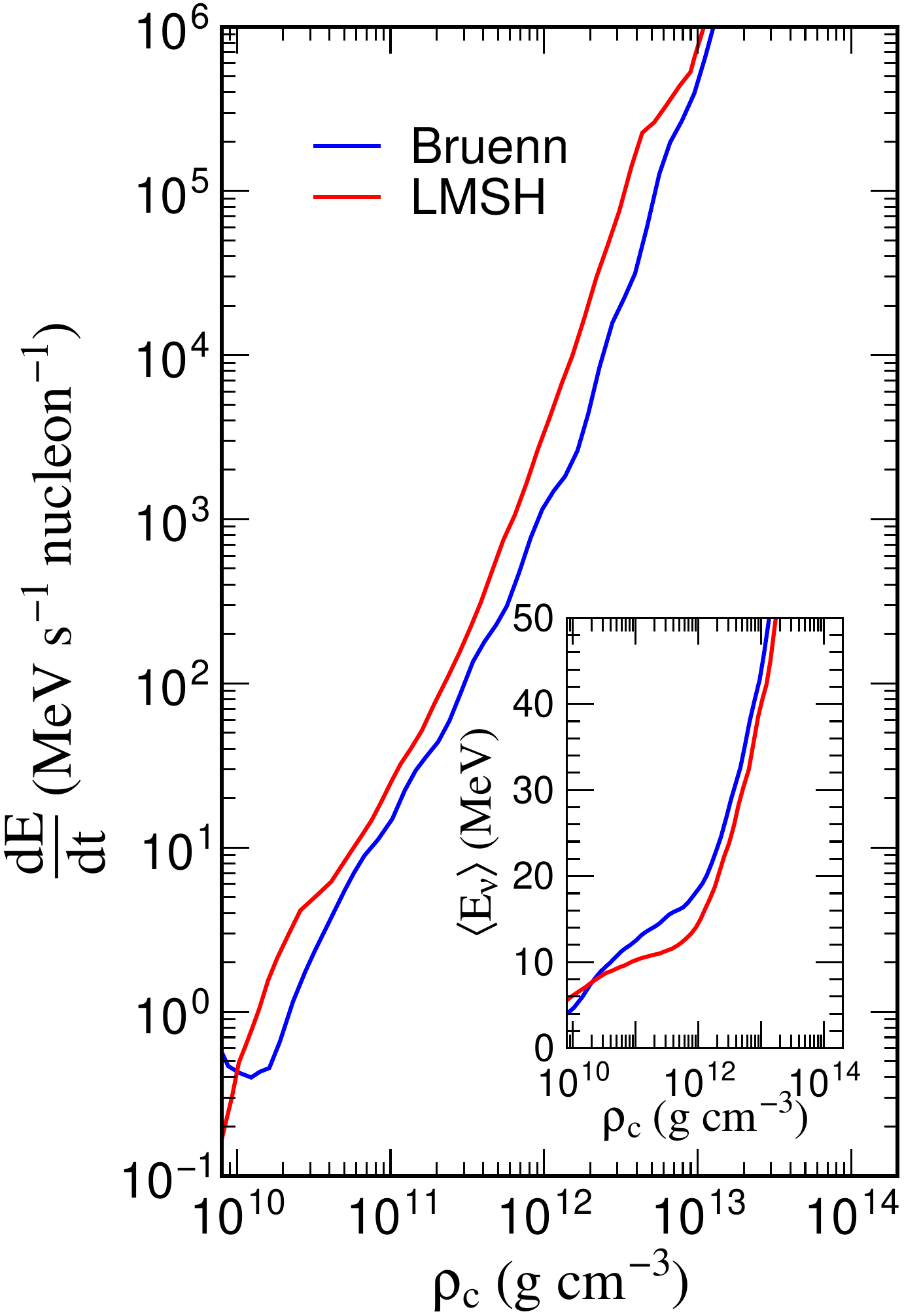}
  \caption{Comparison of a supernova simulation for a 15 $M_\odot$
    star using the shell model weak interaction rates from
    Ref.~\cite{Langanke.Martinez-Pinedo:2003} (labelled LMSH) and the
    Bruenn parametrization which neglects capture on nuclei for $N>40$
    \cite{Bruenn:1985} (labelled Bruenn). The figure shows the central
    core values for $Y_e$ and $Y_lep$ (electrons plus neutrinos)
    (left), the entropy and temperature (middle) and the neutrino
    emission rate (right) as function of core density. The insert in
    the right figure shows the average energy of the emitted neutrinos
    (courtesy of Hans-Thomas Janka).}
  \label{fig:weak-rates-collapse}
\end{figure}

Fig.~\ref{fig:weak-rates-collapse} compares some important quantities
obtained in the simulations of
refs. \cite{Langanke.Martinez-Pinedo.ea:2003,Hix.Messer.ea:2003} with
previous studies which neglected electron captures on heavy nuclei.
Obviously the capture on nuclei is an additional source of
deleptonization, adding to the capture on free protons. This results
in significantly lower values for $Y_e$ at neutrino trapping at
densities around $10^{12}$~g~cm$^{-3}$.  At higher densities the total
lepton fraction $Y_{\text{lep}}$ becomes constant, while the electron
fraction $Y_e$ still decreases. This is related to neutrino tapping
and the formation of the homologous core \cite{Bethe:1990}. In this
regime, continuous electron captures reduce the electron abundance,
but the neutrinos generated by this process interact with matter
mainly by coherent scattering on nuclei with a rate large enough that
their diffusion time scale is longer than the core collapse time
scale. Neutrinos are trapped and add to the total lepton fraction in
the core. But before trapping, the neutrinos can still leave the star
and are an additional cooling mechanism leading to smaller core
entropies than obtained in previous calculations. Lower entropies
reduce the abundance of free protons in the NSE composition, which
increases the importance of capture on nuclei due to their increased
abundances. Neutrinos produced by capture on nuclei have smaller
average energies due to the higher $Q$-value than neutrinos produced by
capture on free protons. Hence the luminosity of electron neutrinos is
increased due to more captures, but their average energies are shifted
to lower values. We stress that the rate for capture on individual
nuclei is noticeably smaller than the capture rate on free
protons. The dominance of capture on nuclei results for the
overwhelmingly higher abundance of nuclei compared to free protons and
are a result of the low entropy, i.e. of the capture process.

\begin{figure}[htbp]
  \centering
  \includegraphics[width=0.70\linewidth]{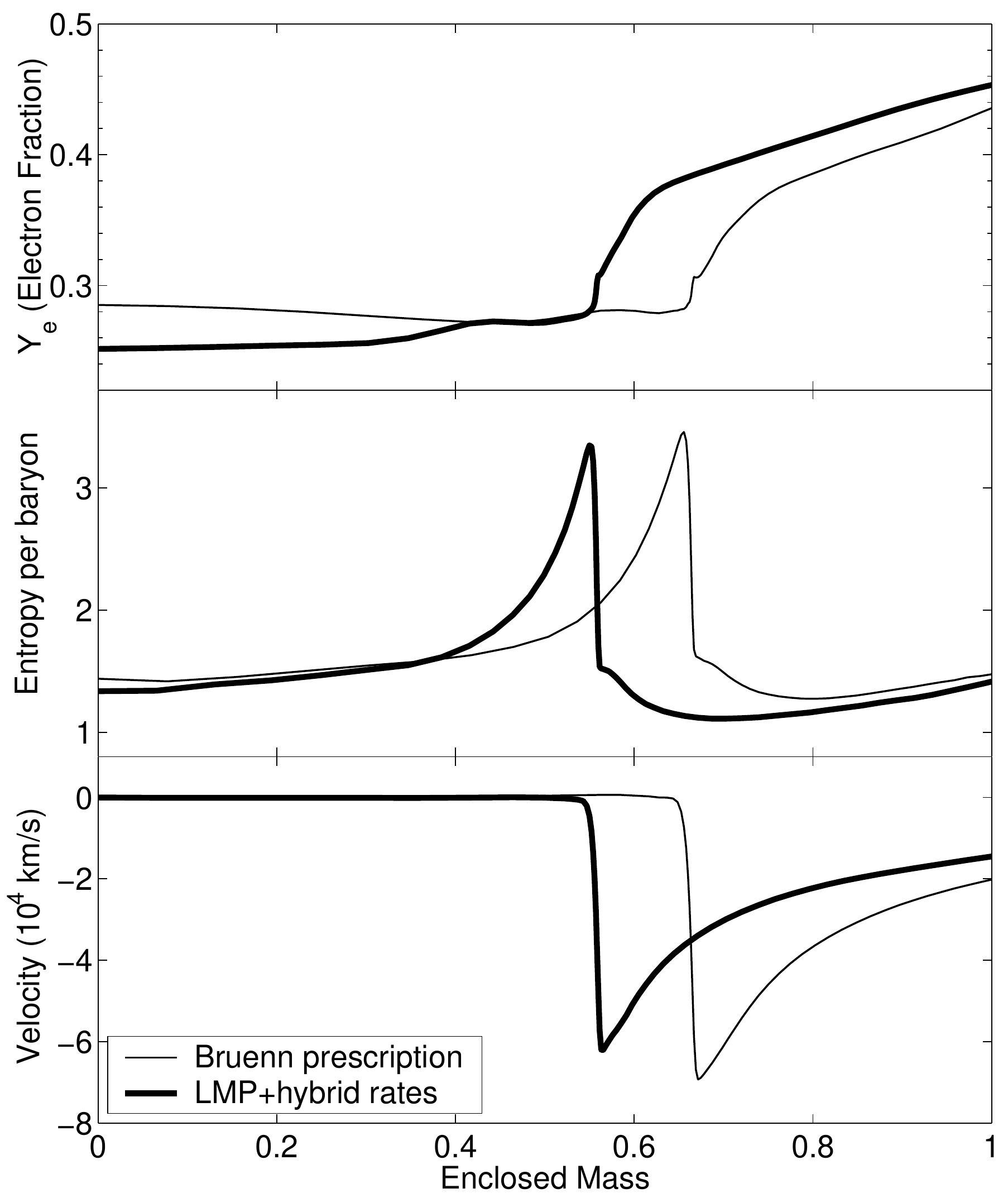}
  \caption{Comparison of $Y_e$ (upper), entropy (middle) and velocity
    profiles (lower panel) at bounce obtained in supernova simulations
    with the shell model rates for nuclei in the mass range
    $A=45$--110
    (\cite{Langanke.Martinez-Pinedo:2001,Langanke.Martinez-Pinedo.ea:2003},
    thick line) and the Bruenn rate parametrization
    (\cite{Bruenn:1985}, thin line). (from
    \cite{Hix.Messer.ea:2003}).}
\label{fig:bounce}
\end{figure}

The fact that electron capture on nuclei proceeds until neutrino
trapping is reached reflects itself also in the core dynamics and
profiles. In the simulations with the improved rates, as shown in
Fig. \ref{fig:bounce} the shock forms with significantly less mass
included (smaller `homologous core' size) and a smaller velocity
difference across the shock. Despite this mass reduction, the radius
from which the shock is launched is actually slightly pushed outwards
due to changes in the density profile.  Despite these significant
alterations also one-dimensional supernova models employing the new
electron capture rates fail to explode.  No noticeable differences in
the simulations are observed if the rate set of Juodagalvis \emph{et
  al.}~\cite{Juodagalvis.Langanke.ea:2010} is used which replaces the
rates for nuclei, for which in \cite{Hix.Messer.ea:2003} the
single-state approximation was used, by rates estimated in the spirit
of the hybrid model.  Multidimensional supernova simulations describe
electron capture now by the rates of
Ref.~\cite{Juodagalvis.Langanke.ea:2010}. However, no dedicated
investigation of the role of electron capture (i.e. in comparison to
the case where capture on heavy nuclei is neglected) has been
performed.

In a recent supernova
simulation~\cite{Fischer.Langanke.Martinez-Pinedo:2013} electron
capture on nuclei has been identified as the dominating
weak-interaction process and the main source of electron neutrinos
during collapse. However, it was shown that pair-deexcitation of
thermally excited nuclear states is an important source of the other
neutrino types (electron anti-neutrinos, muon and tau neutrinos and
their antiparticles).

The contribution of a particular nucleus to the reduction of $Y_e$
during collapse, depends on the product of its abundance and of its
capture rate. Both quantities are time-dependent and have to be
integrated over the duration of the collapse. This study has been
performed by Sullivan \emph{et al.}~\cite{0004-637X-816-1-44} using
rates calculated based on microscopic nuclear models where available.
For the heavy nuclei, for which such rates are not individually
available, they adopted the single-state approximation of
Eq.~(\ref{eq:single-state}).

\begin{figure}[htbp]
  \centering
  \includegraphics[width=0.70\linewidth]{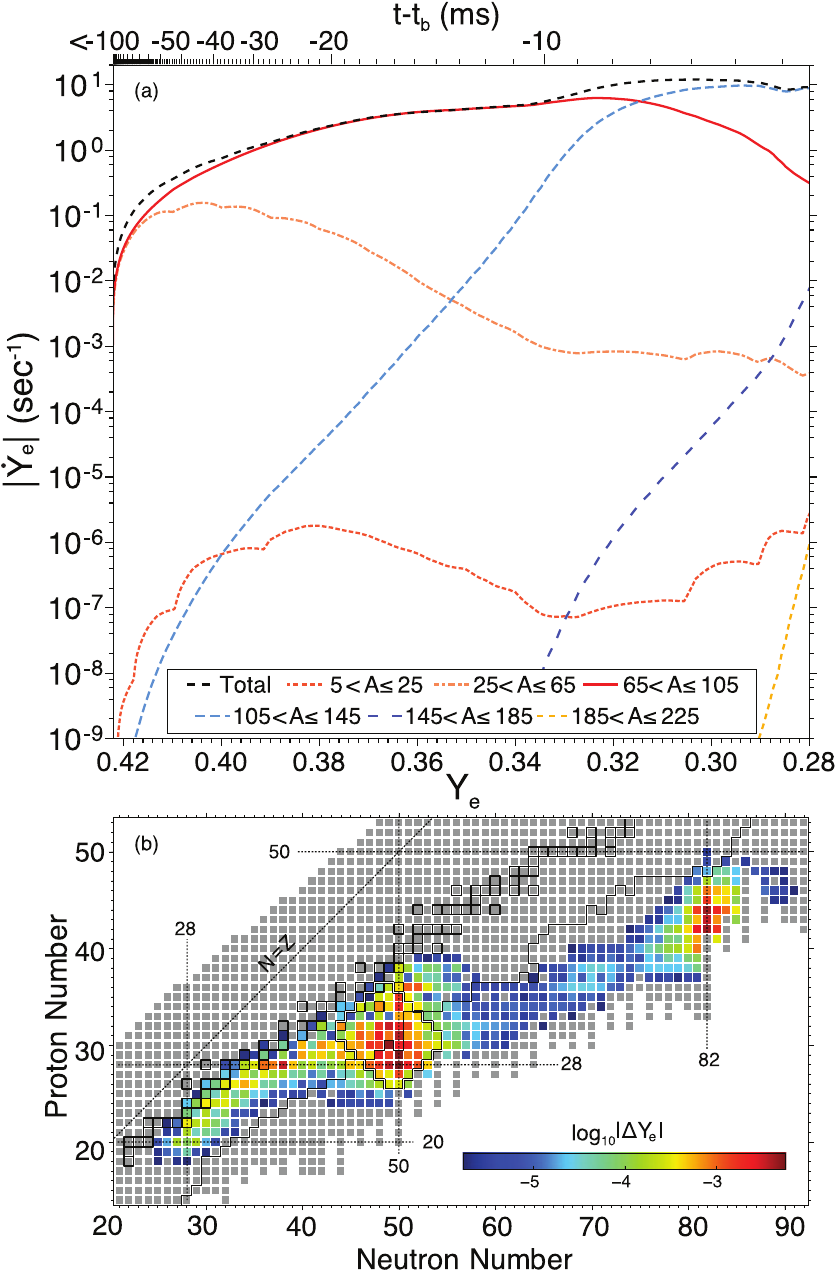}
  \caption{upper panel (a): The contribution of nuclear electron
    capture to the change of $Y_e$ as function of $Y_e$ which
    continuously reduces with time. As reference the upper axis
    vindicates the time until bounce. lower panel (b) The top 500
    nuclei which contribute strongest to electron capture.   (from
    \cite{0004-637X-816-1-44}). \label{fig:top-Ye-nuclei}}
\end{figure}

Fig.~\ref{fig:top-Ye-nuclei} shows in the upper panel which nuclear
ranges contribute to the change of $Y_e$ with time, $\dot{Y}_e$. The
top axis shows the time until bounce. The corresponding densities are
$1.41 \times 10^{11}$, $4.06 \times 10^{11}$,
$1.42 \times 10^{12}$~g~cm$^{-3}$ at $t- t_b = -20, -10, -5$ ms,
respectively. We note that $\dot {Y_e}$ grows with time during
collapse and reaches its maximum after trapping has already set
in. The increase reflects the fact that the electron chemical
potential grows faster than other scales, in particular the average
nuclear $Q$ value of the composition,resulting in strong increases of
the capture rates. The change in the capture rate is mainly driven by
nuclei in the mass range $A=65$-105. Rates calculated within the
hybrid model exist for about 200 nuclei in this range, which, however,
does not cover all nuclei which contribute. The $pf$ shell nuclei, for
which accurate diagonalization shell model rates exist, dominate in
the early collapse. Here capture rates are, however, smaller due to
the smaller electron chemical potentials involved. Nuclei heavier than
$A=105$ contribute or dominate just before and during trapping.

The lower panel of Fig.~\ref{fig:top-Ye-nuclei} identifies how
individual nuclei contribute to $\dot{Y}_e$, determined by integrating
the respective contributions during collapse until trapping
occurs. Due to this study, the relevant nuclei are those around the
$N=50$ shell closure centred in this range from $^{78}$Ni to
$^{82}$Ge.

Sullivan \emph{et al.}~\cite{0004-637X-816-1-44} also investigated
which effect a systematic modification of the electron capture rates
has on the supernova dynamics. When scaling the capture rates for all
nuclei by factors between 0.1 and 10, they observed significant
modifications. A systematic reduction of the rates throttles the
effects which captures on nuclei have during collapse, as outlined
above, driving the results back towards those where capture on nuclei
were neglected. A systematic rate reduction by a factor 10 indeed
increases the enclosed mass at bounce by about $16 \%$, which is a
similar effect as reported in
Ref.~\cite{Hix.Messer.ea:2003}. Sullivan~\emph{et
  al.}~\cite{0954-3899-45-1-014004} and Pascal \emph{et
  al.}~\cite{Pascal.Giraud.ea:2020} argue that the single-state
approximation might overestimate the rates for nuclei close to the
$N=50$ shell gap.  A similar conclusion was drawn from the
measurements of Gamow-Teller distributions for the ground states of
the $N=50$ nuclei $^{86}$Kr~\cite{Titus.Ney.ea:2019} and
$^{88}$Sr~\cite{Zamora.Zegers.ea:2019}.  As discussed above, the
single-state approximation in fact does not consider nuclear structure
effects which should be quite relevant in particular at shell
closures. We note that structure effects are considered in the shell
model rates used in Ref.~\cite{Juodagalvis.Langanke.ea:2010} to set up
a rate table for electron capture under collapse conditions, assuming,
however, NSE for the nuclear composition. It has been shown that the
use of alternative and improved Equations of State has rather small
effects on the supernova
dynamics~\cite{0004-637X-816-1-44,Pascal.Giraud.ea:2020}.  The
dependence of the core composition on different equation of states and
its indirect impact on stellar electron capture rates has been
investigated in Refs.~\cite{Furusawa:2018,Nagakura.Furusawa.ea:2019}.
An improved version of the single-state approximation is presented
in~\cite{Raduta.Gulminelli.Oertel:2017}.  The impact of a reduction of
the $N=50$ shell gap has been explored on
ref.~\cite{Raduta.Gulminelli.Oertel:2016}. We also mention again that
the TQRPA calculations and the hybrid model indicate that the blocking
of the GT strength around $N=50$ is largely overcome at stellar
conditions due to thermal unblocking. Furthermore, both models predict
sizable contributions from forbidden transitions at the astrophysical
conditions at which $N=50$ nuclei are abundant during the collapse.

\subsection{Nucleosynthesis in thermonuclear supernovae}

Thermonuclear, or Type Ia, supernovae are a class of supernovae which
are distinct from the core-collapse version by their explosion
mechanism and also due to their spectral composition (Type Ia spectra
do no exhibit hydrogen lines, in contrast to spectra of core-collapse
or Type II supernovae). In the currently favored model Type Ia
supernovae correspond to the explosion of a White Dwarf in a binary
star system triggered by mass accretion from its companion star when
this enters the Red Giant phase, White Dwarfs are a compact object
produced as the final fate of intermediate mass stars. They are mainly
composed of $N=Z$ nuclei, i.e. $^{12}$C, $^{16}$O, $^{20}$Ne. The
accretion adds to the White Dwarf mass bringing it towards the
Chandrasekhar limit and increases the density in its interior to the
point where carbon burning can be ignited. As the burning occurs in a
highly degenerate environment, the energy set free cannot lead to
expansion, but rather heats the surrounding. This results in a
self-reinforcing acceleration of the burning until degeneracy can be
lifted and the entire White Dwarfs is disrupted. The explosion
mechanism - complete disruption of a White Dwarf in a thermonuclear
runaway - leads to similarity among Type Ia events. For example, the
observed peak magnitude and width of the lightcurves obey a simple
scaling law (Philipps relation~\cite{Phillips:1993}) which makes Type
Ia supernovae to standard candles for cosmological distances. This
fact has been exploited to deduce the current acceleration of our
Universe.

\begin{figure}[htbp]
  \centering
  \includegraphics[angle=90,width=0.31\linewidth]{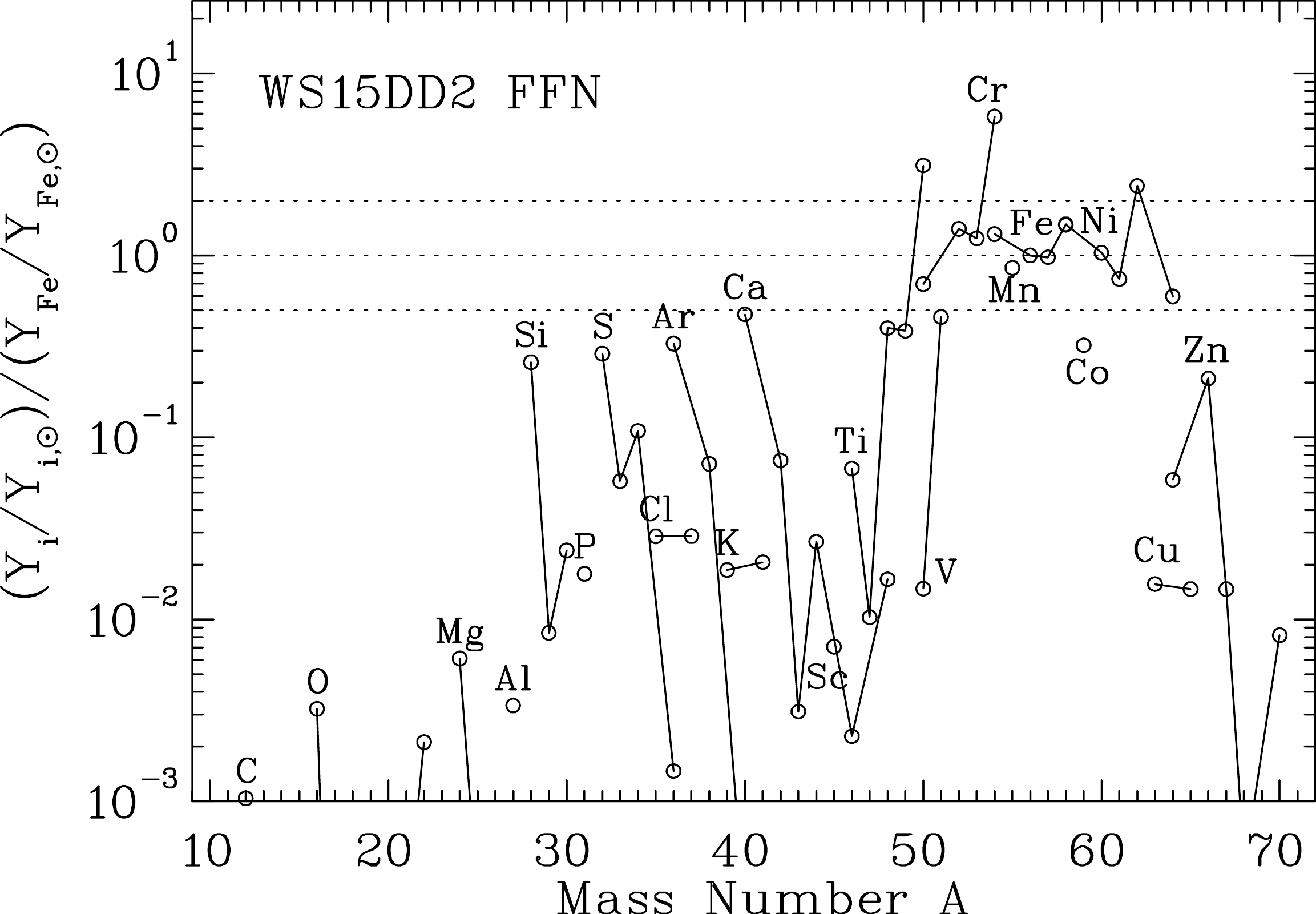}
  \hspace{0.01\linewidth}%
  \includegraphics[angle=90,width=0.31\linewidth]{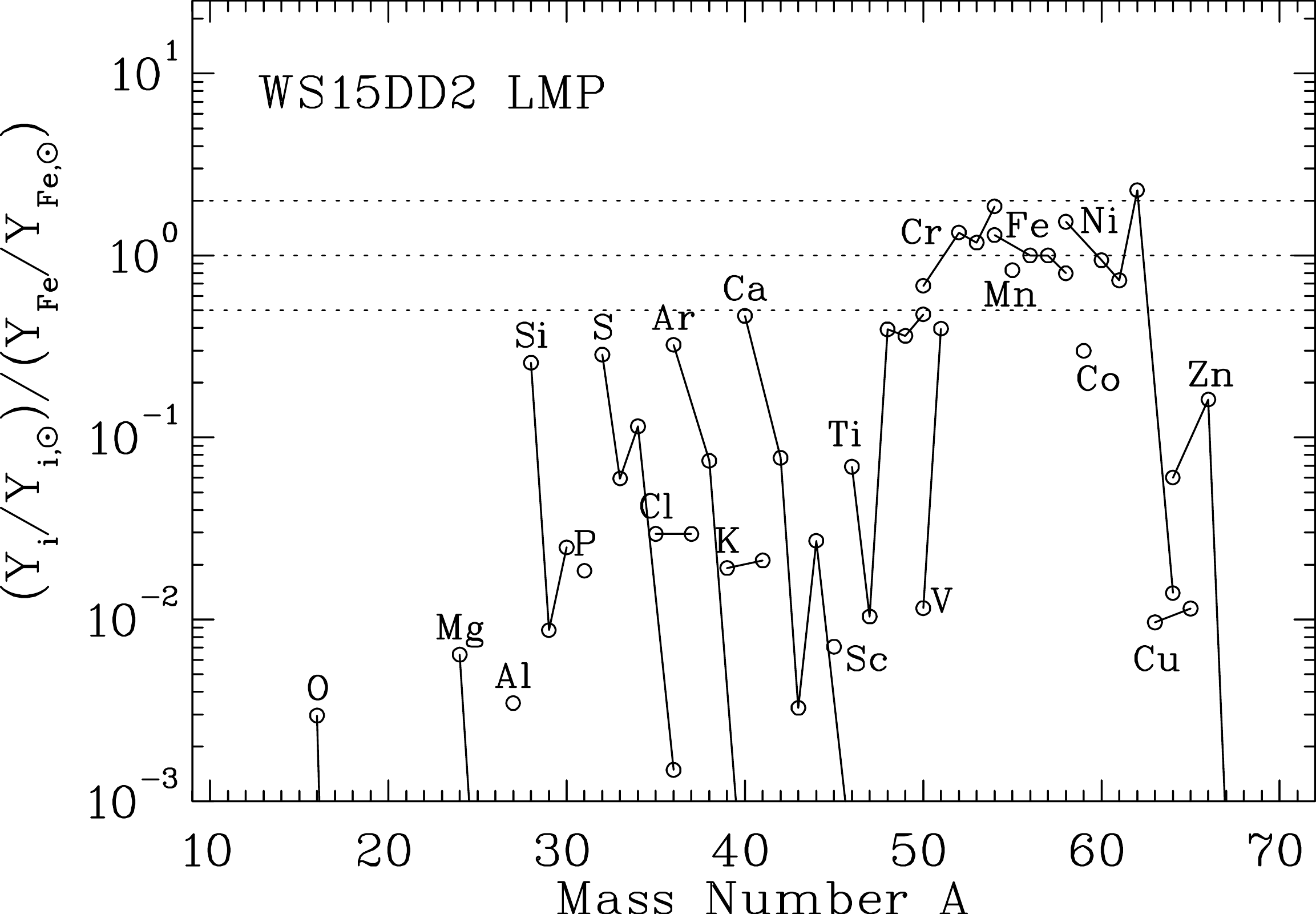}%
  \hspace{0.01\linewidth}%
  \includegraphics[width=0.34\linewidth]{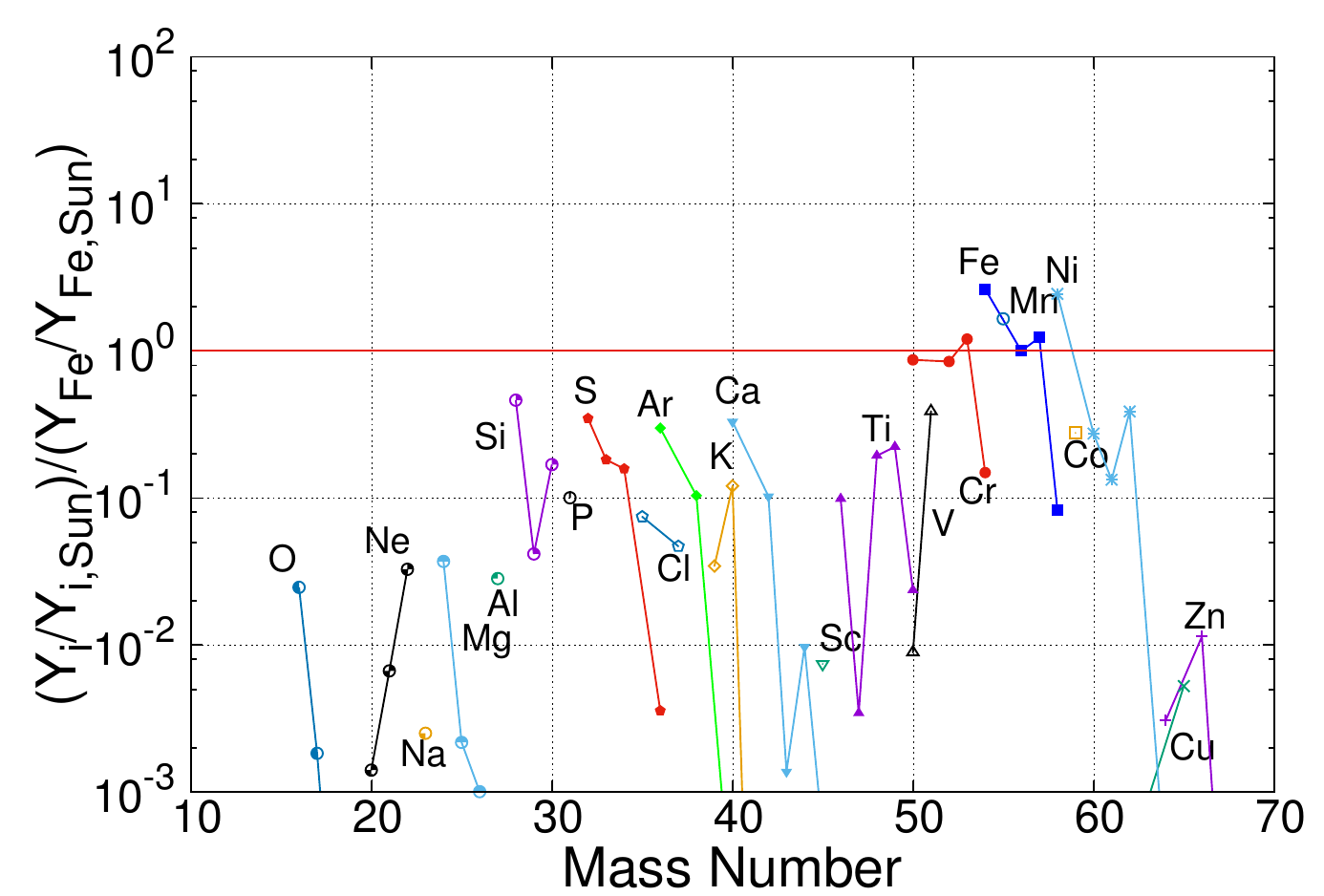}
  \caption{Influence of electron capture rates on type Ia
    nucleosynthesis. The two left panels show yields calculated for
    the WS15 progenitor model of
    Ref. \cite{Nomoto.Thielemann.Yokoi:1984} calculated with the FFN
    (left) and LMP (middle) electron capture rates (courtesy of
    F. Brachwitz, from \cite{Langanke.Martinez-Pinedo:2003}). The
    right panel shows the yields calculated for the W7 progenitor
    model of Ref. \cite{Nomoto.Thielemann.Yokoi:1984} replacing the
    LMP rates with improved shell model rates for selected nuclei in
    the Ni-Fe region (from \cite{Suzuki:2017}).  All yields are
    relative to the solar abundances. The ordinate is normalized to
    $^{56}$Fe).  }
  \label{fig:sn1a-yields}
\end{figure}

After the burning flame has moved through the matter, the inner
material behind the front, with a mass of about 1~M$_\odot$, has
reached temperatures sufficiently high to drive the nuclear
composition into nuclear statistical equilibrium. As the White Dwarf
was composed of $N=Z$ nuclei, mainly $^{56}$Ni is produced.
Deviations towards nuclei with neutron excess occur due to electron
captures in the hot and dense matter behind the front. The impact of
these captures depend, besides the astrophysical conditions of density
(about $10^9$~g~cm$^{-3}$ and temperature ($T \sim 10^9$ K), on the
speed of the flame (i.e. the time for electron captures before the
star is disrupted) and obviously on the rates themselves.  As
discussed above, the diagonalization shell model (LMP) rates are
systematically lower than the FFN rates for $pf$ shell nuclei, in
particular for the nuclei in the Ni-Fe mass range which are of
importance for captures behind the type Ia burning front. Brachwitz
{\it et al.} have performed nucleosynthesis studies in a
one-dimensional supernova simulation based on the well known W15
progenitor model of Ref.~\cite{Nomoto.Thielemann.Yokoi:1984} which
starts from a 1.38~M$_\odot$ C-O White Dwarf~\cite{Brachwitz.Dean.ea:2000,Brachwitz:2001}. The faster FFN rates
lead to a stronger deleptonization in the innermost 0.1~M$_\odot$ core
mass, reaching values down to $Y_e = 0.44$ in the center, while the
slower LMP rates produce $Y_e=0.45$ as the minimum value. As a
consequence, the FFN rates predict an appreciable amount of
neutron-rich nuclei like $^{50}$Ti or $^{52}$Cr, which are strongly
overproduced compared to the solar abundances (left panel of
Fig.~\ref{fig:sn1a-yields}). This overproduction constituted a serious
problem~\cite{Iwamoto.Brachwitz.ea:1999} as roughly half of the
$^{56}$Fe content of the solar abundances are synthesized in type Ia
supernovae and hence all nuclides, produced in type Ia, should not
have overproduction factors larger than 2 as otherwise their relative
abundances are in conflict with observation.  As is shown in the
middle panel of Fig.~\ref{fig:sn1a-yields}, the overproduction is
removed when the slower LMP shell model rates are used. Suzuki has
recently confirmed this finding in a study which replaced the LMP
shell model rates for selected nuclei in the Ni-Fe mass range by those
obtained with the GXPF1 residual interaction which gives better
agreement with the measured GT strength in Ni isotopes~\cite{Suzuki:2017}. Suzuki used a different progenitor model than
Ref. \cite{Brachwitz:2001} (the W7 model of
\cite{Nomoto.Thielemann.Yokoi:1984}). But also his study shows that
the overproduction of neutron-rich nuclei is removed if modern
diagonalization shell model capture rates are used rather than the FFN
rates (right panel in Fig. \ref{fig:sn1a-yields}). Satisfyingly Suzuki
only observes a small difference of $4\%$ in the calculated abundances
based on his shell model rates and on the LMP rates. Detailed studies
of the sensitivity of nucleosynthesis in type Ia supernova can be
found in refs~\cite{Parikh.Jose.ea:2013,Bravo:2019}. 

Electron captures and beta-decays, operating via URCA pairs (see
section \ref{sec:sd-shell-nuclei}), are also important during the
accretion and simmering phases of the evolution of CO WDs before the
type Ia supernova explosion as they determine the neutron excess and
the density at which the thermal runaway
occurs~\cite{Piersanti.Bravo.ea:2017}. Particularly important during
these phases is the $^{13}\mathrm{N}(e^-,\nu_e){}^{13}$C rate whose
value is determined by beta-decay and charge-exchange
data~\cite{Zegers.Brown.ea:2008}.

\subsection{Accreting neutron stars and mergers}

An old isolated neutron star can be described in beta
equilibrium. However, such an equilibrium is broken in the crust if
the star accretes mass from the interstellar medium (ISM) or from a
binary star. For an old neutron star traversing the ISM, a mass of
order $10^{-16}$~M$_\odot$ per year will be accreted as a layer on the
neutron star surface. The temperature of the layer is low and is
usually approximated as $T=0$~\cite{Blaes.Blandford.ea:1990}. In a
binary system the mass flow can be higher leading to repeating burning
of a surface layer with characteristic emission of X-rays with typical
durations up to $\sim 100$ s (X-ray
burster~\cite{Woosley.Heger.ea:2004,Galloway.Muno.ea:2008}). Due to
the re-occurrence of the break-outs, with typical periods of order a
year, the ashes of previous events are pushed to higher densities and
temperatures of order a few $10^8$ K can be reached. These binary
systems can also exhibit rare day-long X-ray bursts (so-called
superbursts). Here carbon flashes, triggered by the fusion of two
$^{12}$C nuclei, heat the neutron star envelope so that hydrogen and
helium burning becomes stable, suppressing the usual shorter x-ray
bursts. These can only occur after the envelope has sufficiently
cooled \cite{Keek.Heger.Zand:2012}. Electron captures play interesting
roles in both accretion scenarios.

The ISM matter which an old isolated neutron star accretes is mainly
hydrogen. At sufficiently high densities hydrogen can start a sequence
of nuclear reactions. However, in contrast to stellar hydrostatic
burning this is not initiated by temperature, but by density
fluctuations triggering so-called pycnonuclear reactions
\cite{Cameron:1959,Gasques.Afanasjev.ea:2005}.  The nuclear reaction
sequence includes hydrogen and helium burning, producing nuclides up
to $^{28}$Si. Particular challenging is the evaluation of the
pycnonuclear triple-alpha reaction rate as neither the $^8$Be
intermediate state nor the Hole state in $^{12}$C can be thermally
reached~\cite{Fushiki.Lamb:1987,Schramm.Langanke.Koonin:1992,Mueller.Langanke:1994}.
The fresh material produced by pycnonuclear reactions rests on original
neutron star crust material, i.e $^{56}$Fe or $^{62}$Ni.

Electron captures can occur once the density reaches a value at which
the electron chemical potential can overcome the nuclear $Q$ value.
For $^{16}$O, which is the main product produced by pycnonuclear
helium reactions, this happens at
$\rho \sim 3 \times 10^{10}$~g~cm$^{-3}$. As beta decay of the
daughter nucleus $^{16}$N is prohibited due to complete filling of the
electron phase space at $T=0$, the daughter nucleus immediately
undergoes a second electron capture to $^{16}$C as the required
density is less than for $^{16}$O.  At the density required for the
double electron capture on $^{16}$O the underlying material of
$^{56}$Fe and $^{62}$Ni has already undergone double electron captures
to $^{56}$Cr, followed to $^{56}$Ti, and $^{62}$Fe, respectively (see
below). As shown in Ref.~\cite{Blaes.Blandford.ea:1990} this leads to
several unstable situations where a denser layer (i.e. $^{16}$C) rests
on less denser layers (i.e. $^{56}$Ti or $^{62}$Fe), resulting in an
overturn of the unstable interfaces. This scenario had been proposed
as a possible explanation for gamma-ray bursts before these were
identified as extra-galactical events with luminosities larger than
observed for supernovae.

Double electron captures are expected also to occur in the crust of an
accreting neutron star in a binary system. When accretion pushes the
original surface layer, made mainly of $^{56}$Fe, to higher densities,
electron captures will transform $^{56}$Fe to $^{56}$Cr once
$\rho > 1.5 \times 10^9$~g~cm$^{-3}$. Haensel and Zdunik have studied
the consequences for the accreted neutron star crust, build on a
single-nucleus ($^{56}$Fe) approach \cite{Haensel.Zdunik:1990} (see
also \cite{Sato:1979,Bisnovatyi-Kogan.Chechetkin:1979}. Upon pushing
the matter to even higher densities, further double electron captures
proceed ($^{56}\mathrm{Cr} \rightarrow {}^{56}$Ti at
$\rho=1.1 \times 10^{10}$~g~cm$^{-3}$,
$^{56}\mathrm{Ti} \rightarrow {}^{56}$Ca at
$\rho=8 \times 10^{10}$~g~cm$^{-3}$,
$^{56}\mathrm{Ca} \rightarrow {}^{56}$Ar at
$\rho = 2.5 \times 10^{11}$~g~cm$^{-3}$), before the density is
reached at which neutrons are emitted from the nucleus
($\rho = 4.1 \times 10^{11}$~g~cm$^{-3}$, neutron drip). Thus, the
double electron capture of $^{56}$Ar is accompanied by the emission of
free neutrons, $^{56}\mathrm{Ar} \rightarrow {}^{52}\mathrm{S} +
4n$. The successive electron captures lowers the charge of the nuclei
so that pycnonuclear fusion reactions, induced by zero-point motion
fluctuations in the Coulomb lattice become possible. The double
electron captures, but in particular pycnonuclear fusion reactions are
considerable heat source, as is discussed in~\cite{Haensel.Zdunik:1990, Yakovlev.Haensel.ea:2013}.

\begin{figure}[htbp]
  \centering
  \includegraphics[width=0.70\linewidth]{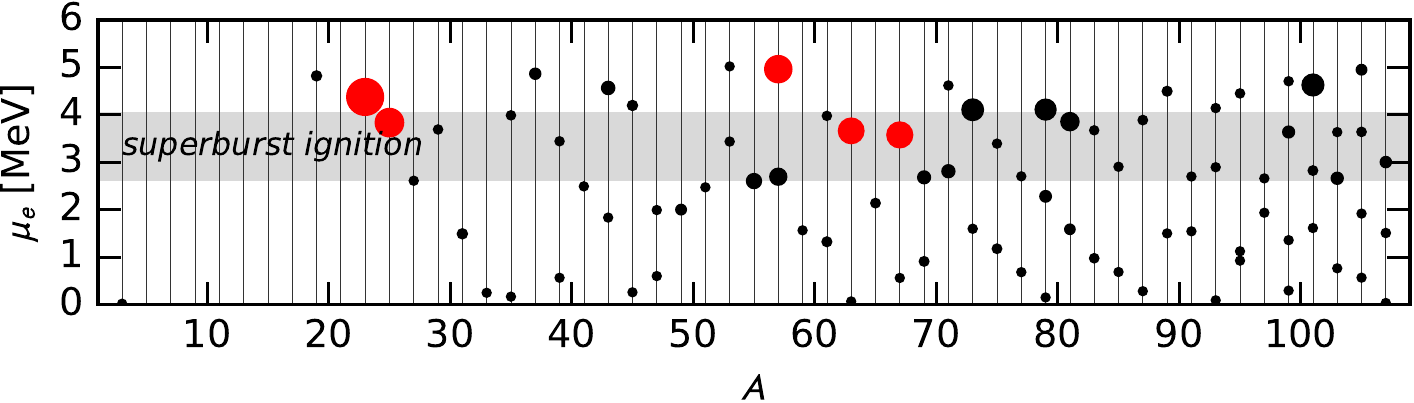}
  \caption{Depth at which URCA pairs of mass number $A$ operate in
    neutron stars.  The size of the data points corresponds to the
    neutrino luminosity of the pair, setting its mass fraction to
    $X=1$. (The top 5 are colored in red.)  The grey band indicates
    constraints for superburst ignition assuming an ignition at a
    column depth between 0.5--$3 \times 10^{12}$~g~cm$^{-2}$.
    (from \cite{Deibel.Meisel.ea:2016}).}
\label{fig:URCA-pairs}
\end{figure}

The crust composition, containing other nuclei than $^{56}$Fe,
complicates the situation. This is also true for the ashes of X-ray
burst events which, due to repeating outbursts, are also successively
pushed to higher densities and run through a similar sequence, to the
one described above, of double electron captures, neutron deliberation
and pycnonuclear fusion reactions \cite{Schatz.Aprahamian.ea:1998}. As
pointed out by Schatz \emph{et al.}~\cite{Schatz.Gupta.ea:2014} the
ashes of former burst events have finite temperatures (a few $10^8$ K)
which, although small compared to typical electron capture $Q$ values,
open up a small energy window at which beta decays of electron capture
daughters can occur. For such an URCA process to occur the electron
capture process has to satisfy two conditions: it must be mediated by
an allowed transition to a state at excitation energies $E_x < T$ and
the beta-decaying nucleus must no have a strong electron capture
branch. On general grounds even-even nuclei, which are the most
abundant nuclei in the crust, do not form URCA pairs but rather
perform double electron captures~\cite{Deibel.Meisel.ea:2016}. On the
other hand, nearly all odd-A nuclei can form URCA pairs. The authors
of Ref.~\cite{Deibel.Meisel.ea:2016} have identified about 85 URCA
pairs. Fig.~\ref{fig:URCA-pairs} shows the neutrino luminosities of
these pairs (setting the mass fraction of the nucleus on which an
electron is captured to $X=1$) and at which depth in the neutron star
they operate.  As pointed out in~\cite{Schatz.Gupta.ea:2014} cooling
by URCA pairs in the crust reduces the heat transport from the crust
into the region of the x-ray burst or superburst ashes which reside at
less dense regions (this region is often called the ocean). This
lowers the steady-state temperature in the ocean. This puts constrains
on the ignition of the $^{12}\mathrm{C}+{}^{12}$C fusion reaction to
start the next burst cycle. This ignition has now to occur at higher
densities~\cite{Deibel.Meisel.ea:2016}.  URCA pairs can also directly
operate in the ocean. However, due to the lower densities nuclei are
less neutron-rich with smaller $Q$ values for electron captures than
those in the crust. As the neutrino luminosity scales with $Q^5$, This
strongly reduces the effectiveness of URCA pairs in the
ocean~\cite{Deibel.Meisel.ea:2016}. 

Due to the simultaneous observation of the gravitational wave and the
electromagnetic signal from GW170817, the merger of two neutron stars
in a binary system has been identified as one of the astrophysical
sites, see e.g.~\cite{Kasen.Metzger.ea:2017}, where the
r-process~\cite{Burbidge.Burbidge.ea:1957,Cameron:1957,Cowan.Thielemann.Truran:1991,Cowan.Sneden.ea:2020}
operates. Particularly important to determine nucleosynthesis in
mergers is the $Y_e$ value of the ejected material that is determined
by weak processes. The initial $Y_e$ profile of the neutron stars can
be determined from beta-equilibrium. However, as the neutron stars
approach each other and finally merge the temperature increases and
neutrino emission becomes very important. An accurate prediction of
the neutrino luminosities requires a description of high density
neutrino-matter interactions~\cite{Burrows.Reddy.Thompson:2006} and
its implementation in neutrino radiation transport as is currently
done in core-collapse supernova~\cite{Janka:2017}. The absorption of
both $\nu_e$ and $\bar{\nu}_e$ together with electron and positron captures
leads to substantial changes on the  $Y_e$ of the ejected material
particularly in the polar
regions~\cite{Wanajo.Sekiguchi.ea:2014,Goriely.Bauswein.ea:2015,Martin.Perego.ea:2018,Radice.Perego.ea:2018a}. These
processes occur when the material is hot and constitutes mainly
of neutrons and protons.

Another important source of material is the so-called secular ejecta
originating from the accretion disk that surrounds the central remnant
produced by the merger. If this is a long-lived neutron star, the
neutrino luminosities are large enough to affect the neutron-to-proton
ratio of the ejected material~\cite{Shibata.Hotokezaka:2019}. If the
central object is a black-hole, the neutron-to-proton ratio is
determined by a dynamical
beta-equilibrium~\cite{Arcones.Martinez-Pinedo.ea:2010} between
electron, positron captures and beta-decays operating in the accretion
disk~\cite{Beloborodov:2003,Siegel.Metzger:2018,Fujibayashi.Shibata.ea:2020}
on hot material that mainly is made of neutrons and protons.  Due to
the current understanding, electron capture on nuclei does not play an
important role for r-process nucleosynthesis in neutron-star merger
events.

\subsection{Fate of intermediate-mass stars}

The final fate of stars depend on their masses at birth. Stars with
masses less than about 8~$M_\odot$ advance through hydrogen and helium
burning. As they suffer significant mass losses by stellar winds their
masses at the end of helium burning is not sufficient to ignite
further burning stages. They end their lives as White Dwarfs, which
are compact objects with a mass limit of 1.44~M$_\odot$ (Chandrasekhar
mass), stabilized by electron degeneracy pressure. Stars with masses
in excess of about 11~M$_\odot$ develop a core at the end of each
burning phase which exceeds the Chandrasekhar mass. As a consequence
they can ignite the full cycle of hydrostatic burning and end their
lives as core-collapse supernovae, leaving either neutron stars or
black holes as remnants. The fate of intermediate-mass stars
(8--11~M$_\odot$) balances on a knife edge between collapsing into a
neutron star or ending in a thermonuclear runaway which disrupts most
of the core~\cite{Jones.Roepke.ea:2016}.  Simulations of such stars
are quite sensitive to astrophysical uncertainties like convective
mixing or mass loss rates \cite{Jones.Roepke.ea:2016}. On the other
hand the major nuclear uncertainty, related to electron capture on
$^{20}$Ne, has recently been removed as this rate, as we have
described above, is now known experimentally at the relevant
astrophysical conditions. We briefly summarize the consequences which
this nuclear milestone has for the fate of intermediate mass stars.

\begin{figure}[htbp]
  \centering
  \includegraphics[width=0.70\linewidth]{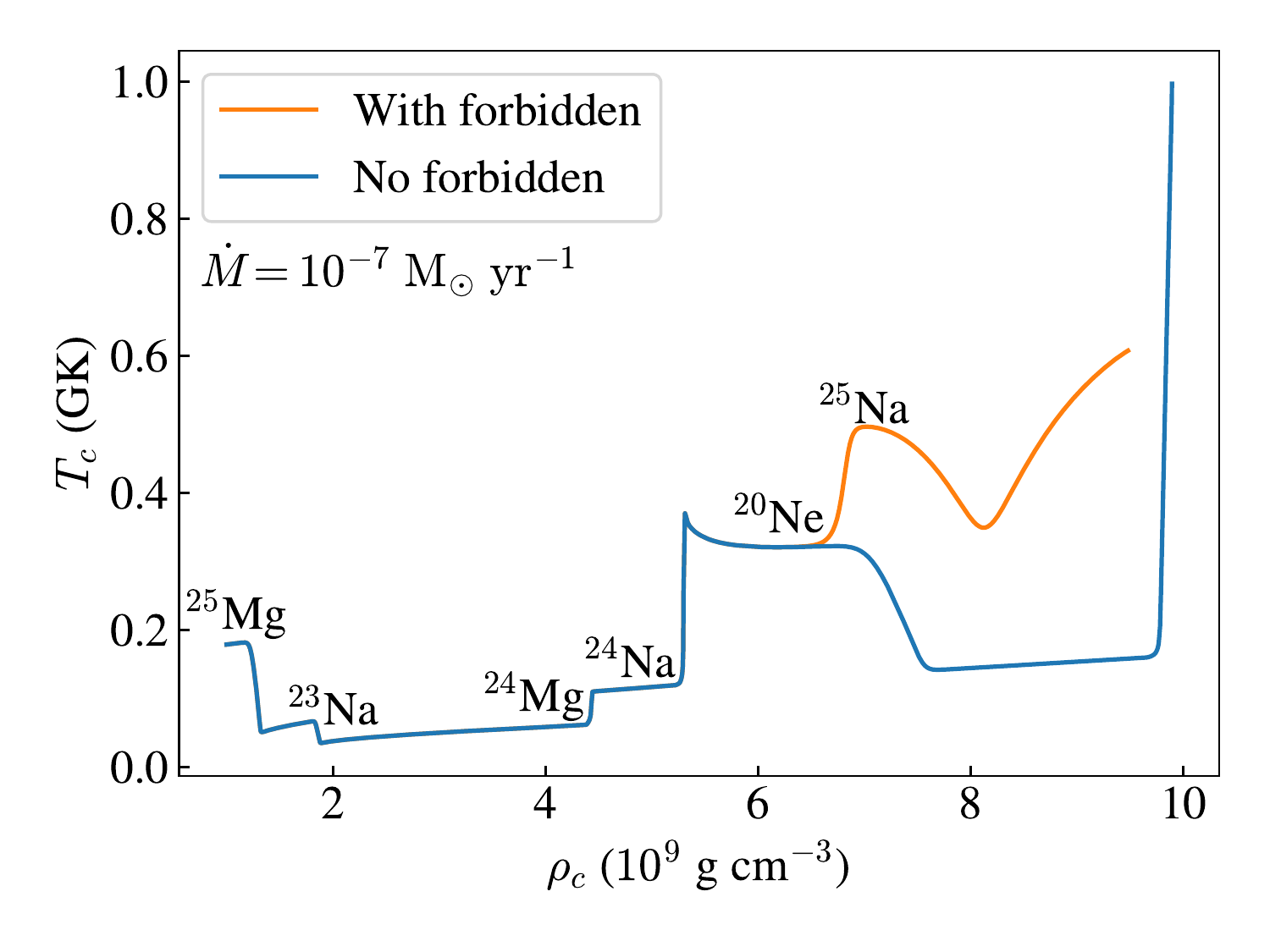}
  \caption{Temperature-density evolution of the 
  ONeMg core of an intermediate mass star. The labels indicate at
  which densities the URCA pairs and the electron captures on
  $^{24}$Mg and $^{20}$Ne operate. The red (blue) lines show the
  evolution with (without) inclusion of the forbidden
  ground-state-to-ground-state contribution to the $^{20}$Ne electron
  capture rate.The calculation assumes that the ONe core 
  accretes a mass 
  of $10^{-7}$~M$_\odot$ per year from ongoing
  hydrostatic burning.
   (from \cite{Kirsebom.Jones.ea:2019}).}
\label{fig:ecapture-profile}
\end{figure}

Intermediate mass stars go through hydrostatic hydrogen, helium and
core carbon burning, but are not massive enough to ignite further
advanced burning stages. In the center of the star a core develops
which mainly consists of $^{16}$O and $^{20}$Ne, with smaller amounts
of $^{23}$Na and $^{24,25}$Mg.  Due to its position on the
Hertzsprung-Russell diagram stars with such an ONe core are referred
to as Super Asymptotic Giant Branch (AGB) stars.  It is important to
note that cores of Super-AGB stars are more dense than their
counterparts after helium burning in more massive stars.  As nuclear
burning has ceased in the ONe core its gravitational collapse is
counteracted by the electron degeneracy pressure. However, the
densities achieved in the core result in electron chemical potentials
large enough to initiate electron capture reactions, which reduce the
pressure against collapse. Here two distinct processes play the
essential role for the development of the core. This is shown in
Fig.~\ref{fig:ecapture-profile} which displays the final
temperature-density evolution of the core center. First, several URCA
pairs ($^{25}$Mg-$^{25}$Na, $^{23}$Na-$^{23}$Ne, $^{25}$Na-$^{25}$Ne)
operate at various phases of this final evolution. These pairs are
efficient cooling mechanism. Second, electron captures also occur on
the abundant $\alpha$-nuclei $^{24}$Mg and $^{20}$Ne once the electron
chemical potentials overcome the capture $Q$ values (the $Q$ value of
$^{16}$O is too high for electron captures at these densities).  But
for these $N=Z$ nuclei, the electron capture daughters ($^{24}$Na and
$^{20}$F) also capture electron which due to their smaller $Q$ values
than in the first captures occur significantly faster than competing
$\beta$-decays. Furthermore the second capture proceeds often to
excited states in their daughters which then decay by $\gamma$ emission
to the ground states and heat the environment (see
Fig.~\ref{fig:ecapture-profile}).  Due to its lower initial $Q$-value
the double electron capture on $^{24}$Mg proceeds at lower densities
than the one on $^{20}$Ne.  Recognizing the low temperature at the
onset of electron capture on $^{20}$Ne (about~300 keV), basically due
to the efficient URCA cooling~\cite{Schwab.Bildsten.Quataert:2017}, the authors
of~Ref. \cite{Martinez-Pinedo.Lam.ea:2014} pointed out that the
transition from the $^{20}$Ne $0^+$ ground state to the $^{20}$F $2^+$
ground state, although second forbidden, could dominate the rate at
core conditions as all other transitions were exponentially suppressed
by either the tail of the electron distribution or a Boltzmann factor
due to thermal excitation.

\begin{figure}[htbp]
  \centering
  \includegraphics[width=0.70\linewidth]{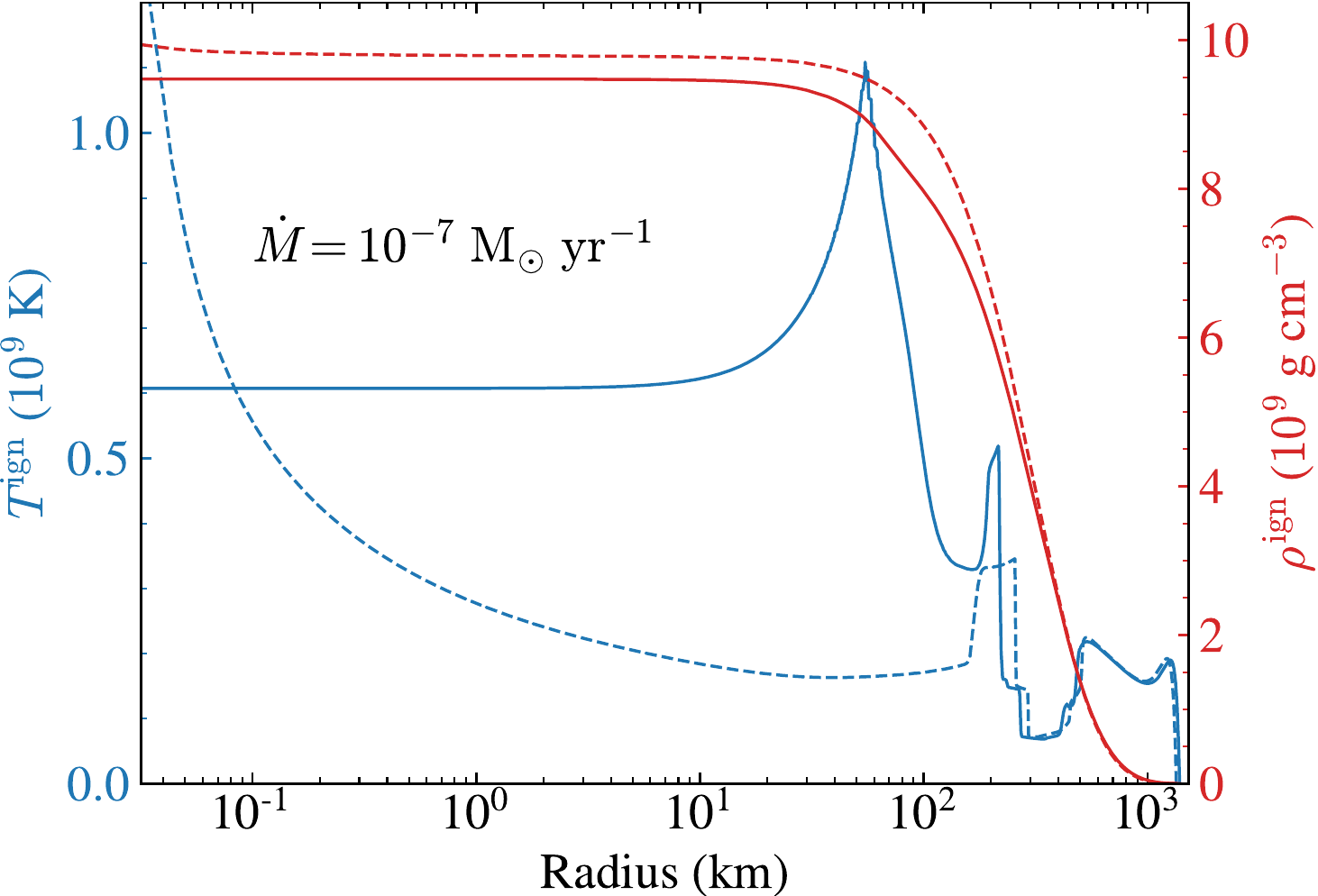}
  \caption{Temperature (blue) and density (red) profiles of an ONe
    core at ignition of oxygen fusion, calculated with (solid lines)
    and without (dashed lines) consideration of the forbidden
    ground-state-to-ground-state transition in the $^{20}$Ne electron
    capture rate. The calculations were performed using the spherical
    MESA code \cite{Paxton.Schwab.ea:2018} and assuming that the ONe
    core accretes a mass of $10^{-7} M_\odot$ per year from ongoing
    hydrostatic burning.  (from \cite{Kirsebom.Jones.ea:2019}).}
\label{fig:ecapture-ignition}
\end{figure}

The inclusion of the `forbidden' ground-state-to-ground-state
transition in the electron capture rate on $^{20}$Ne shifts the onset
of this capture to lower densities before the last epoch of URCA
cooling by the $^{25}$Na-$^{25}$Ne pair (see
Fig.~\ref{fig:ecapture-profile}). This shift in density has, however,
important impact on the fate of the star ending either as
gravitational collapse or thermonuclear explosion.  This fate is
determined by the competition between electron capture and nuclear
energy generation by oxygen fusion~\cite{Jones.Roepke.ea:2016}. If the
ignition of oxygen (requiring temperatures in excess of $10^9$~K)
occurs at low enough densities, the fusion generates sufficient energy
to reverse the collapse and to disrupt the star in a thermonuclear
explosion. At higher densities, the deleptonization behind the burning
front is so rapid that the loss of pressure cannot be recovered by
nuclear burning. In this case the core continues to collapse, ending
as a neutron star. The increase of the $^{20}$Ne electron capture rate
due to the contribution of the forbidden transition seems to shift the
fate towards thermonuclear explosions~\cite{Kirsebom.Jones.ea:2019}.
This is demonstrated in Fig.~\ref{fig:ecapture-ignition} which is
based on a spherical simulation of the core evolution. In the
calculation, performed without the inclusion of the forbidden
contribution to the rate, oxygen is ignited in the center, while in
the case using the experimental $^{20}$Ne capture rate with inclusion
of the forbidden transition, the star develops an isothermal core
with temperatures below the ignition value. In this inner core double
electron capture on $^{20}$Ne continues generating heating which leads
to an off-center ignition of oxygen burning (at a radius of 58 km for
the case shown in Fig. \ref{fig:ecapture-ignition}). We also note that,
without the forbidden contribution, the core reaches a higher density
at ignition.

To determine the fate of the star requires to study the propagation of
the burning front, which needs a 3D hydrodynamical treatment to
resolve the relevant length scales. Such studies have been reported
in~\cite{Kirsebom.Jones.ea:2019} and all simulations, matched to the
parameters of the spherical MESA results ended in thermonuclear
explosions producing an ONeFe White Dwarf remnant.  This has
significant implications for the total nucleosynthesis yields of
intermediate mass stars as thermonuclear explosions eject
about~0.01~M$_\odot$ more mass than gravitational collapse and
intermediate mass stars are much more abundant than heavier stars. A
first exploration shows that the ejecta of thermonuclear explosions
are particularly rich in certain neutron-rich Ca, Ti, Cr isotopes and
in trans-iron elements Zn, Se and
Kr~\cite{Kirsebom.Jones.ea:2019}. This might have interesting
implications for the understanding of the early chemical evolution of
our galaxy~\cite{Jones.Roepke.ea:2019}.

\subsection{Helium flashes in accreting Helium White Dwarfs}

Subluminous B stars are core-helium burning stars with thin hydrogen
envelopes and masses of about 0.5~M$_\odot$ \cite{Heber:2009}.  Often
these stars exist in tight binaries with White
Dwarfs~\cite{Geier.Oestensen.ea:2017}.  When the White Dwarf accretes
matter from the unburned outer layers of its companion star, also some
amount of $^{14}$N is present depending on the initial metallicity of
the donor star.  Electron capture on $^{14}$N is then a decisive
factor for the fate of the accreted material.

Due to its low $Q$ value of 0.667~MeV, electrons are captured on
$^{14}$N once the density of the accreted matter on the WD surface
exceeds a threshold value of about
$1.156\times 10^6$~g~cm$^{-3}$~\cite{Bauer.Schwab.Bildsten:2017}. As
the respective temperatures are rather low (T less than a few
$10^8$~K), the capture solely proceeds by the allowed transition
between the $^{14}$N and $^{14}$C ground states. The respective
transition matrix element is known from the $^{14}$C beta
decay. Coulomb corrections due to environment effects are relatively
minor, but are considered in recent astrophysical
applications~\cite{Bauer.Schwab.Bildsten:2017}.

For the temperatures involved and for densities larger than about
$10^6$~g~cm$^{-3}$, the electron capture rate is larger than the
competing $\beta$ decay and, in the helium-rich environment, the
electron capture is followed by an $\alpha$ capture on
$^{14}$C~\cite{Hashimoto.Nomoto.ea:1986}. The energy generation by
this so-called NCO reaction
($^{14}\mathrm{N}(e^-,\nu_e){}^{14}\mathrm{C}(\alpha,\gamma){}^{18}$O)~\cite{Kaminisi.Arai.Yoshinaga:1975}
--- despite some uncertainties in the $\alpha$-capture rate on $^{14}$C
\cite{Iliadis.Longland.ea:2010} --- is larger than by the triple-alpha
reaction in the relevant temperature-density range.  Thus, it is the
NCO reaction which triggers a rather steep rise of the temperature in
the environment so that as a second step also the triple-alpha
reaction will be ignited. This finally leads to a thermonuclear
instability which is observed as helium flashes.

The evolution of these flashes depend crucially on the accretion mass
flow \cite{Hashimoto.Nomoto.ea:1986}.  If the mass flow is large
($10^{-8}$~M$_\odot$~yr$^{-1}$), the energy released from the
gravitational contraction leads to heating of the environment enabling
the $^{14}$C nucleus to capture an $\alpha$ particle fast. The
electron capture process controls the NCO reaction sequence and no
significant amount of $^{14}$C is being built up. For smaller mass
flows ($10^{-9}$~M$_\odot$~yr$^{-1}$) the energy released by
contraction can be radiated away, keeping the temperature in the core
low. Hence, when the core density exceeds the value for electron
capture the temperature is too low to ignite $\alpha$ captures on
$^{14}$C. This occurs at conditions with higher densities and after
$^{14}$N has been completely converted to $^{14}$C. Simulations also
show that for smaller accretion rates, the core becomes convectively
unstable. The time scale on which the flashes develop depend also on
the accretion rate and are significantly shorter for smaller rates (a
few $10^7$~yr for $10^{-9}$~M$_\odot$~yr$^{-1}$)

\section{Summary}

In his authoritative review on core-collapse supernovae, Hans Bethe
stated in 1990 \cite{Bethe:1990}: ``The theory of electron capture has
gone a full circle and a half.'' He was referring to the fact that in
early models, capture was assumed to occur on free protons. This was
put into question by BBAL~\cite{Bethe.Brown.ea:1979} who noted that
the concentration of free protons during collapse is very low and that
the capture takes place on nuclei with mass numbers $A=60$--80,
changing a proton in the $f_{7/2}$ shell to a neutron in the $f_{5/2}$
orbital by allowed Gamow-Teller transition. Following Bethe, the third
semi-circle is due to Fuller's observation that the neutron $f_{5/2}$
orbitals are occupied at neutron number $N=38$~\cite{Fuller:1982},
blocking Gamow-Teller transitions within the $pf$ shell. Hence at the
time when Bethe wrote his famous article, electron capture in
supernovae was assumed to occur on free protons and capture on nuclei
was switched off for nuclei with $N>38$~\cite{Bruenn:1985}.

As we have summarized in this manuscript, the experimental and
theoretical work of the last two decades implies that this picture is
too simple. Experimental techniques to measure Gamow-Teller strength
distributions based on charge-exchange reactions with progressively
better energy resolutions --- advancing from the pioneering $(n,p)$
reactions to much more refined $(d,{}^2\textrm{He})$ and
$(t,{}^3\textrm{He})$ reactions --- give clear evidence that nuclear
correlations play a decisive role in the total strength, and even
more, for the fragmentation of the GT distribution, thus invalidating
the Independent Particle Model on which the early electron capture
work, which was discussed and reviewed by Bethe in
1990~\cite{Bethe:1990}, were based. In parallel, many-body models
became available which were capable to account for the relevant
nuclear correlations and which describe the experimental GT data quite
well. Importantly, these models, and also experimental data, imply
that the GT strength is not blocked at the shell gap between the $pf$
and $g_{9/2}$ orbitals caused by strong nuclear cross-shell
correlations.  As a major consequence, electron capture takes place on
nuclei during the entire collapse. With this result, the theory of
electron capture has gone now two complete circles.

The evaluation of stellar electron capture for core-collapse
supernovae rests on the fact that, with progressing core density, the
electron chemical potential grows significantly faster than the
average $Q$ value which dominate the core composition. As a
consequence, detailed description of the nuclear strength functions
(i.e. Gamow-Teller) are only needed for nuclei in the Fe-Ni mass range
($A=45$--65), which are most abundant during the early collapse phase,
while for the neutron-rich, heavy nuclei, which dominate later in the
collapse at higher densities, a more overall reproduction of the
strength functions (now, however, including forbidden transitions)
suffice. This is a quite fortunate situation.

For those nuclei, for which the calculation of stellar capture rates
requires detailed descriptions of the allowed strength functions
($pf$- and $sd$-shell nuclei, where the latter occur in burning stages
prior to collapse), diagonalization shell model calculations can be
performed which in general reproduce the measured GT strength
functions quite well. In fact, if the capture rates are calculated
solely from the ground state distributions, the rates obtained from
data and from shell model agree within better than a factor of 2 at
the relevant astrophysical conditions. The theoretical capture rates
(i.e. \cite{Langanke.Martinez-Pinedo:2001,Juodagalvis.Langanke.ea:2010})
consider from excited states also from shell model calculations,
accounting for the fact that each nuclear state has its own individual
strength distribution. There are no indications that the shell model
results for excited states might be less reliable than for the ground
states. However, there is concern that the procedure applied
in~\cite{Langanke.Martinez-Pinedo:2001} might slightly underestimate
the partition function at higher
temperatures~\cite{Misch.Fuller.Brown:2014}.

The fact that cross-shell correlations unblock Gamow-Teller
transitions even in the ground states of nuclei with proton numbers
$Z<40)$ and neutron numbers $N>40$ has been experimentally proven by
experimental data for Gamow-Teller strength distributions and also
from spectroscopic information obtained from transfer reactions. Thus,
the assumption that GT transitions are Pauli blocked for nuclei with
$N>38$ has been disproven by experiment. Modern many-body models like
the diagonalization shell model (for selected nuclei) and the Shell
Model Monte Carlo approach can reproduce such cross-shell
correlations.  The latter approach has been adopted to determine
partial occupation numbers in large model spaces including the shell
gaps at $N=40$ and 50. The capture rates were then calculated within a
`hybrid model' from these occupation numbers within the framework of
the Random Phase Approximation, exploiting the fact that these heavier
nuclei become abundant during the collapse at sufficiently high
densities requiring only the overall, but not the detailed
reproduction of the GT strength functions. Contributions from
forbidden transitions were included, which become progressively
important with increasing density. The hybrid model indicates that the
gaps at $N=40$ and 50 lead to some reduction of the capture rates, but
the rares are clearly large enough so that captures on nuclei dominate
the one on free protons during the entire collapse. This is clearly
borne out in modern supernova simulations, thus closing the second
circle as referred to by Hans Bethe.

The unblocking of the GT strength at the neutron numbers $N=40$ and 50
has also been confirmed by calculations performed within the Thermal
Quasiparticle RPA approach, which consistently considers correlations
up to the 2p-2h level.  As cross-shell correlations require in general
correlations higher than 2p-2h, GT strength, in particular at low
excitation energies, can be missed. This translates into the
observation that at modest temperatures and densities capture rates
obtained within the Thermal QRPA are somewhat smaller than in the
hybrid model.  At higher temperatures and densities the two models
give very similar results, including the neutron-rich nuclei with
$N=50$, which significantly contribute to the capture process at these
astrophysical conditions. Both theoretical approaches imply that at
the respective temperatures of order $T=1$ MeV, configurations from
higher shells, which are strongly reduced in the ground state, are
present in the thermally excited nuclear states and significantly
unblock the GT strength.  This observation is quite important as the
ground state GT distribution for such nuclei has been experimentally
observed to have nearly vanishing strength and the electron capture
rate would nearly be blocked if calculated from the ground state
distribution. While the unblocking appears to be quite solid on
theoretical ground, experimental verification is desirable.

Although core-collapse supernovae are arguably the most important
astrophysical application, electron captures play also a role in other
astrophysical environments. In thermonuclear supernovae the rate of
electron captures on nuclei determine the production yield of
neutron-rich nuclei. As the relevant nuclei are those in the Fe-Ni
mass range, the experimental and theoretical (by diagonalization shell
model calculations) progress have constrained the relevant capture
rates significantly up to a degree that improved description of
details of the GT strength distribution changed the nucleosynthesis
yields by only a few percent. The description of capture rates for
$sd$-shell nuclei, again based on shell model calculations and data,
has reached a similar degree of accuracy which appears to be
sufficient for the simulation of this process for the core evolution
of intermediate mass stars. However, attention has been drawn recently
to the fact that in the low-temperature-low-density environment of
such stellar cores only a few transitions dominate the capture rates
and that in exceptional situations also a forbidden transition can
noticeably contribute to the rate. Such a situation happens for the
capture on $^{20}$Ne where the second forbidden transition from the
$^{20}$Ne ground-state to the $^{20}$F ground state enhances the
capture rate just at the most crucial conditions for the core
evolution. The transition strength has now been measured so that the
entire electron capture rate on $^{20}$Ne is now experimentally
determined in the relevant temperature-density regime.

Double electron captures, initiated on abundant even-even nuclei, are
relevant for the crust evolution of accreting neutron stars. The
process is triggered once the electron chemical potential (i.e. the
core density) is high enough for electrons to overcome the $Q$ value
between the even-even mother nucleus and the odd-odd daughter. As the
$Q$ value of the second capture step on the odd-odd nucleus is smaller
due to nuclear pairing, this energy gain can be transferred into crust
heating.  For simulations of the crust evolution, generally one is not
so much interested in the capture rate (which is often fasted than
competing time scales), but in the portion of the energy gain which is
translated into heat. As this can involve quite exotic neutron-rich
nuclei a detailed determination of this energy portion is a formidable
nuclear structure challenge and current models are likely too
uncertain.

Despite the progress which has been achieved in recent years in the
determination of stellar electron capture rates, further improvements
are certainly desirable and, in specific cases, needed. Additional
precision measurements of Gamow-Teller strength distributions for
$sd$- and $pf$-shell nuclei will lead to further improvements and to
refinements of the shell model calculations, however, it is not
expected that these improvements will have significant impact on
supernova dynamics or nucleosynthesis. It is, however, desirable that
the gap of nuclei (with mass numbers $A=38$--45), for which no shell
model electron capture rates exist, should be filled.  Such
calculations are challenging as they require an accurate description
of cross-shell correlations. They would certainly benefit from some
detailed experimental GT distribution measurements.  A particularly
interesting and important case is $^{40}$Ar, which serves as material
for neutrino detectors like
ICARUS~\cite{Bueno.Gil-Botella.Rubbia:2003,Gil-Botella.Rubbia:2003,Scholberg:2012},
which holds potential for the detection of supernova
neutrinos. Detailed GT$_-$ data from
$(p,n)$~\cite{Bhattacharya.Goodman.Garcia:2009} and
$(^{3}\textrm{He},t)$ charge-exchange data
\cite{Karakoc.Zegers.ea:2014} and M1 data from $(\gamma,\gamma')$
photon scattering reactions~\cite{Li.Pietralla.ea:2006} can serve as
experimental constraints for the determination of charged-current
$(\nu_e,e^-)$ and neutral-current $(\nu,\nu')$ cross sections on
$^{40}$Ar. However, GT$_+$ data, which are relevant for electron
capture and charged-current $(\bar{\nu}_e,e^+)$ cross sections do not
exist yet. In principle, forbidden transitions, not considered in the
shell model electron capture rates for $sd$ and $pf$-shell nuclei, can
contribute to the rates. But such contributions will only be relevant
in core-collapse supernovae at higher temperatures than those at which
these nuclei dominate the core composition. The case of $^{20}$Ne, for
which a second forbidden transition dominates the capture rate at the
relevant conditions during the core evolution of intermediate mass
stars, shows, however, that such exceptional cases can occur in cases
of rather low temperatures where only a few transitions contribute to
the capture rate. No other case has yet been identified, however,
caution is asked for.

The shell gaps at neutron numbers $N=40$ and 50 do not block electron
capture on nuclei in current supernova models. In both cases, this is
based on modern many-body models which at $N=40$ overcome the gap by
nucleon correlations, while for $N=50$ thermal excitations are the
main unblocking mechanism (plus contributions from other multipoles
than Gamow-Teller). For $N=40$, the finding is supported by
experimental data, although yet quite limited.  It would be desirable
if the data pool could be enlarged. It is particularly tempting that
recent developments open up the measurements of GT distributions for
unstable neutron-rich nuclei, based on charge-exchange reactions
performed in inverse kinematics. Such additional data would certainly
be welcome to further constrain models. At $N=50$, theoretical models
imply that cross-shell correlations induced by thermal excitation
render ground state GT distributions not applicable for the
calculation of capture rates at the finite temperatures which exist in
the astrophysical environment when these heavier and very neutron-rich
nuclei dominate the capture process.  Although the two models which
have been applied to $N=50$ nuclei agree rather well in their rate
predictions, improvements of the models are conceivable.  On one hand,
the finite-temperature QRPA model should be extended to non-spherical
nuclei and, in the midterm, also to include higher correlations like
in second QRPA approach.  On the other hand, the Shell Model Monte
Carlo approach is uniquely suited to study nuclear properties at the
finite temperatures of relevance. It might be interesting to calculate
the GT strength function at those temperatures directly within the
SMMC approach. This presupposes the handling of a numerically
ill-defined inverse Laplace transformation. First steps in this
direction have been taken in Ref.~\cite{Radha.Dean.ea:1996}.

Of course, it is always conceivable that observations or astrophysical
simulations of supernovae or other astrophysical objects point to the
need of particular electron capture rates which then require specific
experimental and theoretical attention.

In summary, the description of stellar electron capture has come a
long and winding way.  The experimental and theoretical progress of
recent years has probably firmly established that electron capture
proceeds on nuclei during core-collapse supernovae. The circle, as
attributed to by Hans Bethe, might have come to an end.

\section*{Acknowledgements}

The authors gratefully acknowledge support and assistance by their
colleagues Sam M.~Austin, B.~A.~Brown, E.~Caurier, S.~Couch,
D.~J.~Dean, J.~Engel, T.~Fischer, Y.~Fujita, A.~Heger, H.-T.~Janka,
A.~Juodagalvis, R.~W.~Hix, E.~Kolbe, O.~Kirsebom, A.~Mezzacappa,
F.~Nowacki, E.~O'Connor, A.~Poves, J.~Sampaio, H.~Schatz, A.~Spyrou,
T.~Suzuki, F.-K.~Thielemann, S.~E.~Woosley, Y.~Zhi, and A.~P.~Zuker.
G.M.P. acknowledges the support of the Deutsche Forschungsgemeinschaft
(DFG, German Research Foundation) -- Project-ID 279384907 -- SFB 1245
``Nuclei: From Fundamental Interactions to Structure and Stars'' and
the ``ChETEC'' COST Action (CA16117), funded by COST (European Cooperation
in Science and Technology).  R.Z. gratefully acknowledges support by
the US National Science Foundation under Grants PHY-1913554 (Windows
on the Universe: Nuclear Astrophysics at the NSCL), PHY-1430152 (JINA
Center for the Evolution of the Elements), and PHY-1927130
(AccelNet-WOU: International Research Network for Nuclear Astrophysics
[IReNA])

\section*{References}
\bibliographystyle{iopart-num}
\bibliography{refs}

\end{document}